\documentclass[aps,prd,superscriptaddress,amsmath,amssymb,amsthm,nofootinbib,eqsecnum,preprintnumbers]{revtex4}
\usepackage{xcolor}
\usepackage{url}
\usepackage[colorlinks=true,
    linkcolor=blue,
    citecolor=blue,
    urlcolor=blue]{hyperref}
 \usepackage{amsmath}
\usepackage{graphicx}
\usepackage{epstopdf}
\usepackage{float}
\usepackage{hyperref}
\usepackage{color}
\usepackage[T1]{fontenc}
\usepackage[utf8]{inputenc}
\usepackage[toc,page]{appendix}
\usepackage[usenames,dvipsnames]{xcolor}
\usepackage[normalem]{ulem}

\usepackage{tikz}
\usetikzlibrary{shapes,positioning,shadows,graphs.standard,automata,arrows}
\usepackage{pgfplots}
\pgfplotsset{compat=newest, width=2.669cm, height=2.669cm, scale only axis=true,enlargelimits=false}
\pgfplotsset{tick label style={font=\tiny}}
\pgfplotsset{every major tick/.append style={major tick length=3pt}}
\pgfplotsset{every minor tick/.append style={minor tick length=1.5pt}}
\usepgfplotslibrary{groupplots}
\usepackage{subcaption}
\usepackage{caption}

\providecommand{\renewoperator}[3]{%
\renewcommand*{#1}{\mathop{#2}#3}}

\makeatletter
\providecommand*{\diff}%
{\@ifnextchar^{\DIfF}{\DIfF^{}}}
\def\DIfF^#1{%
\mathop{\mathrm{\mathstrut d}}%
\nolimits^{#1}\gobblespace}
\def\gobblespace{%
\futurelet\diffarg\opspace}
\def\opspace{%
\let\DiffSpace\!%
\ifx\diffarg(%
\let\DiffSpace\relax
\else
\ifx\diffarg[%
\let\DiffSpace\relax
\else
\ifx\diffarg\{%
\let\DiffSpace\relax
\fi\fi\fi\DiffSpace}

\renewoperator{\Re}{\mathrm{Re}}{\nolimits}
\renewoperator{\Im}{\mathrm{Im}}{\nolimits}

\usepackage{bm}

\newcommand{\be}{\begin{equation}}
\newcommand{\ee}{\end{equation}}
\newcommand{\ba}{\begin{eqnarray}}
\newcommand{\ea}{\end{eqnarray}}

\newcommand{\beq}{\begin{equation}}
\newcommand{\eeq}{\end{equation}}
\newcommand{\beqa}{\begin{eqnarray}}
\newcommand{\eeqa}{\end{eqnarray}}

\def\ha{{\hat a}}
\def\hb{{\hat b}}
\def\d{{\rm d}}

\usepackage{xcolor}
\usepackage{scalerel}

\usepackage{tikz}
\usetikzlibrary{svg.path}
\definecolor{darkmagenta}{HTML}{8B008B}
\definecolor{orcidlogocol}{HTML}{A6CE39}
\tikzset{
  orcidlogo/.pic={
    \fill[orcidlogocol] svg{M256,128c0,70.7-57.3,128-128,128C57.3,256,0,198.7,0,128C0,57.3,57.3,0,128,0C198.7,0,256,57.3,256,128z};
    \fill[white] svg{M86.3,186.2H70.9V79.1h15.4v48.4V186.2z}
                 svg{M108.9,79.1h41.6c39.6,0,57,28.3,57,53.6c0,27.5-21.5,53.6-56.8,53.6h-41.8V79.1z M124.3,172.4h24.5c34.9,0,42.9-26.5,42.9-39.7c0-21.5-13.7-39.7-43.7-39.7h-23.7V172.4z}
                 svg{M88.7,56.8c0,5.5-4.5,10.1-10.1,10.1c-5.6,0-10.1-4.6-10.1-10.1c0-5.6,4.5-10.1,10.1-10.1C84.2,46.7,88.7,51.3,88.7,56.8z};
  }
}

\newcommand\orcidicon[1]{\href{https://orcid.org/#1}{\mbox{\scalerel*{
\begin{tikzpicture}[yscale=-1, scale=0.05, transform shape]
\pic{orcidlogo};
\end{tikzpicture}
}{|}}}}

\begin{document}

\title{ModMax electrodynamics and holographic magnetotransport}

\author{Jos{\'e} Barrientos\,$^{\orcidicon{0000-0003-3445-8151}}$}
\email{jbarrientos@academicos.uta.cl}
\affiliation{Sede Esmeralda, Universidad de Tarapac{\'a}, Avenida Luis Emilio Recabarren 2477, Iquique, Chile}
\affiliation{Institute of Mathematics of the Czech Academy of Sciences, {\v Z}itn{\'a} 25, 115 67 Prague 1, Czech Republic}

\author{Nicol\'as C\'aceres\,$^{\orcidicon{0009-0005-7381-0244}}$}
\email{ntcaceres@puc.cl}
\affiliation{Facultad de Física, Pontificia Universidad Cat\'olica de Chile, Avenida Vicu{\~n}a Mackenna 4860, Santiago, Chile}

\author{Felipe Diaz\,$^{\orcidicon{0000-0003-4107-7162}}$}
\email{fdiaz@itmp.msu.ru}
\affiliation{Institute for Theoretical and Mathematical Physics, Lomonosov Moscow State University, 119991 Moscow, Russia}

\author{Ulises Hernandez-Vera\,$^{\orcidicon{0009-0009-0193-778X}}$}
\email{uhernandez.vera@gmail.com}
\affiliation{Instituto de Matem\'aticas, Universidad de Talca, Casilla 747, Talca 3460000, Chile}


\begin{abstract}
We study magnetotransport in a holographic model where ModMax nonlinear electrodynamics is coupled to Einstein anti--de Sitter gravity. To incorporate momentum relaxation, we introduce spatially linear free scalar fields that break translational symmetry, resulting in an anisotropic medium. Using linear response theory, we compute the DC conductivity matrix in the presence of an external magnetic field, expressing the conductivities in terms of horizon data. Our results demonstrate how the nonlinear ModMax parameter modifies charge transport, particularly influencing the Hall angle and Nernst signal. The nonlinear corrections introduce distinct deviations in both longitudinal and Hall conductivities while preserving the characteristic temperature scaling of strange metals, offering new insights into strongly coupled systems with nonlinear electromagnetic interactions. Notably, the Nernst signal reproduces that of high-$T_c$ cuprate superconductors showing a superconducting dome and a normal phase, with the ModMax deformation parameter tuning critical and onset temperatures. In the strongly nonlinear regime, we find evidence of an exotic state dominated by quasiparticle excitations in the dual material.

\end{abstract}
\maketitle
\tableofcontents
\section{Introduction}

Holography \cite{tHooft:1993dmi, Susskind:1994vu}, particularly the AdS/CFT correspondence \cite{Maldacena:1997re, Witten:1998qj, Gubser:1998bc}, provides a powerful framework for studying strongly correlated systems. It is a unique approach in which open questions become more computationally manageable and conceptually transparent. 
In particular, exploiting the fact that certain macroscopic phenomena can be captured without detailed knowledge of the underlying microscopic physics---a property often referred to as UV-independence---holography has been fruitfully applied to condensed matter theory (CMT). It does so by mapping complex quantum many-body systems onto gravitational theories in higher-dimensional spacetimes. For comprehensive reviews and an extensive list of references, see \cite{Hartnoll:2009sz, Zaanen:2015oix, Hartnoll:2016apf}.

A particularly remarkable achievement in this area is the ability to model high-$T_c$ superconductors using holographic techniques \cite{Hartnoll:2008vx,Hartnoll:2008kx, Horowitz:2008bn}, see \cite{Horowitz:2010gk} for a review. This begins with introducing a charged topological black hole in the bulk,\footnote{{The term \textit{topological black holes} has historically been used to describe black holes whose horizons exhibit nontrivial topology (see, for example, \cite{Mann:1997iz}).}} which places the boundary theory on a Minkowski background geometry and at finite density by sourcing a global $U(1)$ symmetry with an associated chemical potential. If the black hole is nonextremal, the dual field theory is also in a thermal state,\footnote{For a detailed analysis of the zero-temperature limit of the holographic superconductor, see \cite{Horowitz:2009ij}.} as the Hawking temperature of the black hole is identified with the temperature of the boundary theory \cite{Witten:1998zw}. The system also requires to dress the black hole horizon with quantum fields, so the model is further coupled to a massive, complex, charged scalar field that condenses near the black hole horizon, leading to the spontaneous breaking of the Abelian gauge symmetry \cite{Gubser:2005ih, Gubser:2008px} and signaling the onset of superconductivity in the boundary theory. The framework allows us to study superconductivity in strongly coupled regimes, where traditional Bardeen–Cooper–Schrieffer (BCS) theory may fail. 
The bulk theory has been extensively developed to capture a range of phenomena associated with 
$(2+1)$-dimensional superconductors and strange metals. These extensions include higher-curvature corrections \cite{Gregory:2009fj,Zeng:2014uoa}, nonminimal scalar couplings \cite{Zhou:2015dha,Jiang:2017imk, Cisterna:2017qrb, Cisterna:2018mww,Cisterna:2018hzf,Cisterna:2019uek,Caldarelli:2013gqa, Baggioli:2021ejg, Hernandez-Vera:2024zui}, the incorporation of fermionic excitations \cite{Chen:2009pt,Faulkner:2009am,Faulkner:2010da,Faulkner:2010tq,Faulkner:2011tm,Faulkner:2013bna,Grandi:2021bsp, Grandi:2021jkj, Grandi:2023jna,Bahamondes:2024zsm}, addition of topological terms in the gauge sector \cite{Lopez-Arcos:2013uga, Andrade:2017ghg, Alejo:2019utd}, and modifications to scalar field profiles to model lattice effects in the boundary theory \cite{Donos:2013eha,Andrade:2013gsa, Donos:2014uba} (see e.g., \cite{Baggioli:2021xuv} for a review). 
A particularly notable extension involves incorporating nonlinearities in the gauge sector.
Nonlinear electrodynamics (NLE) refers to classical extensions of Maxwell’s theory that are significant in strong-field regimes while smoothly reducing to Maxwell’s electrodynamics in the weak-field limit \cite{Sorokin:2021tge}. The origins of these theories can be traced back to the works conducted by Born and Infeld \cite{Born:1934gh}, who aimed to remove the divergence associated with the self-energy of the electron, and Euler and Heisenberg \cite{Heisenberg:1936nmg}, who developed a one-loop effective action for quantum electrodynamics incorporating vacuum polarization effects from virtual electron-positron pairs. 
NLE Lagrangians can be consistently incorporated into the holographic framework, thereby extending the range of dual field theories accessible within this approach. This inclusion is well-motivated as in the IR limit, the string theory partition function acquires higher-curvature corrections, and Kaluza-Klein reductions naturally yield couplings between nonlinear gauge sectors and gravity \cite{Gibbons:1995cv}. Additionally, nonlinear kinetic terms for vector fields arise from the quantization of string actions \cite{Fradkin:1985qd}.
Within the gauge/gravity correspondence, nonlinear electrodynamics has been extensively employed (see, e.g., \cite{Miskovic:2010ey, Miskovic:2010ui, Dey:2016pei, Cano:2022ord, Arenas-Henriquez:2022ntz, Bravo-Gaete:2022mnr, Bueno:2022ewf, Alvarez:2022upr, Santos:2023mee, Bravo-Gaete:2023iry, Santos:2023flb, Colipi-Marchant:2023awk, Bravo-Gaete:2025vyd, Barbosa:2025smt}) as it allows for nontrivial modifications of the bulk geometry. Several notable analytic solutions incorporating nonlinear gauge fields can be found in \cite{Soleng:1995kn, Ayon-Beato:1998hmi, Gonzalez:2009nn, Bronnikov:2000vy, Ayon-Beato:2000mjt, Cataldo:2000we, Hassaine:2007py,Hassaine:2008pw,Maeda:2008ha, Cembranos:2014hwa, Barrientos:2016ubi, Cisterna:2016nwq, EslamPanah:2022ihg}. In the context of holographic superconductors, nonlinear electrodynamics has also been widely explored to access different physical regimes and to alter transport properties in the boundary theory (see, for example, \cite{Jing:2010zp, Jing:2011vz, Gangopadhyay:2012gx, Liu:2015lit,Baggioli:2016oju, Wang:2018hwg, Chakraborty:2019vld, An:2020tkn, Lai:2021mxa}).

In four dimensions, two of Maxwell theory’s most notable properties are the electromagnetic duality of its field equations and the conformal invariance of its action.
Recently, the goal of preserving these symmetries beyond the linear regime led to the development of ModMax theory \cite{Bandos:2020jsw, Kosyakov:2020wxv}; a one-parameter extension of Maxwell electrodynamics in four dimensions.
ModMax emerges in the conformal limit of Dirac-Born-Infeld (DBI) nonlinear electrodynamics and also arises in higher-rank gauge field generalizations within string and M-theory frameworks \cite{Bandos:2020hgy} (see also \cite{Nastase:2021ffq}).
As we will show in \autoref{Sec:Thermo}, the ModMax action reduces on shell to standard Maxwell theory in configurations that either lack magnetic fields or exhibit a linear relation between the Lorentz scalar and pseudoscalar invariants [see Eq. \eqref{SModMax}]. Consequently, it is essential to introduce magnetic fields in the bulk to probe the holographic implications of the theory’s nonlinear structure.

Introducing magnetic fields in the AdS/CMT correspondence significantly enriches the structure of the boundary theory, allowing for the investigation of more realistic condensed matter phenomena, such as the Hall effect \cite{Hartnoll:2007ip, Hartnoll:2007ai, Blake:2014yla, Amoretti:2015gna,Amoretti:2020mkp,Amoretti:2021fch,Amoretti:2021lll,Ge:2023yom}, the Meissner effect \cite{Domenech:2010nf,Montull:2011im,Montull:2012fy,Salvio:2012at,Salvio:2013jia,Natsuume:2022kic}, and the Nernst effect \cite{Hartnoll:2007ih, Kim:2015wba}. 
The Hall angle quantifies the deflection of charge carriers under the influence of a magnetic field, indicating how much the electric current deviates from the direction of the applied electric field. In cuprate strange metals, the Hall angle exhibits an anomalous temperature dependence \cite{PhysRevLett.67.2088, TYLER19971185}, a feature that has been successfully reproduced in holographic scalar-vector models (see for instance \cite{Blake:2014yla,Blauvelt:2017koq, Ahn:2023ciq}).
The Nernst effect is a thermoelectric phenomenon observed in conducting materials, where an electric field is generated perpendicular to both an applied magnetic field and a thermal gradient. 
It is quantified by the Nernst coefficient, which measures the transverse voltage produced in response to these perturbations. Remarkably, it can be used to probe high-$T_c$ superconductors as the Nernst effect shows that the normal phase of cuprate superconductor is aberrant \cite{xu2000vortex,wang2005field,li2007low, wang2006nernst}. In \cite{wang2006nernst}, it is shown that the Nernst signal becomes bell-shaped for these materials and shows a linear decay after the onset temperature. The sign of the Nernst signal indicate different phases for high-$T_c$ superconductors; if the signal is positive indicates that the system is in a vortex-liquid phase with mobile vortices that carry entropy, and a temperature gradient causes them to move generating perpendicular magnetic fields contributing to the signal, and if the signal becomes negative is usually associated with quasiparticle excitations \cite{PhysRevB.70.054503}. 

Despite the simplicity of ModMax theory---stemming from the maximal symmetry of its Lagrangian---its holographic implications remain largely unexplored.\footnote{For recent holographic applications, see \cite{Tartaglione:2022zye, Rathi:2023vhw}.} Indeed, since the theory preserves the Maxwellian $SO(2)$ duality rotation symmetry, the dual field theory automatically respects particle-vortex duality \cite{Hartnoll:2007ai} that allows to describe Drude-like conductivities \cite{damle1997nonzero}. Given its tractable structure, ModMax provides a compelling framework to probe nonlinear effects in the gauge/gravity correspondence. In this work, we couple ModMax electrodynamics to anti--de Sitter (AdS) gravity and investigate the magnetotransport properties of the dual field theory. We find that the theory extends the landscape of holographic magnetotransport beyond Maxwell electrodynamics and may offer insights into exotic phases of quantum matter. Of particular interest is the highly nonlinear regime, which has not been previously studied in this context. We also discuss how ModMax nonlinearities can be leveraged to more realistically model high-$T_c$ cuprate superconductors, as the deformation parameter $\gamma$ allows for distinct behavior between hole-doped and electron-doped systems.

This paper is organized as follows. In  \autoref{Sec:Thermo}, we introduce the gravitational bulk theory consisting of AdS gravity coupled to ModMax nonlinear electrodynamics, and study topological dyonic black hole solutions along with their holographic thermodynamics. We show that the theory reduces on shell to standard Maxwell electrodynamics for the static configurations considered. In \autoref{subsec:isotropic}, we analyze time-dependent perturbations in the hydrodynamic limit to compute the DC conductivities using Kubo formulas, finding that the resulting magnetotransport describes a boundary theory in a perfect Hall state. In \autoref{subsec:MomRelax}, we include scalar fields with linear transverse profiles to break translational symmetry and model lattice effects in the dual theory. We compute the Hall angle and the Nernst signal, showing that the system exhibits behaviors similar to those observed in high-$T_c$ superconductors, with the deformation parameter enabling access to novel regimes beyond the reach of standard Maxwell electrodynamics. Finally, in \autoref{Sec:Conclusions}, we
conclude with open questions and potential future directions.

\section{Dyonic Black Holes and Holographic Thermodynamics}\label{Sec:Thermo}

We are interested in constructing dyonic black holes by considering NLE theories coupled to AdS gravity. In this framework, nonlinear corrections to the bulk $U(1)$ gauge field modify the boundary transport properties, offering new avenues for exploring holographic phenomena.
A particularly simple yet intriguing NLE theory is ModMax electrodynamics, which has attracted growing attention in gravitational contexts, especially in black hole physics. When coupled to general relativity, it gives rise to new classes of charged black holes, including spherically symmetric solutions that generalize the dyonic Reissner-Nordstr\"{o}m metric \cite{Flores-Alfonso:2020euz, Amirabi:2020mzv}. Notably, a Melvin-Bonnor-ModMax background and its (A)dS generalizations have enabled the embedding of Schwarzschild and C-metric black holes into an electromagnetic universe, representing the first black hole solutions embedded in an electromagnetic universe in the context of nonlinear electrodynamics \cite{Barrientos:2024umq}.
In these static configurations, the electromagnetic invariants satisfy a proportionality relation. This property ensures that all such Maxwell solutions automatically solve the ModMax field equations, effectively reducing the Einstein-ModMax system to the Einstein-Maxwell system with a rescaled Newton constant. 
As we will show, a consequence of this feature is that key physical properties---such as black hole thermodynamics, phase transitions, and potentially their holographic behavior---remain unaffected, aside from modifications to the effective couplings.

The rich symmetry structure and inherent nonlinearities of the theory make it an appealing framework for exploring holographic models at finite density with nonlinear corrections. We therefore consider ModMax electrodynamics minimally coupled to four-dimensional AdS gravity, described by the renormalized action,
\begin{equation}\label{SEMM}
    S_{\rm ren} = S_{\rm EH}+S_{\rm MM} + S_{\rm bdry}\,, 
\end{equation}
where the Einstein-Hilbert action is
\begin{equation}
    S_{\rm EH} = \frac{1}{16\pi G}\int_{\cal M}{\rm d}^4x \sqrt{-g}\left(R+\frac{6}{\ell^2}\right)\,,
\end{equation}
with $\ell$ denoting the AdS radius. The ModMax contribution reads, 
\begin{equation}
\begin{aligned}\label{SModMax}
    S_{\rm MM} =-{}& \frac{1}{4\pi G}\int_{\cal M}{\rm d}^4x \sqrt{-g}{\cal L}_{\rm MM} \\  
    =-{}& \frac{1}{8\pi G}\int_{\cal M}{\rm d}^4x \sqrt{-g}\left({\cal S}\cosh\gamma - \sqrt{{\cal S}^2 + {\cal P}^2}\sinh\gamma\right)\,,
\end{aligned}
\end{equation}
where ${\cal S} = \tfrac{1}{2} F_{\mu\nu} F^{\mu\nu}$ and ${\cal P} = \tfrac{1}{4} \epsilon_{\mu\nu}{}^{\lambda\rho} F_{\lambda\rho} F^{\mu\nu}$ are the Lorentz scalar and pseudoscalar invariants, respectively. Here, $F_{\mu\nu} = 2 \partial_{[\mu} A_{\nu]}$ is the field strength of the $U(1)$ gauge field $A_\mu$\,, and $\epsilon_{\mu\nu\rho\sigma}$ is the Levi-Civita tensor. 
The dimensionless ModMax deformation parameter $\gamma$ is restricted to a positive number as causality and unitarity require. This bound also guarantees the convexity of the Lagrangian with respect to the electric field \cite{Bandos:2020jsw}. 
The boundary action is added to ensure a well-defined variational principle and finiteness of the on shell action \cite{deHaro:2000vlm} and is given by
\begin{equation}
    S_{\rm bdry} = \frac{1}{8\pi G}\int_{\partial\cal M}{\rm d}^3x \sqrt{-h}\left[K-\frac{2}{\ell} + \frac{\ell}{2}{\cal R}(h)\right]\,,
\end{equation}
where $h_{ij}$ is the induced metric at $r=\infty$ with scalar curvature ${\cal R}(h)$\,, and $K = h^{ij}K_{ij}$ is the trace of the extrinsic curvature along the holographic coordinate. Note that we do not include a boundary term for the matter sector, as the ModMax Lagrangian vanishes sufficiently fast on shell near the boundary for the class of configurations considered in this work. 

The field equations read,
\begin{equation}
\begin{aligned}\label{EOM}
     \nabla_{\mu}\left({\cal L}_{\cal S}F^{\mu\nu} +\frac12 \epsilon_{\lambda\rho}{}^{\mu\nu}{\cal L}_{\cal P}F^{\lambda\rho}\right) ={}&0\,, \\ 
R_{\mu\nu}-\frac12  g_{\mu\nu}R - \frac{3}{\ell^2}g_{\mu\nu} - 8\pi G T_{\mu\nu} ={}& 0\,,
\end{aligned}
\end{equation}
where 
\begin{equation}
    {\cal L}_{\cal S}:= \frac{\partial {\cal L}_{\rm MM}}{\partial \cal S}\,,\qquad {\cal L}_{\cal P}:= \frac{\partial {\cal L}_{\rm MM}}{\partial \cal P}\,,
\end{equation}
and the (traceless) stress tensor reads
\begin{equation}\label{Tmunu}
8\pi G T_{\mu\nu} := -\frac{2}{\sqrt{g}}\frac{\delta S_{\rm ren}}{\delta g^{\mu\nu}} =   4{\cal L}_{\cal S}F_{\mu}{}^{\alpha}F_{\nu\alpha} + 2\left({\cal P}{\cal L}_{\cal P} - {\cal L}_{\rm MM}\right)g_{\mu\nu}\,.
\end{equation}

An exact dyonic black hole solution to the field equations is described by the line element,
\begin{equation}\label{defm}
    {\rm d}s^2 = -f(r) {\rm d}t^2 + \frac{{\rm d}r^2}{f(r)} + r^2 \gamma_{\ha\hb}^{(k)}{\rm d}x^\ha {\rm d}x^\hb\,,
\end{equation}
where
\begin{equation}
    f(r) = k+\frac{r^2}{\ell^2} - \frac{2m}{r} + \frac{(q_{\rm E}^2+q_{\rm M}^2)}{r^2}e^{-\gamma}\,, 
\end{equation}
and $\gamma_{\ha\hb}^{(k)}$\,, with indices $\hat{a} = 1,2$\,, is the metric of the codimension-2 Killing horizon surface  $\Gamma$ parametrized by $k=0,1,-1$ describing a flat, spherical, hyperbolic topology, respectively, i.e.,   \begin{equation}
    {\rm d}\Omega_k^2 \equiv \gamma_{\ha\hb}^{(k)}{\rm d}x^\ha {\rm d}x^\hb = \begin{cases}
        {\rm d}\theta^2 + \sin^2\theta {\rm d}\phi^2\,, &k=1\,, \\ {\rm d}x^2+{\rm d}y^2\,,&k=0\,, \\ {\rm d}\rho^2 + \sinh^2\rho {\rm d}\varphi^2\,, &k=-1\,,
    \end{cases}
\end{equation}
and, modulo constant terms, the gauge field reads,
\begin{equation}
    A = \begin{cases}
    \frac{q_{\rm E} }{r}e^{-\gamma}{\rm d}t +q_{\rm M} \cos\theta {\rm d}\phi\,,&k=1\,, \\  \frac{q_{\rm E} }{r}e^{-\gamma}{\rm d}t +q_{\rm M} x {\rm d}y\,,&k=0\,, \\ \frac{q_{\rm E} }{r}e^{-\gamma}{\rm d}t +q_{\rm M} \cosh\rho {\rm d}\varphi\,, &k=-1\,.
    \end{cases}
\end{equation}

The solution has an electric and magnetic charge that can be computed using Gauss's law,\footnote{They can also be obtained by computing the electromagnetic flux across the black hole horizon \cite{Ferrero:2020twa}, finding the same results.}
\begin{equation}\label{QEQM}
    Q_{\rm E} = \frac{1}{4\pi G}\oint_{\Sigma_\infty} \star E = \frac{\omega_k }{4\pi G}q_{\rm E}\,,\qquad Q_{\rm M} = \frac{1}{4\pi G}\oint_{\Sigma_\infty} F = \frac{\omega_k }{4\pi G}q_{\rm M}\,,
\end{equation}
where $\star$ indicates Hodge-dualization, $\omega_k$ is the volume of the horizon two-surface, which $\omega_1 = 4\pi$ and divergent otherwise, and
\begin{equation}
    E^{\mu\nu}:= \frac{\partial \cal L}{\partial F_{\mu\nu}}\,.
\end{equation}
This explicitly shows the model's electromagnetic duality $q_{\rm E} \leftrightarrow q_{\rm M}$\,. 

The quasilocal stress tensor,
\begin{equation}
    \tau_{ij} := -\frac{2}{\sqrt{-h}}\frac{\delta S_{\rm ren}}{\delta h^{ij}} =-\frac{1}{8\pi G}\left( K_{ij}-Kh_{ij}+\frac{2}{\ell}h_{ij} - \ell G_{ij}(h)\right)\,,
\end{equation}
gives the holographic stress tensor as \cite{Myers:1999psa}
\begin{equation}
    \langle T_{ij}\rangle = \lim_{r\to \infty}\frac{r}{\ell}\tau_{ij} = \frac{m}{8\pi G}{\rm diag}\left[2\ell^{-2},1,{\mathcal V}\right]\,,\qquad {\mathcal V} \equiv \begin{cases}
         \sin^2\theta\,,&k=1\,,\\ 1\,,&k=0\,,\\\sinh^2\rho&k=-1\,,
    \end{cases}
\end{equation}
that is traceless and covariantly conserved, which can be written as the one of a perfect conformal fluid,
\begin{equation}
    \langle T_{ij}\rangle = \frac{\epsilon}{2}\left(g_{ij}^{(0)}+3u_iu_j \right)\,, 
\end{equation}
where the fluid three-velocity
\begin{equation}
    {u^i\partial_i = \frac{1}{\sqrt{-g_{tt}^{(0)}}}\partial_t}\,,\qquad u_iu_jg^{ij}_{(0)}=-1\,.
\end{equation}

Similarly, the electromagnetic current of the dual theory is defined as
\begin{equation}
\begin{aligned}
    \langle J^{i}\rangle :={}& \frac{1}{\sqrt{-g^{(0)}}}\frac{\delta S_{\rm ren}}{\delta A_i^{(0)}}  = -\frac{1}{\sqrt{-g^{(0)}}}\lim_{r\to\infty}\left[\frac{\sqrt{-g}}{2\pi G}\left({\cal L}_{\cal S}F^{ri} + \frac12  {\cal L}_{\cal P}\epsilon_{r\alpha\beta\gamma}g^{rr}g^{\gamma i}F^{\alpha\beta}\right)\right]  \\ ={}& \frac{q_{\rm E}}{4\pi G \ell^2}\delta^i{}_t\,,
\end{aligned}
\end{equation}
where 
\begin{equation}
    A^{(0)} = \lim_{r\to\infty}A_i {\rm d}x^i =\begin{cases}
        q_{\rm M}\cos\theta{\rm d}\phi\,,&k=1\,,\\ q_{\rm M}x{\rm d}y\,,&k=0\,,\\ q_{\rm M} \sinh\rho {\rm d}\varphi\,,&k=-1\,,
    \end{cases}
\end{equation}
is the boundary gauge field, and
\begin{equation}
g^{(0)}_{ij} = \lim_{r\to\infty}\frac{\ell^2}{r^2}h_{ij}
\end{equation}
is the boundary metric.
One can check that the holographic quantities satisfy the Ward identities
\begin{equation}
\begin{aligned}
    0={}&\nabla_i^{(0)}\langle J^i\rangle\,, \\0={}&\nabla^i_{(0)}\langle T_{ij}\rangle+\langle J^i\rangle F^{(0)}_{ij}\,,
\end{aligned}
\end{equation}
with $\nabla_i^{(0)}$ the Levi-Civita covariant derivative with respect to $g^{(0)}$\,, and $F^{(0)}_{ij}$ the curvature of the boundary gauge field.

We can compute the holographic energy by using the fact that if the boundary metric has a Killing vector field $\hat\xi=\hat\xi^i\partial_i$ satisfying ${\cal L}_{\hat\xi} g^{(0)}_{ij} = 0 = {\cal L}_{\hat\xi} A^{(0)}$\,, then we have the following identity:
\begin{equation}
\nabla_i^{(0)}\left[\left(\langle T^i{}_j\rangle + \langle J^i\rangle A^{(0)}_j\right)\hat\xi^j\right]=0\,.
\end{equation}
Then, we can compute conserved quantities by integrating the conserved current $j^i = \langle T^i{}_j\rangle + \langle J^i\rangle A^{(0)}_j$ over a two-dimensional spatial hypersurface $\hat{\Sigma}_\infty$ as
\begin{equation}
    Q_\xi = \oint_{\hat{\Sigma}_\infty}{\rm d}^2x \sqrt{-\hat{\sigma}}\hat{u}_i j^i\,,
\end{equation}
where the spacelike surface $\hat{\sigma}$ is described by using the Arnowitt-Deser-Misner (ADM) decomposition of the boundary metric
\begin{equation}
    {\rm d}s_{(0)}^2 = -\hat{N}^2 {\rm d}t^2+ \hat{\sigma}_{\hat{a}\hat{b}}\left({\rm d}x^{\hat{a}} +\hat{N}^{\hat{a}}{\rm d}t\right)\left({\rm d}x^{\hat{b}} +\hat{N}^{\hat{b}}{\rm d}t\right)\,,
\end{equation}
and $\hat{\sigma}_{\hat{a}\hat{b}}$ is the induced metric on $\hat{\Sigma}_{\infty}$\,, with future-directed unit normal vector 
\begin{equation}
    \hat{u} \equiv \hat{u}^i\partial_i = \frac{1}{\hat{N}}\left(\partial_t -\hat{N}^{\hat{a}}\partial_{\hat{a}}\right)\,.
\end{equation}
In the case at hand, the only nontrivial conserved quantity is the holographic energy associated with $\hat{\xi} = \partial_t$ that gives
\begin{equation}\label{HoloM}
    M = \oint_{\hat\Sigma_\infty} {\rm d}^2y\sqrt{-\hat{\sigma}}\hat{u}_i \left(\langle T^i{}_t \rangle + \langle J^i\rangle A_t^{(0)}\right)\hat\xi^t = \oint_{\hat\Sigma_\infty}{\rm d}^2y\sqrt{-\hat{\sigma}} \langle T^t{}_t \rangle =\frac{\omega_k}{4\pi}\frac{m}{G}\,.
\end{equation}

One can easily check that the energy matches the one found using conformal methods \cite{Ashtekar:1999jx}. Consider a Weyl rescaling of the metric $\hat{g} = \hat\Omega^2 g$ to remove the boundary divergences of the line element and construct a conserved charge,
\begin{equation}
    Q(\xi) = \frac{\ell}{8\pi G}\lim_{\hat\Omega\to0}\oint_{\hat\Sigma_\infty} {\rm d}{\hat S}^\mu\hat{n}^\alpha \hat{n}^\beta \hat{W}^\nu{}_{\alpha\mu\beta}\xi_\nu\,,
\end{equation}
where ${\hat W}^\mu{}_{\alpha\beta\gamma}$ is the Weyl tensor of $\hat g$\,, $\hat{n}_\mu = \frac{\ell\partial_\mu \hat\Omega}{\sqrt{|\hat\Omega|}}$ is the outward pointing normal to the boundary, and ${\rm d}{\hat S}^\mu = \ell^2 {\rm d}\Omega^2_k \delta^\mu_t$ is the spacelike surface element tangent to $\hat\Omega = 0$\,. We find for $\hat\Omega = r^{-1}$\,,
\begin{equation}
    Q(\partial_t) = \frac{\omega_k}{4\pi }\frac{m}{G} = M\,,
\end{equation}
in agreement with the holographic energy \eqref{HoloM}.

The electric and magnetic charges can also be computed with these boundary quantities by 
\begin{equation}
    Q_{\rm E} = -\oint_{\hat\Sigma_\infty}{\rm d}^2x\sqrt{-\sigma}\hat{u}_i \langle J^i\rangle = \frac{1}{4\pi G}\oint_{\hat\Sigma_\infty}\star E = \frac{\omega_k}{4\pi G}q_{\rm E}
\end{equation}
and the magnetic charge
\begin{equation}
    Q_{\rm M} = \frac{1}{4\pi G}\oint_{\hat\Sigma_\infty}F_{(0)} = \frac{\omega_k}{4\pi G}q_{\rm M}\,,
\end{equation}
in agreement with \eqref{QEQM}.

A particularly interesting solution regime corresponds to considering the ultranonlinear regime $\gamma \gg q_{\rm E}$\,. In this limit, the geometry recovers that of AdS-Schwarzschild's black holes, and the gauge field becomes purely magnetic, ceasing to contribute to the on shell action. This would correspond to a magnetic stealth charge over the uncharged black hole. Notice that in the dual theory, the gauge potential $A^{(0)}$ and the vacuum expectation value of the corresponding $U(1)$ current, $\langle J^i\rangle$\,, remain both constant in this limit. In this regime, we notice that the solution resembles the one found perturbatively by Wald for describing black holes in magnetic backgrounds generated by accretion disks \cite{Wald:1974np}. 

To proceed with the thermodynamic analysis, we consider the Euclidean continuation of the solution, where the Hawking temperature reads
\begin{equation}\label{temp1}
    T = \frac{f'(r_+)}{4\pi} = \frac{3 r_+^4+\ell ^2 \left(kr_+^2-Q^2\right)}{4 \pi  r_+^3 \ell ^2} = \frac{3 r_+^2+\ell ^2 \left(k-e^{\gamma } \left(\Phi_{\rm E}^2+\Phi_{\rm M}^2\right)\right)}{4 \pi \ell^2r_+}\,,
\end{equation}
where 
\begin{equation}
Q^2 \equiv e^{-\gamma}(q_{\rm E}^2+q_{\rm M}^2)\,,
\end{equation}
with a conjugate Bekenstein-Hawking entropy, 
\begin{equation}
    S = \frac{{\rm Area}(\Gamma)}{4G} = \frac{\omega_k}{4G}r_+^2\,.
\end{equation}

Using \eqref{HoloM}, we can check that the thermodynamic quantities satisfy the first law of black hole thermodynamics
\begin{equation}
    {\rm d}M = T{\rm d}S + \Phi_{\rm E} {\rm d}Q_{\rm E} + \Phi_{\rm M} {\rm d}Q_{\rm M} + V{\rm d}P\,,
\end{equation}
where 
\begin{equation}
        \Phi_{\rm E} = \left(\frac{\partial M}{\partial q_{\rm E}} \right)_{q_{\rm M},\beta,P}=\frac{q_{\rm E}}{r_+}e^{-\gamma}\,,\qquad \Phi_{\rm M} = \left(\frac{\partial M}{\partial q_{\rm M}} \right)_{q_{\rm E},\beta,P}=\frac{q_{\rm M}}{r_+}e^{-\gamma}\,,
\end{equation}
are the electric and magnetic potential differences between the black hole horizon and conformal infinity, respectively, and 
\begin{equation}
    P = \frac{3}{8\pi G\ell^2}\,,\qquad V = \frac{1}{3}\omega_k r_+^3\,,
\end{equation}
are, respectively, the vacuum pressure and thermodynamic volume \cite{Kastor:2009wy}.
The thermodynamic quantities also satisfy the Smarr relation,
\begin{equation}
    M = 2TS-2PV+\Phi_{\rm E} Q_{\rm E} + \Phi_{\rm M} Q_{\rm M}\,,
\end{equation}
and its quadratic generalization
\begin{equation}
M^2=\frac{S}{4\pi G}+\frac{\pi G Q^4}{4S}+\frac{Q^2}{2}+\frac{GS}{2\pi \ell^2}\left(Q^2+\frac{S}{\pi G}+\frac{S^2}{2\pi^2 \ell^2}\right)\,.
\end{equation}

In the Euclidean regime, the renormalized action takes the form,
\begin{equation}
    S_{\rm ren}^{\rm E} = \frac{\beta \omega_k}{8\pi G} \left[m-\frac{r_+^3}{\ell^2} - \frac{e^{-\gamma} }{r_+}\left(q_{\rm E}^2 - q_{\rm M}^2\right)\right]= \beta( M-\Phi_{\rm E} Q_{\rm E}) - S\,,
\end{equation}
where $\beta = 1/T$ is the inverse of the Hawking temperature. We can identify the renormalized Euclidean action with the Gibbs potential as $S_{\rm ren}^{\rm E} =\beta \mathcal{G} $ as
\begin{equation}
\begin{aligned}
    E ={}& \left[\left(\frac{\partial }{\partial \beta}\right)_{Q_{\rm E},Q_{\rm M}} - \frac{Q_{\rm E}}{\beta}\left(\frac{\partial}{\partial Q_{\rm E}}\right)_{\beta,Q_{\rm M}}\right]S_{\rm ren}^{\rm E} = \frac{\omega_k}{4\pi}\frac{m}{G} =M\,, \\ S={}&\left[\beta\left(\frac{\partial}{\partial \beta}\right)_{\beta,Q_{\rm M}} -1\right]S_{\rm ren}^{\rm E} = \frac{\omega_k}{4G}r_+^2 = \frac{{\rm Area}(\Gamma)}{4G}\,, \\ \Phi_{\rm E} ={}& -\frac{1}{\beta}\left(\frac{\partial S_{\rm ren}^{\rm E}}{\partial Q_{\rm E}}\right)_{\beta,Q_{\rm M}} = \frac{q_{\rm E}}{r_+}e^{-\gamma}\,.
\end{aligned}
\end{equation}

Notice that there is no need to add a term for the magnetic charge to relate the Euclidean on shell action with the Gibbs free energy, showing that the $q_{\rm M}$ is a constant external parameter of the dual theory.

The specific heat of the solution is
\begin{equation}
    C_P = \left(\frac{\partial M}{\partial T}\right)_{q_{\rm M},q_{\rm E},P} = \left[\frac{3 r_+^4+\ell ^2 \left(kr_+^2-Q^2\right)}{3 r_+^4-\ell ^2 \left(kr_+^2-3 Q^2\right)}\right]2S\,,
\end{equation}
which is always positive for positive temperature, i.e., $C_P(T\geq 0)\geq 0$\,, and the solution is thermodynamically stable and can always reach thermal equilibrium with a heat bath.

\begin{figure}
\begin{center}
\includegraphics[scale=0.6]{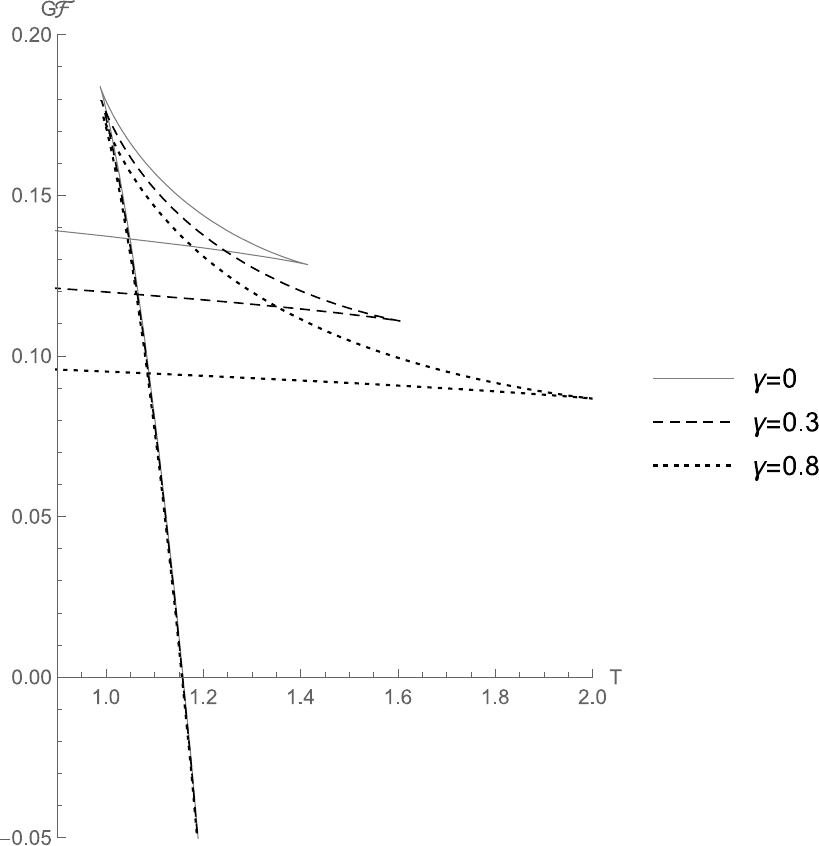}
\caption{Free energy $G{\cal F}$ as a function of temperature for the black hole solution with $k=1\,, \ell=\sqrt{3}\,, e^{\gamma}Q^2=q_{\rm E}^2+q_{\rm M}^2=1.125\times 10^{-2}$\,.}\label{figuresswallowtail}
\end{center}
\end{figure}

We can also study the canonical ensemble by coupling the system to energy and charge reservoirs at fixed temperature and the intensive variable (the electric potential). Then, the thermodynamic potential is the Helmholtz free energy $\cal F$\,. In this case, we need to add a boundary term with a well-posed variational principle, as we are now fixing the gauge curvature rather than the potential \cite{Caldarelli:1999xj}. We find that the term is proportional to the Hawking-Ross term generalized to NLE \cite{Barrientos:2022bzm} such that
\begin{equation}
\begin{aligned}
    \tilde{S}_{\rm ren}^{\rm E} ={}& {S}_{\rm ren}^{\rm E} -\frac{1}{4\pi G}\int_{\partial \cal M}{\rm d}^3x \sqrt{h}n_\mu\left({\cal L}_{\cal S}F^{\mu\nu}+\frac12 {\cal L}_{\cal P}\epsilon_{\alpha\beta}{}^{\mu\nu}F^{\alpha\beta}\right) A_\nu\,  \\  ={}&{\frac{\beta\omega_k}{16\pi G}\left[kr_+-\frac{r_+^3}{\ell^2}+3\frac{e^{-\gamma}\left(q_{\rm E}^2+q_{\rm M}^2\right)}{r_+}\right]} \\ ={}& \beta M - S   = \beta {\cal F}
\end{aligned}
\end{equation}
is a well-defined action in the canonical ensemble and can be associated with the Helmholtz free energy that shows a swallowtail (see \autoref{figuresswallowtail} for an example) that appears in classical catastrophe theory, just as for AdS-Reissner-Nordstr\"{o}m black holes \cite{Chamblin:1999tk}. 

An interesting process in classical thermodynamics is the so-called Joule-Thomson (JT) expansion \cite{10.1088/978-1-627-05532-1} that has been studied in various black hole solutions \cite{Okcu:2016tgt, Okcu:2017qgo, Mo:2018qkt, Cisterna:2018jqg, Barrientos:2022uit}. The JT effect is an isenthalpic process in which temperature change is produced when a gas is allowed to expand from a high-pressure region to a low-pressure region, through a valve or porous plugs. 
As a result, it is possible to obtain heating or cooling effects due to this process, where both are controlled by an inversion point characterized by the so-called JT coefficient. This inversion point is defined as the point where the inversion curves intersect the isenthalpic curves in the $T-P$ plane. The change of temperature with respect to pressure can be described by the JT coefficient defined as \cite{REI65}
\begin{equation}\label{mujt}
\mu_{\rm JT}=\left(\frac{\partial T}{\partial P}\right)_H=\frac{1}{C_P}\left[T\left(\frac{\partial V}{\partial T}\right)_P-V\right]\,.
\end{equation}
The $\mu_{\rm JT}$ sign determines whether heating or cooling will occur. In the JT expansion, the pressure change is negative, but the temperature change can be positive or negative. For $\mu_{\rm JT}>0$\,, one has the cooling region in the $T-P$ plane, whereas $\mu_{\rm JT}<0$ determines the heating region in the $T-P$ plane. Replacing the thermodynamic quantities into Eq. \eqref{mujt} one finds
\begin{equation}\label{muJT}
    \mu_{\rm JT} = \left(\frac{8\pi G P r_+^4 + 2kr_+^2 - 3Q^2}{8\pi G Pr_+^4 + kr_+^2 - Q^2}\right)\frac{V}{S}\,.
\end{equation}
By setting 
$\mu_{\rm JT}=0$\,, we can define the inversion pressure, $P_{\rm i}$\,, which is the specific point in the black hole's pressure gradient where the system transitions from cooling (or heating) to heating (or cooling). Substituting this condition into Eq. \eqref{temp1} for the corresponding inversion temperature $T_{\rm i}$\,, we obtain the following equation for the inversion curves
\begin{equation}\label{Ti}
T_{\rm i}=\sqrt{\frac{GP_{\rm i}}{2\pi}}\frac{16\pi GP_{\rm i}Q^2-k\sqrt{k^2+24\pi GP_{\rm i}Q^2}+k^2}{\left(\sqrt{k^2+24\pi GP_{\rm i}Q^2}-k\right)^{3/2}}\,.
\end{equation}
Now, we can plot isenthalpic curves in the $T-P$ plane. The event horizon can be determined using Eq. \eqref{defm}, and this result can be substituted into Eq. \eqref{temp1} to yield the isenthalpic curves in the $T-P$ plane.
\begin{figure}
\begin{center}
\includegraphics[scale=1.16]{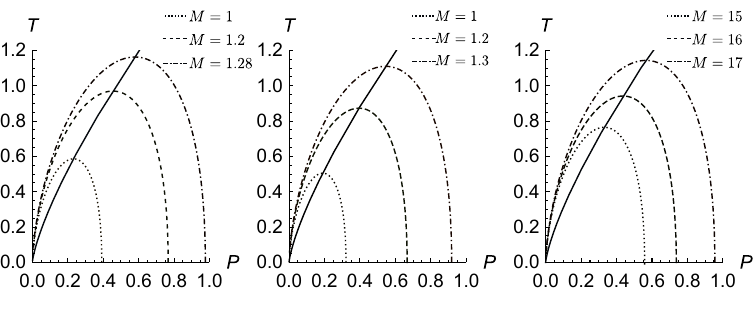}
\caption{The solid curves represent inversion curves with $Q=5$\,, while the isenthalpic curves are shown in dotted, dashed, and dot-dashed for different values of $M$\,. The left, center, and right plots correspond, respectively, to $k=-1,0,1$\,.}\label{figureJT}
\end{center}
\end{figure}
Considering the black hole's mass as equivalent to the enthalpy in the extended phase space \cite{Kastor:2009wy}, the isenthalpic curves for various mass values are plotted in \autoref{figureJT}. The inversion curve intersects the maximum points of the isenthalpic curves, dividing them into cooling and heating regions. For $P<P_{\rm i}$\,, the slope of the isenthalpic curve is positive, indicating cooling during expansion. Conversely, for $P>P_{\rm i}$\,, the slope becomes negative beneath the inversion curve, signifying heating. Notice that if one considers the large-$\gamma$ regime, the JT coefficient \eqref{muJT} and inversion curves \eqref{Ti} become constants, implying the absence of a transition point. That is, the isenthalpic curves no longer intersect the inversion curves, and no JT effect occurs in this limit, mirroring the behavior of the AdS-Schwarzschild solution. 

We observe that these solutions are thermodynamically equivalent to AdS black holes charged under Maxwell electrodynamics coupled to the Einstein-Hilbert action. This equivalence arises not only because the black hole solutions \eqref{defm} are isometric to AdS-Reissner-Nordstr\"{o}m topological black holes with charges being screened by the deformation parameter $\gamma$\,, but also due to the existence of an on shell relation at the level of the free energy as well. For this configuration, the scalar $\cal S$ and the pseudoscalar $\cal P$ are given by
\begin{equation}
\mathcal{S} = \frac{q_{\rm M}^2 - e^{-2\gamma} q_{\rm E}^2}{r^4}\,, \qquad \mathcal{P} = \frac{2 e^{-\gamma} q_{\rm E} q_{\rm M}}{r^4}\,,
\end{equation}
revealing that both quantities are proportional. This implies that the ModMax Lagrangian \eqref{SModMax} can be recast as
\begin{equation}
S_{\rm MM} = -\frac{1}{8\pi G_{\rm eff}} \int_{\cal M} \mathrm{d}^4 x \sqrt{-g} \mathcal{S}\,,
\end{equation}
which corresponds to pure Maxwell theory and
\begin{equation}
G_{\rm eff} = \frac{G}{\cosh\gamma - \left(\frac{e^{2\gamma}q_{\rm E}^2 + q_{\rm M}^2}{e^{2\gamma}q_{\rm E}^2 - q_{\rm M}^2}\right) \sinh\gamma}\,,
\end{equation}
defines an effective Newton's constant parametrized by the deformation parameter and charges.
In this setup, the theory is on shell equivalent to pure Maxwell theory. This equivalence holds for all configurations where either $\cal P$ is proportional to $\cal S$ or for solutions without magnetic charges. Consequently, the geometric and holographic properties of such solutions are identical to those of pure Maxwell theory, and the nonlinear effects of ModMax theory remain invisible. To probe these nonlinearities, we introduce nontrivial bulk perturbations that break this on shell equivalence. These perturbations source time-dependent operators in the dual field theory, allowing us to compute linear response functions and extract transport coefficients that now encode the nonlinear structure of ModMax theory.


\section{Holographic Transport Coefficients}\label{Sec:Transport}

In this section, we compute the magnetotransport of the holographic ModMax theory by analyzing linear perturbations around a charged black hole background with an $\mathbb{R}^2$ horizon topology. We first introduce isotropic fluctuations of the gauge and metric fields and employ Kubo formulas in the hydrodynamic limit to derive the conductivities in terms of retarded Green’s functions evaluated in a thermal equilibrium state. Next, we break translational symmetry in the transverse spatial directions by introducing scalar fields, which induce momentum relaxation and allow for a finite DC conductivity. By perturbing all dynamical fields, we analytically determine the full set of transport coefficients---particularly the Hall angle and the Nernst signal---in terms of horizon data, thereby capturing the nonlinear effects of ModMax electrodynamics.

\subsection{Linear response induced by isotropic fluctuations}\label{subsec:isotropic}
To ensure spatial translational invariance, introduce a uniform transverse magnetic field, and meaningfully describe Hall effects in the boundary theory, we consider the planar dyonic black hole solution ($k=0$)\,. We perform the following coordinate transformation:
\begin{equation}
    r = \frac{\ell^2}{z}\,,\qquad x^a = \frac{X^{\hat a}}{\ell}\,,
\end{equation}  
with $x^{\hat a} = (x,y)$ and $X^{\hat a}=(X,Y)$\,, which brings the metric into the form  
\begin{equation}\label{MetricPoincare}
    {\rm d}s^2 = \frac{\ell^2}{z^2} \left( -h(z)\,{\rm d}t^2 + \frac{{\rm d}z^2}{h(z)} + {\rm d}X^2 + {\rm d}Y^2 \right)\,,
\end{equation}  
where the blackening function reads 
\begin{equation}
    h(z) = 1 - \tilde{m} z^3 + \tilde{Q}^2 z^4\,,\qquad \tilde{m} \equiv \frac{2m}{\ell^2}\,,\qquad \tilde{Q}^2 \equiv \frac{Q^2}{\ell^6}\,.
\end{equation}  

Following \cite{Hartnoll:2007ai,Hartnoll:2009sz}, for the sake of simplicity, we place the black hole horizon at $z=1$ by fixing the mass parameter as $\tilde{m} = 1 + \tilde{Q}^2$\,, and then, we introduce isotropic, time-dependent fluctuations to the gauge and metric fields,
\begin{equation}
   \delta A_{{\hat{a}}}=\frac{1}{2 \pi}\int {\rm d}\omega \delta a_{{\hat{a}}}(z) e^{-i\omega t}\,, \qquad \delta h_{t{\hat{a}}}=\frac{1}{2\pi}\int {\rm d}\omega {\delta h_{t{\hat{a}}}(z)}{z^{-2}} e^{-i\omega t}\,,
\end{equation}
which reduces the linearized field equations to the following system:
\begin{equation}
\begin{aligned}
    &4Q^2z^3\left[q_{\rm E}h\delta a'_X-e^{\gamma}q_{\rm M}\left(i\omega \delta a_Y + q_{\rm M} \delta h_{tX}\right)\right]+\left(q_{\rm E}^2+e^{2\gamma}q_{\rm M}^2\right)\left(z h \delta h''_{tX}-2h\delta h'_{tX}\right)=0\,, \\ &4Q^2z^3\left[q_{\rm E}h\delta a'_Y-e^{\gamma}q_{\rm M}\left(i\omega \delta a_X + q_{\rm M} \delta h_{tY}\right)\right]+\left(q_{\rm E}^2+e^{2\gamma}q_{\rm M}^2\right)\left(z h \delta h''_{tY}-2h\delta h'_{tY}\right)=0\,,  \\  &4Q^2z^2\left(iq_{\rm E} \omega \delta a_Y + q_{\rm E} q_{\rm M}\delta h_{tX} + e^{\gamma}q_{\rm M} h\delta a'_X\right) +i\omega\left(q_{\rm E}^2 + e^{2\gamma}q_{\rm M}^2\right) \delta h'_{tY}=0\,,  \\ &4Q^2z^2\left(iq_{\rm E} \omega \delta a_X - q_{\rm E} q_{\rm M}\delta h_{tY} - e^{\gamma}q_{\rm M} h\delta a'_Y\right) +i\omega\left(q_{\rm E}^2 + e^{2\gamma}q_{\rm M}^2\right) \delta h'_{tX}=0\,, 
\end{aligned}
\end{equation}
where, for simplicity, we have set $\ell=1$ throughout this analysis and primes denote partial differentiation with respect to the holographic coordinate $z$\,.

To solve the system of perturbation equations, we impose ingoing wave boundary conditions at the horizon, requiring the fields to behave as
\begin{equation}
    \delta a_{{\hat{a}}} \sim a_{{\hat{a}}}(z)h(z)^{\frac{i\omega}{4\pi T}}\,, \qquad \delta h_{t{\hat{a}}} \sim h_{{\hat{a}}}(z)h(z)^{1+\frac{i\omega}{4\pi T}}\,,
    \label{5.6}
\end{equation}
with $T =  h'(1)/4\pi$ the black hole Hawking temperature. 

We then perform a hydrodynamic expansion in small frequency $\omega \to 0$ near the boundary, expanding the radial profiles as
\begin{equation}
\begin{aligned}
    a_{{\hat{a}}}(z)={}&A_{{\hat{a}}0}+\omega A_{{\hat{a}}1}+\omega^2 A_{{\hat{a}}2}+\dots\,,\\
    h_{{\hat{a}}}(z)={}&H_{{\hat{a}}0}+\omega H_{{\hat{a}}1}+\omega^2 H_{{\hat{a}}2}+\dots\,, 
\end{aligned}
\end{equation}
and match the asymptotic behavior with the near-horizon expansion. Solving the equations order-by-order in $\omega$\,, we obtain at leading order the system,
\begin{equation}
\begin{aligned}
    q_{\rm E} H_{{\hat{a}}0} + e^\gamma A'_{{\hat{a}}0} ={}& 0\,,\\   2e^{-\gamma}\left[Q^2z^4+e^{\gamma}\left(2h-3\right)\right]H'_{{\hat{a}}0} + zh H''_{{\hat{a}}0} ={}& 0\,,
\end{aligned}
\end{equation}
whose general solution is given by
\begin{equation}
\begin{aligned}
H_{{\hat{a}}0}={}&\gamma_{{\hat{a}}}\,, \\  A_{{\hat{a}}0}={}&\alpha_{{\hat{a}}}-q_{\rm E}e^{-\gamma}\gamma_{{\hat{a}}}z\,,
\end{aligned}
\end{equation}
with $\gamma_{{\hat{a}}}$ and $\alpha_{{\hat{a}}}$  integration constants. At the next order in the hydrodynamic expansion, the system reads,
\begin{equation}
\begin{aligned}\label{Firstomega}
   H_{{\hat{a}}1}''+2H_{{\hat{a}}1}'\partial_z \log\psi={}&C_{H{\hat{a}}}\,, \\ A_{{\hat{a}}1}'+q_{\rm E}e^{-\gamma}H_{{\hat{a}}1}={}&C_{A{\hat{a}}}\,,
\end{aligned}
\end{equation}
where $\psi\equiv z^{-1}h(z)$\,, and $C_{H{\hat{a}}}$ and $C_{A{\hat{a}}}$ are functions that depend on the zeroth-order solutions, and are given in Appendix \ref{Appendix:IntConst}. A first integral for the equations for $H_{{\hat{a}}1}$ is 
\begin{equation}
    H_{{\hat{a}}1}'=\frac{C_{{\hat{a}}}}{\psi^2}+\frac{1}{\psi^2}\int_{0}^{z}{\rm d}uC_{H{\hat{a}}}(u)\psi(u)^2\,,
\end{equation}
where the constants $C_{\hat{a}}$ are fixed by imposing regularity at the horizon, and the coefficients $\alpha_{{\hat{a}}}$ can be written as functions of $\gamma_{{\hat{a}}}$\,, as shown in Appendix \ref{Appendix:IntConst}. Integrating once more, we find
 \begin{equation}
 \begin{aligned}
     H_{{\hat{a}}1}={}&\gamma_{{\hat{a}}}-i\int_{0}^{z}{\rm d}u\gamma_{{\hat{a}}} P_{{\hat{a}}}(u)\,,\\ 
     A_{{\hat{a}}1}={}&\alpha_{{\hat{a}}}-q_{\rm E}e^{-\gamma}\int_{0}^{z} {\rm d}u\left[H_{{\hat{a}}1}(u)-i\left(\frac{\gamma_{x}Q_{{\hat{a}}}(u)+\gamma_{y}R_{{\hat{a}}}(u)}{h(z)}\right)\right]\,, 
 \end{aligned}
 \end{equation}
where the functions \( P_{\hat{a}} \)\,, \( Q_{\hat{a}} \)\,, and \( R_{\hat{a}} \) can be found in Appendix \ref{Appendix:IntConst}. We do not proceed further in the boundary expansion as we take the hydrodynamic limit. Instead, we directly relate the boundary data to the integration constants. 

To this end, we use the \( z \)-independent solutions of the field equations and introduce two additional constants, \( \delta_{\hat{a}} \)\,, ensuring a total of four integration constants that fully account for the four boundary data \( H_{\hat{a}0} \) and \( A_{\hat{a}0} \) such that
\begin{equation}
\begin{aligned}
    \delta h^0_{\hat{a}} ={}&-i\omega {\delta_{\hat{a}}}(\omega){q_{\rm M}}^{-1} + \gamma_{\hat{a}}(\omega)\,, \\
    \delta a^0_{\hat{a}} ={}& \delta_{{\hat{a}}}(\omega) + \alpha_{\hat{a}}(\omega)\,. 
\end{aligned}
\end{equation}
In the hydrodynamic expansion, it suffices to determine up to the linear order in the fluctuations. Then, the solutions read,
\begin{equation}
\begin{aligned}
    \delta a_{X}&=\delta_{X}+h^{\frac{i\omega}{4\pi T}}\left\{\alpha_{X}-\gamma_{X} e^{-\gamma}q_{\rm E}\left[z-i\omega \int_{0}^{z}{\rm d}u \left(\frac{P_{X}(u)}{\psi^2(u)}(z-u)+\frac{\gamma_{X}Q_{X}+\gamma_{Y}R_{X}(u)}{h}\right)\right]\right\}\,,  \\
    \delta a_{Y}&=\delta_{Y}+h^{\frac{i\omega}{4\pi T}}\left\{\alpha_{Y}-\gamma_{Y} e^{-\gamma}q_{\rm E}\left[z-i\omega \int_{0}^{z}{\rm d}u \left(\frac{P_{Y}(u)}{\psi^2(u)}(z-u)+\frac{\gamma_{X}Q_{Y}+\gamma_{Y}R_{Y}(u)}{h}\right)\right]\right\}\,,  
\end{aligned}
\end{equation}
for the gauge-field perturbations, and 
\begin{equation}
\begin{aligned}
\delta h_{tX}(\omega,z) ={}& -\frac{i\omega}{q_{\rm M}}\delta_{Y}+h^{1+\frac{i\omega}{4\pi T}}\left[\delta h^{0}_{X}+\frac{i\omega}{q_{\rm M}}\delta_{Y}-i\omega\delta h^{0}_{X}\int_{0}^{z}{\rm d}u\frac{P_{X}(u)}{\psi^{2}(u)}\right]\,,\\
    \delta h_{tY}(\omega,z) ={}& \frac{i\omega}{q_{\rm M}}\delta_{X}+h^{1+\frac{i\omega}{4\pi T}}\left[\delta h^{0}_{Y}+\frac{i\omega}{q_{\rm M}}\delta_{X}-i\omega\delta h^{0}_{Y}\int_{0}^{z}{\rm d}u\frac{P_{Y}(u)}{\psi^{2}(u)}\right]\,, 
\end{aligned}
\end{equation}
for the metric perturbations. 

We now evaluate the on shell action \eqref{SEMM} to second order in perturbations. Substituting the previously derived solutions, the action is expressed entirely in terms of boundary data. Functional differentiation with respect to the sources then yields the linear response functions. In what follows, we focus on the current-current correlator, which captures the modifications to the system's electrical response induced by the nonlinear corrections to Maxwell theory. The retarded Green's function is given by
\begin{equation}
    G_{J_{\hat{a}} J_{\hat{b}}}^R(\omega)=-i \int {\rm d}^2 x\, {\rm d} t e^{i \omega t} \theta(t)\left\langle\left[J_{\hat{a}}(t), J_{\hat{b}}(0)\right]\right\rangle=-i\omega\frac{q_{\rm E}(q_{\rm E}^2+q_{\rm M}^2)}{q_{\rm M}(q_{\rm E}^2+e^{2\gamma}q_{\rm M}^2)\kappa}\epsilon_{{\hat{a}}{\hat{b}}}\,,
\end{equation}
with $\epsilon_{XY}=-\epsilon_{YX}=1$ the antisymmetric tensor and $\kappa\equiv8 \pi G$\,. 
Applying the Kubo formula, we extract the electrical conductivity in thermal equilibrium from the retarded current-current correlator,
\begin{equation}
    \sigma_{\hat{a}\hat{b}}=-\lim_{\omega\to 0}\frac{\text{Im}( G_{J_{\hat{a}} J_{\hat{b}}}^R(\omega))}{\omega}=\frac{q_{\rm E}(q_{\rm E}^2+q_{\rm M}^2)}{q_{\rm M}(q_{\rm E}^2+e^{2\gamma}q_{\rm M}^2)\kappa} \epsilon_{{\hat{a}}{\hat{b}}}= \frac{\rho(B^2+\kappa^2\rho^2)}{B(B^2 e^{2\gamma}+\kappa^2\rho^2)}\epsilon_{{\hat{a}}{\hat{b}}}\,,
\end{equation}
where $\rho = q_{\rm E}/\kappa$ is the electric charge density and $B = q_{\rm M}$ the magnetic field. Notice that the electrical conductivity is finite at zero temperature, indicating a metallic behavior. From this expression, we observe that in the limit $\gamma \to 0$\,, one finds 
\begin{equation}\label{SigmaxyHallPerfect}
\lim_{\gamma\to0}\sigma_{\hat{a}\hat b} = \epsilon_{\hat{a}\hat{b}} \frac{\rho}{B}\,,
\end{equation}
recovering the result found in \cite{Hartnoll:2009sz} for pure Maxwell theory. As in the standard Maxwell case, the electrical conductivity matrix in ModMax theory exhibits no diagonal components, indicating the absence of longitudinal conductivity. Instead, only the Hall conductivity is present, and the Hall angle asymptotes $\theta_{\rm H}\to\pm \frac{\pi}{2}$\,, characteristic of a perfect Hall state (i.e., dissipationless transport) in the dual field theory. 

Moreover, the nonvanishing components of the conductivity matrix become exponentially suppressed as $\gamma$ increases, indicating a transition to a perfect insulating phase in the large-$\gamma$ regime. In this limit, the bulk geometry reduces to the AdS planar Schwarzschild black hole, while the gauge field becomes a purely magnetic stealth configuration. Surprisingly, in contrast to standard AdS-Schwarzschild black holes---which are known to exhibit nonvanishing holographic electrical conductivity (see, e.g., \cite{Donos:2014cya})---the solution considered here behaves as a perfect insulator under isotropic perturbations.

We now proceed to compute the remaining magnetotransport properties, beginning with the retarded Green's function associated with the momentum density-current correlator
\begin{equation}
G^R_{J_{\hat{a}} T_{t\hat{b}}}(\omega) = -i \int {\rm d}^2x\, {\rm d}t\, e^{i \omega t} \theta(t) \left\langle [J_{\hat a}(t), T_{t{\hat b}}(0)] \right\rangle 
= i\omega  \frac{3 (1 + e^{-\gamma}(q_{\rm E}^2 + q_{\rm M}^2))}{4 q_{\rm M} \kappa}\epsilon_{\hat a \hat b}\,,
\end{equation}
and we find the thermoelectric conductivity matrix through Kubo’s formula, i.e., 
\begin{equation}
\alpha_{\hat a \hat b} = -\frac{1}{T} \lim_{\omega \to 0} \frac{\operatorname{Im} G^R_{J_{\hat a} T_{t \hat b}}(\omega)}{\omega}
= -  \frac{3 (1 + e^{-\gamma}(q_{\rm E}^2 + q_{\rm M}^2))}{4 q_{\rm M} \kappa T}\epsilon_{\hat a \hat b}\,.
\end{equation}
Next, we consider the momentum-momentum density correlator,
\begin{equation}
\begin{aligned}
G^R_{T_{t\hat a} T_{t\hat b}}(\omega) &= -i \int {\rm d}^2x\, {\rm d}t\, e^{i \omega t} \theta(t) \left\langle [T_{t\hat a}(t), T_{t\hat b}(0)] \right\rangle \\
&= i\omega \frac{2 e^{-\gamma} \left(-3  e^{\gamma} + q_{\rm E}^2 + q_{\rm M}^2\right)^2}{(q_{\rm E}^2 + q_{\rm M}^2) \kappa}\delta_{\hat a\hat b}  - i \omega \frac{9 e^{-2\gamma} q_{\rm E} \left( e^{\gamma} + q_{\rm E}^2 + q_{\rm M}^2\right)^2}{q_{\rm M} (q_{\rm E}^2 + q_{\rm M}^2) \kappa}\epsilon_{\hat a\hat b}\,.
\end{aligned}
\end{equation}
Applying Kubo’s formula once more, we obtain the thermal conductivity matrix,
\begin{equation}
\begin{aligned}
\bar{\kappa}_{\hat a\hat b} ={}& -\frac{1}{T} \lim_{\omega \to 0} \frac{\operatorname{Im} G^R_{T_{t\hat a} T_{t\hat b}}(\omega)}{\omega} \\
={}& \delta_{\hat a\hat b} \frac{2  e^{-\gamma} \left(3 e^{\gamma} - q_{\rm E}^2 - q_{\rm M}^2\right)^2}{(q_{\rm E}^2 + q_{\rm M}^2) \kappa T}  +  \epsilon_{\hat a\hat b} \frac{9 e^{-2\gamma} q_{\rm E} \left( e^{\gamma} + q_{\rm E}^2 + q_{\rm M}^2\right)^2}{q_{\rm M} (q_{\rm E}^2 + q_{\rm M}^2) \kappa T}\,, 
\end{aligned}
\end{equation}
which domain the thermal transport in the absence of electromagnetic fields.

In the limit $\gamma \to 0$\,, we recover full agreement with the standard Maxwell case \cite{Hartnoll:2009sz}. Interestingly, in the opposite limit $\gamma \to \infty$\,, the system exhibits nontrivial thermoelectric holographic conductivities and a divergent thermal conductivity. This behavior leads to dispersionless momentum transport in the uncharged black hole with the previously discussed magnetic stealth charge configuration. 

As a final remark, when analyzing systems in the presence of a magnetic background, it is important to subtract the contributions from magnetization currents  \cite{Hartnoll:2007ih,Hartnoll:2016apf}. To account for this, we define the magnetization and energy magnetization as follows:
\begin{equation}
\begin{aligned}
    \hat{M} = -\frac{\partial S_{\text{ren}}}{\partial B}\,, \qquad
    \hat{M}_{{\rm E}} = -\frac{\partial S_{\text{ren}}}{\partial B_{{\rm E}}}\,,
\end{aligned}
\end{equation}
where $B$ and $B_{\rm E}$ are determined from the asymptotic form of the perturbations. Specifically, in this case
\begin{equation}
\begin{aligned}
    \delta A_Y &\rightarrow X B \quad \text{as} \quad z \rightarrow 0\,, \\
    \delta h_{tY} &\rightarrow X B_{\rm E} \quad \text{as} \quad z \rightarrow 0\,,
\end{aligned}
\end{equation}
with \( B = q_{\rm M} \) and \( B_{\rm E} \) a constant. By imposing these boundary conditions, truncating the perturbative solutions accordingly, and substituting into the expressions for the magnetizations, we obtain
\begin{equation}\label{magnetizations}
    \hat{M} = \frac{B}{2\kappa}e^{-\gamma}\,,\qquad
    \hat{M}_{\rm E} = \frac{1}{2} \mu {\hat M}\,,
\end{equation}
where the chemical potential is given by \( \mu = -e^{-\gamma} q_{\rm E} \)\,, such that $\alpha_{\ha\hb}\to\alpha_{\ha\hb} + \frac{\hat M}{T}\epsilon_{\ha\hb}$ and $\bar\kappa_{\ha\hb}\to\bar\kappa_{\ha\hb} + \frac{2({\hat M}_{\rm E}-\mu {\hat M})}{T}\epsilon_{\ha\hb}$\,. Then, the corrected thermoelectric and thermal conductivity matrices are given by
\begin{equation}
\begin{aligned}
\alpha_{\hat a\hat b} ={}& -\frac{q_{\rm M}e^{-\gamma}}{2\kappa T}\left[1+\frac{3}{2}\left(\frac{e^\gamma+q_{\rm E}^2+q_{\rm M}^2}{q_{\rm M}^2} \right)\right]\epsilon_{\hat a\hat b}\,, \\  \bar\kappa_{\hat a\hat b}={}& \frac{1}{\kappa T}\left[-\frac{2e^{-\gamma}\left(q_{\rm E}^2+q_{\rm M}^2-3e^{\gamma}\right)^2}{q_{\rm E}^2+q_{\rm M}^2}\delta_{\hat a \hat b} + \frac{e^{-2\gamma}q_{\rm E}}{2q_{\rm M}}\left(\frac{18(e^{\gamma}+q_{\rm E}^2+q_{\rm M}^2)^2}{q_{\rm E}^2+q_{\rm M}^2} + q_{\rm M}^2\right)\epsilon_{\hat a \hat b}\right]\,.
\end{aligned}
\end{equation}

As observed, the conductivity matrix is symmetric and finite, but the electric conductivity exhibits only transverse Hall components. To construct a more realistic dual model, it is necessary to break translational invariance, which also introduces longitudinal components. This can be achieved by introducing transverse scalar fields, as proposed in \cite{Andrade:2013gsa}, which induce momentum relaxation and effectively incorporate latticelike effects in the boundary theory.\footnote{Similarly, a holographic $Q$-lattice was constructed in \cite{Donos:2013eha} using a scalar field with a plane-wave profile.} In order to have analytic expressions for the DC conductivities, we employ the method developed in \cite{Donos:2014cya}, which allows for the computation of magnetotransport coefficients purely in terms of horizon data. In the following subsection, we introduce scalar fields with a linear transverse profile to explicitly break translational invariance, thereby generating nontrivial Hall and Nernst effects.

\subsection{Hall and Nernst effects in the presence of momentum relaxation}\label{subsec:MomRelax}
As aforementioned, the conductivity matrix \eqref{SigmaxyHallPerfect} is finite, with only the off-diagonal components being nonzero, indicating that the system realizes a perfect Hall state. To move beyond this regime, we introduce scalar fields that explicitly break translational symmetry, following the approach of \cite{Andrade:2013gsa}. The inclusion of scalar operators corresponds in the dual theory to introducing an external relevant perturbation that makes the conformal field theory massive depending on the scalar boundary conditions \cite{Witten:2001ua,Papadimitriou:2007sj}, which, in this case, the modification enables momentum relaxation and leads to a richer transport structure. 

The analysis of DC transport is carried out using the method introduced in \cite{Donos:2014cya}, which allows for an analytic computation of the DC conductivities in terms of horizon data by constructing electric $J^\ha$ and heat $Q^\ha$ conserved currents which at linear order satisfy a generalized Ohm's law.

Let us consider the action\footnote{We do not include boundary terms for the scalar fields, as we consider scalars with a linear transverse profile that does not introduce additional divergences in the on shell action. This setup suffices for computing the DC conductivities via the method of evaluating conserved currents at the horizon, as described in \cite{Donos:2014cya}. However, when constructing the optical conductivities or transport coefficients in the hydrodynamic limit, as discussed in \autoref{subsec:isotropic}, the perturbations generate new divergences that must be renormalized. The holographic renormalization of scalar fields coupled to gravity---under various boundary conditions---has been thoroughly analyzed in \cite{Papadimitriou:2007sj,Anabalon:2015xvl,Caceres:2023gfa}.}
\begin{equation}
    S_{\rm ren}^{\rm A} = S_{\rm ren}  + S_{\rm A}\,,
\end{equation}
where $S_{\rm ren}$ is given in \eqref{SEMM} and
\begin{equation}
    S_{\rm A} = -\frac{1}{2} \int_{\cal M}{\rm d}^4x \sqrt{-g}g^{\mu\nu}\delta^{\hat a \hat b}\partial_\mu \psi_{\hat a} \partial_\nu \psi_{\hat b}\,,
\end{equation}
with $\psi_{\hat a} = (\psi_x,\psi_y)$ denoting the scalar fields. The equations of motion for the metric and gauge fields remain the same as in the system described in \eqref{EOM}, but now the scalars satisfy the massless wave equation,
\begin{equation}
    \Box \psi_{\hat a} = 0\,,
\end{equation}
and the Einstein equations now become
\begin{equation}
    R_{\mu\nu}-\frac12 R g_{\mu\nu} -\frac{3}{\ell}g_{\mu\nu} = 8\pi G \left[T_{\mu\nu} +\frac{1}{2}g_{\mu\nu}\left(\partial_\mu \psi_{\ha}\partial_\nu\psi_\hb-\frac12g_{\mu\nu}\partial_\alpha \psi_\ha \partial^\alpha\psi_\hb\right)\delta^{\ha\hb}\right]\,,
\end{equation}
where $T_{\mu\nu}$ is defined in \eqref{Tmunu}.

Assuming scalar fields with transverse linear profile, i.e.,
\begin{equation}
    \psi_{\hat a} = \alpha x_\ha\,,
\end{equation}
the line element \eqref{defm} with $k=0$ is still a solution of the new system, but now the blackening function becomes
\begin{equation}
    f(r) = \frac{r^2}{\ell^2}-\frac{2m}{r} + \frac{Q^2}{r^2} - \frac{1}{2} \alpha^2\,.
\end{equation}

We now examine small perturbations of the fields, expressed as
\begin{equation}\label{anispert}
\begin{aligned}
      g_{t\hat a}={}&r^2\epsilon H_{t\hat a}(r)\,, &&g_{r\hat a}=r^2 \epsilon H_{r\hat a}(r)\,,\\
       A_{\hat a}={}&\epsilon(a_{\hat a}(r)-E_{\hat a}t)\,,  &&\psi_{\hat a}=\alpha x_{\hat a}+\epsilon{\alpha^{-1}}{\chi_{\hat a}(r)}\,,
\end{aligned}
\end{equation}
where $a_{\hat a}$\,, $H_{t\hat a}$\,, and $\chi_{\hat a}$ are independent fluctuation fields and $\epsilon$ is an infinitesimal perturbation parameter. The ModMax equations define a conserved current along the holographic radial direction, i.e., $\partial_r J_{\hat a} = 0$\,, whose linearization around \eqref{anispert} leads to the nontrivial components,
\begin{equation}\label{corriente}
   J_{\hat a} = \frac{e^{\gamma}\left(q_{\rm E}^2+q_{\rm M}^2\right)}{q_{\rm E}^2+e^{2\gamma}q_{\rm M}^2}\left(fa_{\hat a}^{\prime}+e^{-\gamma}q_{\rm E} H_{t \hat a}+q_{\rm M}f \epsilon_{\hat a}{}^{\hat b}H_{r\hat b}\right)\,.
\end{equation}

To ensure regularity of the fluctuations, appropriate boundary conditions must be imposed both at the horizon and at asymptotic infinity. It is convenient to work in Eddington-Finkelstein coordinates $(v,r)$ where
\begin{equation}
    v = t + \int \frac{\d r}{f(r)}\,,
\end{equation}
which leads to the perturbed metric
\begin{equation}
        {\rm d}s^2 = -f {\rm d} v^2 +  2\,{\rm d} v {\rm d} r + r^{2}\delta_{\hat a \hat b}\d x^{\hat a}\d x^{\hat b}  +\epsilon\left( H_{t\hat a} {\rm d} v
        - \frac{H_{t \hat a}}{f} {\rm d} r +  H_{r \hat a} {\rm d} r \right)\d x^{\hat a}\,,
    \end{equation}
where we impose
\begin{equation}
H_{t\ha} = f H_{r\ha}\,,
\end{equation}
and
\begin{equation}
    a_{\ha}=-E_{\ha}\int\frac{\d r}{f(r)}\,,
\end{equation}
in order to ensure regularity of the metric and gauge field perturbations throughout the spacetime. 

Solving the Einstein field equations on this perturbed background, we find that the metric perturbations at the horizon radius $r_+$ read,
\begin{equation}
\begin{aligned}
     H_{tx}={}&-\frac{4e^{\gamma}(q_{\rm E}^2+q_{\rm M}^2)(E_{x}e^{-\gamma}q_{\rm E}r_+^2  \alpha^2+q_{\rm M}E_{y}(4e^{-\gamma}(q_{\rm E}^2+q_{\rm M}^2)+r_+^2 \alpha^2))}{16q_{\rm E}^4q_{\rm M}^2+(4q_{\rm M}^3+e^{\gamma}q_{\rm M}r_+^2 \alpha^2)^2+q_{\rm E}^2(32q_{\rm M}^4+8e^{\gamma}q_{\rm M}^2r_+^2 \alpha^2+r_+^4 \alpha^4)}\,,\\ 
       H_{ty}={}&-\frac{4e^{\gamma}(q_{\rm E}^2+q_{\rm M}^2)(E_{y}e^{-\gamma}q_{\rm E}r_+^2  \alpha^2-q_{\rm M}E_{x}(4e^{-\gamma}(q_{\rm E}^2+q_{\rm M}^2)+r_+^2 \alpha^2))}{16q_{\rm E}^4q_{\rm M}^2+(4q_{\rm M}^3+e^{\gamma}q_{\rm M}r_+^2 \alpha^2)^2+q_{\rm E}^2(32q_{\rm M}^4+8e^{\gamma}q_{\rm M}^2r_+^2 \alpha^2+r_+^4\alpha^4)}\,.
\end{aligned}
\end{equation}
Thus, by substituting this expression into (\ref{corriente}), we find that the conductivity matrix, 
\begin{equation}
    \sigma_{\ha\hb} \equiv \frac{\partial J_\hb}{\partial E^\ha}\,,\qquad E^\ha \equiv (E_x,E_y)\,,
\end{equation}
is given by
\begin{equation}
\begin{aligned}
  \sigma_{xx}={}& \frac{  r_+^2\alpha^2e^{\gamma}\left(q_{\rm E}^2+q_{\rm M}^2\right)(4e^{-\gamma}(q_{\rm E}^2+q_{\rm M}^2)+r_+^2\alpha^2)}{\alpha^2r_+^2\left[\alpha^2r_+^2\left(q_{\rm E}^2+e^{2\gamma}q_{\rm M}^2\right)+8e^{\gamma}q_{\rm M}^2\left(q_{\rm E}^2+q_{\rm M}^2\right)\right]
+ 16q_{\rm M}^{2}(q_{\rm E}^2+q_{\rm M}^2)^2}\,,\\
  \sigma_{xy}={}&\frac{8q_{\rm E}q_{\rm M}e^{\gamma}\left(q_{\rm E}^2+q_{\rm M}^2\right)^2\left(2e^{-\gamma}(q_{\rm E}^2+q_{\rm M}^2)+r_+^2\,\alpha^2\right)}{(q_{\rm E}^2+e^{2\gamma}q_{\rm M}^2)(\alpha^2r_+^2\left[\alpha^2r_+^2\left(q_{\rm E}^2+e^{2\gamma}q_{\rm M}^2\right)+8e^{\gamma}q_{\rm M}^2\left(q_{\rm E}^2+q_{\rm M}^2\right)\right]
+ 16q_{\rm M}^{2}(q_{\rm E}^2+q_{\rm M}^2)^2)}\,, 
 \end{aligned}
 \end{equation}
 and
  \begin{equation}
  \sigma_{yx}=-\sigma_{xy}\,, \qquad \sigma_{yy}=\sigma_{xx}\,.
\end{equation}

We observe that the conductivity matrix reduces to \eqref{SigmaxyHallPerfect} when $\alpha = 0$\,, and to the result corresponding to pure Maxwell theory (cf. \cite{Donos:2014cya,Cisterna:2019uek}) when $\gamma = 0$\,. Notably, the deformation parameter $\gamma$ does not affect the anomalous temperature dependence of the transport coefficients, implying that the dual metals retain the same thermodynamic behavior as those described by Maxwell theory coupled to AdS gravity. However, $\gamma$ does modify the numerical values of the coefficients, introducing new regimes that allow for the exploration of exotic quantum phases governed by the ModMax deformations.

We can now compute the Hall angle, defined by 
\begin{equation}
    \theta_{\rm H} = \arctan \left(\frac{\sigma_{xy}}{\sigma_{xx}}\right)\,,
\end{equation}
which describes the deflection of charge carriers due to a magnetic field. We find
\begin{equation}\label{tanhTheta}
\tan\theta_{\rm H}=\frac{8q_{\rm E}q_{\rm M}(q_{\rm E}^2+q_{\rm M}^2)}{(q_{\rm E}^2+e^{2\gamma}q_{\rm M}^2)\alpha^2r_+^2}\left[\frac{2(q_{\rm E}^2+q_{\rm M}^2)+e^{\gamma}\alpha^2r_+^2}{4(q_{\rm E}^2+q_{\rm M}^2)+e^{\gamma}\alpha^2r_+^2}\right]\,.
\end{equation}
Notice that the expression inside the square brackets in \eqref{tanhTheta} is bounded between zero and one. Therefore, the leading behavior of the Hall angle is governed by
\begin{equation}
\tan\theta_{\rm H}\sim\frac{8q_{\rm E}q_{\rm M}(q_{\rm E}^2+q_{\rm M}^2)}{(q_{\rm E}^2+e^{2\gamma}q_{\rm M}^2)\alpha^2r_+^2}\,,
\end{equation}
which, in the low-temperature regime, scales as the square of the inverse of the temperature, $\tan\theta_{\rm H}\sim T^{-2}$\,, which is the anomalous temperature
dependence of the Hall angle of cuprate strange metals \cite{PhysRevLett.67.2088, TYLER19971185}. { Nonetheless, our model yields a conductivity matrix with the same temperature scaling as in pure Maxwell theory, with the deformation parameter merely shifting its numerical values. In contrast, strange metals display, in addition to this scaling of the Hall angle, an anomalous behavior in the longitudinal resistivity, which becomes linear in temperature. To reconcile these two features within a holographic framework, \cite{Blake:2014yla} argues that one must include an additional contribution to the conductivity matrix that affects the longitudinal resistivity but leaves the Hall angle unchanged.}

\begin{center}
\begin{figure}[h!]
\centering
\includegraphics[scale=0.9]{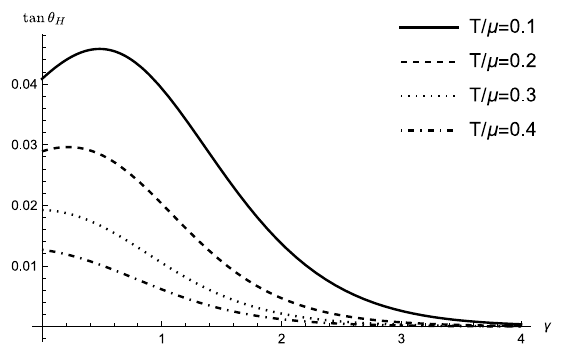}
\caption{{The tangent of the Hall angle as a function of the ModMax deformation parameter $\gamma$\,, shown for various values of the normalized temperature $T/\mu$\,, with the scalar intensity fixed at $\alpha = 10$\,, the electromagnetic charges at $q_{\rm E}=1$\,, $q_{\rm M}=0.5$, and the AdS radius at $\ell=1$ for all curves.}}
\label{fig:hall}
\end{figure}
\end{center}

Furthermore, in the limit $\gamma\to\infty$\,, the Hall angle vanishes for all values of temperature, electric and magnetic charges, and scalar intensity $\alpha$\,. This implies that the electric current flows strictly parallel to the applied electric field, with no transverse deflection, indicating the absence of Hall response despite the presence of a magnetic charge in the boundary. The system thus exhibits purely Ohmic behavior, with no transverse forces acting on the charge carriers. As shown in \autoref{fig:hall}, the Hall angle depends nonlinearly on the ModMax parameter. Notably, in the small-$\gamma$ regime, a perfect Hall state can also be realized for small values of the charges.

To compute the thermoelectric conductivity matrices $\alpha_{\ha\hb}$ and $\bar\alpha_{\ha\hb}$\,, we follow the holographic framework of \cite{Donos:2014cya} where a conserved heat current is constructed from the holographic stress tensor and the electric current as
\begin{equation}
    Q^\ha = T^{t\ha} - \mu J^\ha\,,
\end{equation}
which in our model reduces to 
\begin{equation}
   Q_\ha = f^2\partial_rH_{r\ha} - A_t J_\ha = -4\pi T H_{t\ha}(r_+)\,,
\end{equation}
where in the last equality we have used the fact that the current is conserved along the holographic coordinate and evaluated it at $r=r_+$\,. 
Then the thermoelectric conductivities, 
\begin{equation}
    \bar{\alpha}_{\ha \hb} = \frac{1}{T}\frac{\partial Q_\hb}{\partial E^\ha}\,,
\end{equation}
are 
\begin{equation}
\begin{aligned}
\bar\alpha_{xx}={}&\frac{16\pi \alpha^2 e^{-2\gamma} \left(q_{\rm E}^2 + q_{\rm M}^2\right) q_{\rm E} r_+^2}
{\alpha^2r_+\left[\alpha^2r_+\left(q_{\rm E}^2r_+^2+e^{2\gamma}q_{\rm M}^2\right)+8e^{\gamma}q_{\rm M}\left(q_{\rm E}^2r_++q_{\rm M}^3\right)\right]
+ 16q_{\rm M}^{2}(q_{\rm E}^2+q_{\rm M}^2)^2}\,,\\
\bar\alpha_{xy}={}&\frac{16\pi e^{-\gamma}(q_{\rm E}^2+q_{\rm M}^2)q_{\rm M}^2\left[\alpha^2 r_++4(q_{\rm E}^2+q_{\rm M}^2)\right]}{\alpha^2r_+\left[\alpha^2r_+\left(q_{\rm E}^2r_+^2+e^{2\gamma}q_{\rm M}^2\right)+8e^{\gamma}q_{\rm M}\left(q_{\rm E}^2r_++q_{\rm M}^3\right)\right]+16q_{\rm M}^2(q_{\rm E}^2+q_{\rm M}^2)^2}\,,
\end{aligned}
\end{equation}
together with
\begin{equation}
\bar\alpha_{xy}=-\alpha_{xy}\,,\qquad \bar\alpha_{yy}=\alpha_{xx}\,,
\end{equation}
where we have already subtracted the magnetization effects as explained below \eqref{magnetizations}. We also obtain $\alpha_{\ha\hb}$ using the perturbations, 
\begin{equation}
\begin{aligned}
     g_{t\ha} ={}& -tf\zeta_\ha +\epsilon H_{t\ha}\,, && g_{r\ha} = r^2\epsilon H_{r\ha}\,, \\ A_\ha ={}& -t E_\ha + \zeta_\ha \epsilon a(r)\,, && \psi_\ha = \alpha x_\ha + \epsilon \alpha^{-1}\chi_\ha (r)\,, 
\end{aligned}
\end{equation}
 where the linear in time perturbations comprise the only
holographic sources at the boundary and $\zeta_\ha$ can be identified with the temperature gradient that sources the heat current \cite{Donos:2014cya}. Repeating the same procedure as before, we find that the thermoelectric conductivity,
\begin{equation}
\alpha_{\ha\hb} = \frac{1}{T}\frac{\partial J_\hb}{\partial \zeta^\ha}\,,
\end{equation}
satisfies $\alpha_{\ha\hb} = \bar\alpha_{\ha\hb}$\,, showing that the conductivity matrix is symmetric.

\begin{figure}[htbp]
  \centering
  \begin{subfigure}[b]{0.46\textwidth}
    \includegraphics[width=\linewidth]{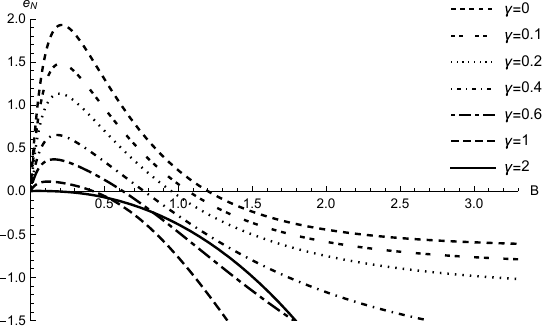}
    \caption{The Nernst signal in terms of the magnetic field for different values of the ModMax parameter and $T = 0.5\,, q_{\rm E}=1.5\,,\alpha=3.5\,, \ell=1$\,.}
    \label{fig:NernstGamma}
  \end{subfigure}
  \hfill
  \begin{subfigure}[b]{0.46\textwidth}
    \includegraphics[width=\linewidth]{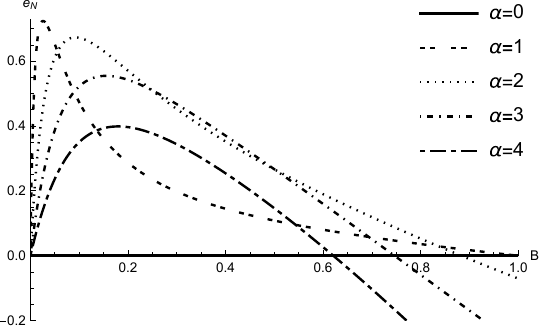}
    \caption{The Nernst signal in terms of the magnetic field for different values of the scalar intensity $\alpha$ with $T = 0.5\,, q_{\rm E}=1.5\,,\gamma=0.5\,,\ell=1$\,.}
    \label{fig:NernstAlpha}
  \end{subfigure}
  \caption{Nernst signal as a function of the magnetic field for various values of the bulk parameters.}
  \label{fig:NernstSignal}
\end{figure}

Finally, we compute the Nernst response, defined as the electric field induced by a thermal gradient, as follows:
\begin{equation}
    E^\ha = -\vartheta^{\ha\hb} \nabla_\hb T\,,
\end{equation}
where
\begin{equation}\label{NernstTheta}
    \vartheta^{\ha}{}_{\hb} = -\rho^{\ha\hat{c}}\alpha_{\hat{c}\hb}\,,
\end{equation}
and $\rho = (\sigma)^{-1}$\,, the resistivity matrix. The Nernst signal corresponds to the transverse response $e_N\equiv \vartheta^x{}_{y}$\,. The Nernst effect can actually be used to probe high-$T_c$ superconductors \cite{wang2006nernst}, as for regular metals the Nernst signal is linear in the magnetic field, while for cuprate superconductors it exhibits a pronounced bell-shaped dependence in $B$ signaling unconventional vortex dynamics or pseudogap phenomena. 
The Nernst signal for the model reads
\begin{equation}
    e_N = -\frac{{\cal M}_{(1)}{\cal M}_{(2)}}{{\cal N}_{(1)}{\cal N}_{(2)}}\,,
\end{equation}
with
\begin{equation}
\begin{aligned}
     {\cal M}_{(1)}={}& 16\pi e^{-2\gamma}r_+^2q_m \alpha^2\left(e^{2\gamma}q_{\rm M}^2+q_{\rm E}^2\right)^2 \,,  \\  {\cal M}_{(2)}={}& \alpha^2r_+^2q_{\rm M} e^{2\gamma}\left(\alpha^2r_++q_{\rm M}^2+q_{\rm E}^2\right)\left(q_{\rm E}^2+e^{2\gamma}q_{\rm M}^2\right)\\&+4 e^{3 \gamma } q_{\rm M}^3 \left(q_{\rm M}^2+q_{\rm E}^2\right) \left(\alpha ^2 r_++q_{\rm M}^2+q_{\rm E}^2\right)-16 q_{\rm E}^2 \left(q_{\rm M}^2+q_{\rm E}^2\right)^2 \\  &+4 e^{\gamma } q_{\rm E}^2 \left(q_{\rm M}^2+q_{\rm E}^2\right) \left(\alpha ^2 r_+ (q_{\rm M}-2 r_+)+q_{\rm M} \left(q_{\rm M}^2+q_{\rm E}^2\right)\right)\,, \\  {\cal N}_{(1)} ={}& 16q_{\rm M}^2\left(q_{\rm M}^2+q_{\rm E}^2\right)^2 + \alpha^2r_+\left[8e^{\gamma}q_{\rm M}\left(q_{\rm M}^3+r_+ q_{\rm E}^2\right)+e^{2\gamma}r_+q_{\rm M}^2\alpha^2+r_+^3q_{\rm E}^2+\alpha^2\right]\,, \\  {\cal N}_{(2)} ={}& \left(4q_{\rm E}^3+e^\gamma r_+^2q_{\rm E} \alpha^2\right)^2 + q_{\rm M}^2\left[16q_{\rm E}^2\left(q_{\rm M}^2+2q_{\rm E}^2\right) + 8e^\gamma r_+^2q_{\rm E}^2\alpha^2 + e^{4\gamma}r_+^4\alpha^4\right]\,. 
\end{aligned}
\end{equation}
As shown in \autoref{fig:NernstSignal}, the signal exhibits a nonlinear, bell-shaped behavior at small magnetic fields. As the field strength increases, the response gradually transitions to an approximately linear decay on $B$ and becomes negative, showing a $B$-linear contribution from quasiparticles.\footnote{See \cite{kontani2002nernst} for a description of high-$T_c$ superconductors in terms of quasiparticle excitations without assuming thermally excited vortices.} Indeed, the curve shown in \autoref{fig:NernstSignal} reproduces the Nernst signal observed in the LSCO family of cuprate superconductors (see Fig. 23 of \cite{wang2006nernst}) with the critical and onset temperatures tuned by different values of the deformation parameter. The signal vanishes for $\alpha=0$ as can be seen in \autoref{fig:NernstAlpha}, as expected. 
As previously mentioned, this behavior is characteristic of high-temperature superconductors, where the critical temperature serves as an effective order parameter marking the transition between the nonlinear and linear regimes of the Nernst signal \cite{wang2006nernst}. Indeed, we find that at large values of $B$\,, while keeping $\gamma$ small enough, the signal asymptotes a constant negative value, i.e.,
\begin{equation}
    e_N \sim -\frac{\pi}{4}\frac{ \alpha ^2 r_+^2}{q_{\rm E}^2}e^{3 \gamma } + {\cal O}\left(B^{-2}\right)\,,
\end{equation}
for large-$B$\,. 
We also notice that the ModMax deformation modifies the amplitude of the curve in the low-$B$ regime, giving the possibility to model different cuprate superconductors. 

Moreover, if the deformation parameter becomes much stronger than the electric charge, the Nernst signal always becomes negative, indicating that the dual metal in the highly nonlinear regime of the bulk ModMax theory is in some exotic state dominated by quasiparticle excitations abusting the superconducting dome.\footnote{Although we are not studying the superconducting phase of the dual material, we adopt the nomenclature of \cite{wang2006nernst}, where the Nernst signal of superconducting samples exhibits a Gaussian-like dome within the superconducting phase and turns negative in the normal phase at the critical temperature. Moreover, there exists an onset temperature beyond which a quasiparticle contribution causes the signal to grow linearly with the magnetic field after the critical temperature.} 
As shown in \cite{wang2006nernst}, high-$T_c$ cuprate superconductors are confined to a vortex-liquid state below the critical temperature. As the magnetic field increases, the Nernst signal becomes dominated by a negative quasiparticle contribution when the nonlinear effects are strong enough (as can be seen in the solid curve in \autoref{fig:NernstGamma}). A similar qualitative behavior appears in the temperature dependence of the Nernst signal, which lacks a sharp transition, which also emerges in our holographic model. {Furthermore, upon reaching an onset temperature, the signal begins to increase slightly and linearly with
the magnetic field, eventually becoming positive. This linear growth is absent in our holographic model, but we expect it to appear in an extended framework that incorporates a dual superconducting state.}

\begin{center}
\begin{figure}[h!]
\centering
\includegraphics[scale=0.9]{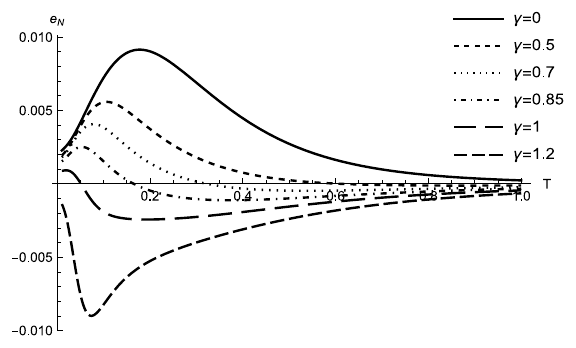}
\caption{{Dependence of the Nernst signal on the temperature $T$ for various values of $\gamma$\,, with fixed parameters $q_{\rm E} = 0.1$\,, $q_{\rm M} = 0.1$\,, and $\alpha = 0.3$\,.}}
\label{fig:NernstTemperature}
\end{figure}
\end{center}

{As illustrated in \autoref{fig:NernstTemperature}, the dependence of the Nernst signal on the temperature reflects this smooth crossover and subsequently starts to rise near the onset temperature. Then, varying the deformation parameter $\gamma$ allows for broader Nernst profiles, suggesting that both the critical and onset temperatures of the dual metallic phase can be effectively tuned within the model.}

As a final comment, we notice that the Nernst coefficient, defined as $\nu \equiv e_N/B$\,, 
becomes field independent  in the small-$B$ regime, i.e.,
\begin{equation}
   \nu \sim \left[\frac{16 \pi  e^{-2 \gamma }}{\alpha ^2   r_+^2}-\frac{16 \pi  \alpha ^2   r_+^2}{\left(4 q_{\rm E}^2+\alpha ^2 e^{\gamma }   r_+^2\right)^2}\right] + {\cal O}(B)\,, 
\end{equation}
 as expected. 

\section{Conclusions}\label{Sec:Conclusions}

In this paper, we compute the transport coefficients of a $(2+1)$-dimensional strongly coupled field theory with holographic duality to Einstein gravity coupled to ModMax NLE. We begin by analyzing the (holographic) thermodynamics of static nonlinearly charged black holes, identifying an on shell equivalence with Maxwell theory for specific configurations. We introduce time-dependent perturbations in all fields and employ linear response theory to derive the DC magnetotransport properties to go beyond this trivial point.

Our results reveal that while the structure of the conductivity matrices resembles that of Maxwell theory, the nonlinear corrections significantly alter their numerical values. To incorporate lattice effects at the boundary, we introduce scalar fields and compute the transport matrices using conserved currents, yielding simple analytical expressions in terms of horizon data. Within this framework, we observe both the Hall and Nernst effects in the dual theory.

Notably, the Hall angle exhibits exponential suppression with the ModMax parameter $\gamma$ at fixed charge, while for small-$\gamma$\,, a perfect Hall state can emerge. In the low-temperature regime, the Hall angle follows the anomalous temperature dependence characteristic of cuprate strange metals. Furthermore, the Nernst signal displays a qualitative profile reminiscent of high-$T_c$ superconductors, featuring a superconducting dome and a normal phase, with the onset and critical temperatures modulated by the ModMax parameter.

A particularly unexplored yet intriguing regime is the ultranonlinear $\gamma\to\infty$\,. In this limit, the black hole solution reduces to a magnetic stealth configuration over an AdS-Schwarzschild background; a universal feature of all black holes sourced by ModMax NLE. Despite this simplification, the dual boundary theory retains a nontrivial magnetic current, significantly influencing transport properties. 

 In this limit, the Hall angle becomes strongly suppressed, which indicates enhanced effective dissipation arising from nonlinear electrodynamic effects. Furthermore, the Nernst signal becomes negative in this regime, revealing an exotic quasiparticle-dominated phase in the dual theory; a particularly noteworthy result as such states are typically inaccessible in conventional AdS/CFT frameworks due to the strongly coupled nature of the boundary theory.

While the ModMax parameter $\gamma$ does not alter the temperature dependence of magnetotransport, it provides a powerful tool to probe exotic phases in the dual theory. By tuning $\gamma\,$\,, we can effectively control critical and onset temperatures, offering a pathway to model realistic strongly correlated materials---such as strange metals or high-$T_c$ superconductors---within this holographic framework. This is significant because it shows how nonlinear electrodynamics can lead to new transport regimes not allowed in linear theories, and gives insight into how holography may hint at exotic phases of matter.
It would be interesting to explore vortex-driven phases of the Nernst signal by analyzing the hydrodynamic properties of the dual field theory \cite{Hernandez:2017mch,Amoretti:2022acb}, and to study the entropy and viscosity associated with individual vortices, following the approach of \cite{behnia2022nernst}.
 
A natural extension of the present model is to construct holographic superconductors using standard techniques \cite{Hartnoll:2008vx, Hartnoll:2008kx, Horowitz:2008bn}, by introducing mass and charge for the bulk scalar fields, allowing them to condense below a critical temperature. It would also be interesting to explore alternative boundary conditions that capture the holographic Meissner effect \cite{Natsuume:2022kic}, a hallmark of superconductivity. These enhancements would enable a more systematic investigation of the role of nonlinearities in boundary phenomena and contribute to the development of more realistic holographic models of superconductors. Although ModMax is a relatively simple model with a high degree of symmetry, analytic solutions beyond the probe limit are not expected, as even in the case of standard Maxwell theory numerical methods are typically required. {While our model already reproduces several qualitative features of the Nernst signal observed in cuprate superconductors, some experimental samples studied in \cite{wang2006nernst} exhibit a linear increase in the signal with magnetic field beyond the normal phase; an effect we have not yet fully captured. Nevertheless, given the signs of superconducting behavior at the boundary and the tunability of the critical temperature via the deformation parameter $\gamma$\,, we expect these features to emerge more clearly in the extended model.}

Another avenue of exploration is to consider richer black hole solutions that arise from gravity coupled to ModMax \cite{Flores-Alfonso:2020nnd, BallonBordo:2020jtw,Ayon-Beato:2024vph, Barrientos:2024umq}. These solutions provide more intricate bulk geometries that could play a significant role in the dual transports (see \cite{Meert:2024dud} for a detailed example of how rotation modifies the conductivity matrix). Particularly intriguing are the accelerating, nonlinearly charged black holes found in \cite{Barrientos:2022bzm,Hale:2025veb}, as acceleration have already yielded a variety of interesting results in the holographic context \cite{Arenas-Henriquez:2022www, Arenas-Henriquez:2023hur, Cisterna:2023qhh, Tian:2023ine, Bunney:2024xic, Arenas-Henriquez:2024ypo, Luo:2024cwm, Li:2025rzl}. So far, transport properties in these geometries have only been explored in the probe limit using NLE \cite{Roychowdhury:2024oih}, where metallic signatures have already emerged. 
We leave the study of acceleration effects on magnetotransport for future work.

Finally, a particularly intriguing direction is the inclusion of fermionic excitations in the holographic framework, with special attention to the effects of nonlinearities of the bulk gauge sector. While fermionic degrees of freedom have been extensively studied in holographic superconductors without magnetic fields \cite{Chen:2009pt, Faulkner:2009am, Faulkner:2010da, Faulkner:2010tq, Faulkner:2011tm, Faulkner:2013bna, Grandi:2021bsp, Grandi:2021jkj, Grandi:2023jna, Bahamondes:2024zsm}, incorporating them in models with magnetic fields and nonlinear electrodynamics could potentially reproduce phenomena such as the giant Nernst signal observed in experiments \cite{bel2003giant}. We leave these promising avenues for future work.
\acknowledgments  
We thank Gabriel Arenas-Henriquez, Sebasti\'an Bahamondes, Ali Fatemiabhari, Oriana Labrin,  Constanza Quijada, David Rivera-Betancour, and Leonardo Sanhueza for comments and discussions on related topics. 
U.H. would like to thank Charles University for its hospitality during the final
stage of this project. 
The work of J.B. is supported by FONDECYT Postdoctorado Grant No. 3230596. The work of N.C. is supported by Beca Doctorado Nacional (ANID) 2023 Scholarship No. 21231876. The work of F.D. has been supported by FONDECYT Grants No. 1230853 and No. 1231779. U.H. is partially supported by ANID Grant No.  21231297 and INES Project from Universidad de Talca.
\appendix
\section{Integration Constants}\label{Appendix:IntConst}
Here, we provide the functions and coefficients that solve the isotropically perturbed system of \autoref{subsec:isotropic}. 

The $C_{A{\ha}}$ and $C_{H{\ha}}$ coefficients appearing in the differential equations in the first order in the hydrodynamic perturbations \eqref{Firstomega} read
\begin{equation}
\begin{aligned}
    C_{AX}={}&\frac{i  e^{-2 \gamma}}{4 q_{\rm M} h(z)} \Bigg[4 q_{\rm E} \bigl(-e^{\gamma} \alpha_Y + q_{\rm E} z \gamma_Y\bigr)\Bigg.\\ 
    & - \frac{e^{\gamma} \bigl(4 e^{\gamma} q_{\rm M} (q_{\rm E}^2 + q_{\rm M}^2) z^2 \alpha_X + e^{2 \gamma} q_{\rm M}^2 \gamma_Y 4\pi T + q_{\rm E} \bigl(-4 q_{\rm M} (q_{\rm E}^2 + q_{\rm M}^2) z^3 \gamma_X + q_{\rm E} \gamma_Y 4\pi T\bigr)\bigr) h'(z)}{(q_{\rm E}^2 + q_{\rm M}^2) z^2 4\pi T}\Bigg]\,,\\ 
    C_{AY}={}&\frac{i  e^{-2 \gamma}}{4 q_{\rm M} h(z)} \Bigg[4 q_{\rm E} \bigl(e^{\gamma} \alpha_X - q_{\rm E} z \gamma_X\bigr)\Bigg. \\ 
    & + \frac{e^{\gamma} \bigl(-4 e^{\gamma} q_{\rm M} (q_{\rm E}^2 + q_{\rm M}^2) z^2 \alpha_Y + e^{2 \gamma} q_{\rm M}^2 \gamma_X 4\pi T + q_{\rm E} \bigl(4 q_{\rm M} (q_{\rm E}^2 + q_{\rm M}^2) z^3 \gamma_Y + q_{\rm E} \gamma_X 4\pi T\bigr)\bigr) h'(z)}{(q_{\rm E}^2 + q_{\rm M}^2) z^2 4\pi T}\Bigg]\,,\\
  C_{HX} ={}& -\frac{i e^{-2 \gamma}}{q_{\rm M} z^2 h(z)^2 4\pi T} \Big[q_{\rm M} \left(-3 e^{\gamma} + (q_{\rm E}^2 + q_{\rm M}^2) z^4\right)^2 \gamma_X  + 9 e^{2 \gamma} q_{\rm M} \gamma_X h(z)^2 \Big.\\
    &\left.+ z \left(3 q_{\rm E} (q_{\rm E}^2 + q_{\rm M}^2) z^4 \gamma_Y + e^{\gamma} \left(-4 (q_{\rm E}^2 + q_{\rm M}^2) z^3 \alpha_Y + 3 q_{\rm E} \gamma_Y\right)\right) 4\pi T\right. \\
    & \Big.- e^{\gamma} h(z) \left(18 e^{\gamma} q_{\rm M} \gamma_X - 10 q_{\rm M} (q_{\rm E}^2 + q_{\rm M}^2) z^4 \gamma_X + 3 q_{\rm E} z \gamma_Y 4\pi T\right)\Big]\,,\\
C_{HY} ={}& -\frac{i e^{-2 \gamma}}{q_{\rm M} z^2 h(z)^2 4\pi T}\Big[ 
        q_{\rm M} \left(-3 e^{\gamma} + (q_{\rm E}^2 + q_{\rm M}^2) z^4\right)^2 \gamma_Y  - 18 e^{2 \gamma} q_{\rm M} \gamma_Y h(z)\Big. \\  &+ 9 e^{2 \gamma} q_{\rm M} \gamma_Y h(z)^2  - 3 q_{\rm E} (q_{\rm E}^2 + q_{\rm M}^2) z^5 \gamma_X 4\pi T  + e^{\gamma} z (10 q_{\rm M} (q_{\rm E}^2 + q_{\rm M}^2) z^3 \gamma_Y h(z) 
       \\  & \Big.+ \left(4 (q_{\rm E}^2 + q_{\rm M}^2) z^3 \alpha_X - 3 q_{\rm E} \gamma_X + 3 q_{\rm E} \gamma_X h(z)\right) 4\pi T )
    \Big]\,.
\end{aligned}
\end{equation}
The $C_{\ha}$ coefficients are  obtained by imposing regularity on the horizon of the fluctuations, and for the first integral for the equations for $H_{\ha 1}$ are    
\begin{equation}
\begin{aligned}
C_X ={}& \frac{3 i (e^{\gamma} + q_{\rm E}^2 + q_{\rm M}^2) \gamma_X}{ q_{\rm E}^2 + q_{\rm M}^2-3 e^{\gamma} } 
    + \frac{3 i e^{-2 \gamma} q_{\rm E} (q_{\rm E}^2 + q_{\rm M}^2) \gamma_Y}{q_{\rm M}} + \frac{e^{-\gamma} \left(3 i q_{\rm E} \gamma_Y-4 i (q_{\rm E}^2 + q_{\rm M}^2) \alpha_Y\right)}{q_{\rm M}}\,, \\ 
C_Y ={}& 
    \frac{3 i (e^{\gamma} + q_{\rm E}^2 + q_{\rm M}^2) \gamma_Y}{q_{\rm E}^2 + q_{\rm M}^2-3 e^{\gamma}}-\frac{3 i e^{-2 \gamma} q_{\rm E} (q_{\rm E}^2 + q_{\rm M}^2) \gamma_X}{q_{\rm M}}-\frac{e^{-\gamma} \left(3 i q_{\rm E} \gamma_X-4 i (q_{\rm E}^2 + q_{\rm M}^2) \alpha_X \right)}{q_{\rm M}} \,.
\end{aligned}
\end{equation}
The $\alpha_{\ha}$ coefficients can be written as functions of $\gamma_{\ha}$ as
\begin{equation}
\begin{aligned}
\alpha_Y ={}& \frac{e^{-\gamma}}{4 (q_{\rm E}^2 + q_{\rm M}^2)} \left[
    -3 e^{2\gamma} q_{\rm M} \gamma_X 
    + 3 q_{\rm E} (q_{\rm E}^2 + q_{\rm M}^2) \gamma_Y  + e^{\gamma} \bigl(
        q_{\rm M} (q_{\rm E}^2 + q_{\rm M}^2) \gamma_X 
        + 3 q_{\rm E} \gamma_Y
    \bigr)\right]\,, \\ 
\alpha_X ={}& \frac{e^{-\gamma}}{4 (q_{\rm E}^2 + q_{\rm M}^2)} \left[
    3 q_{\rm E} (q_{\rm E}^2 + q_{\rm M}^2) \gamma_X 
    + 3 e^{2\gamma} q_{\rm M} \gamma_Y + e^{\gamma} \bigl(
        3 q_{\rm E} \gamma_X 
        - q_{\rm M} (q_{\rm E}^2 + q_{\rm M}^2) \gamma_Y
    \bigr)\right]\,.
\end{aligned}
\end{equation}
Finally, integrating for the second time the fluctuations, we find
\begin{equation}
\begin{aligned}
    R_X ={}& \frac{1}{4} (1 - z) \left( \frac{3 e^{\gamma} q_{\rm M} (1 + z)}{q_{\rm E}^2 + q_{\rm M}^2} - q_{\rm M} (-3 + z + 4z^2) \right)\,, \\  
    Q_X ={}& \frac{1}{4} (1 - z) \left( -e^{-\gamma} q_{\rm E} \left(1 + (1 - 4z)^2 z \right) + 3 q_{\rm E} \left(-\frac{16 (1 - z) z^2}{q_{\rm E}^2 + q_{\rm M}^2-3 e^{\gamma} } + \frac{1 + z}{q_{\rm E}^2 + q_{\rm M}^2} \right) \right)\,, \\  
    P_X ={}& \frac{e^{-\gamma} (1 - z)^2 \left(3 e^{2\gamma} (2 + z) + (q_{\rm E}^2 + q_{\rm M}^2)^2 (1 + z + z^2 + 4z^3) - e^{\gamma} (q_{\rm E}^2 + q_{\rm M}^2)(9 + z(8 + 7z))\right)}{q_{\rm E}^2 + q_{\rm M}^2-3 e^{\gamma} }\,,\\  
    R_Y ={}& \frac{1}{4} q_{\rm E} (1 - z) \left( \frac{48 (1 - z) z^2}{3 e^{\gamma} - q_{\rm E}^2 - q_{\rm M}^2} + \frac{3 (1 + z)}{q_{\rm E}^2 + q_{\rm M}^2} - e^{-\gamma} \left(1 + (1 - 4z)^2 z \right) \right)\,, \\  
    Q_Y ={}& \frac{1}{4} q_{\rm M} (1 - z) \left(z + 4z^2 - \frac{3 e^{\gamma} (1 + z)}{q_{\rm E}^2 + q_{\rm M}^2} -3\right)\,, \\  
    P_Y ={}& \frac{e^{-\gamma} (1 - z)^2 \left(3 e^{2\gamma} (2 + z) + (q_{\rm E}^2 + q_{\rm M}^2)^2 (1 + z + z^2 + 4z^3) - e^{\gamma} (q_{\rm E}^2 + q_{\rm M}^2)(9 + z(8 + 7z))\right)}{q_{\rm E}^2 + q_{\rm M}^2-3 e^{\gamma}}\,.  
\end{aligned}
\end{equation}
Finally, we express the integration constant $\delta_{\ha}$ and $\gamma_{\ha}$ in terms of the boundary data as
\begin{equation}
\begin{aligned}
\delta_X &= -\frac{i e^{-2\gamma}}{16 q_{\rm M} (q_{\rm E}^2 + q_{\rm M}^2)^2} \Big(
    9 e^{4\gamma} \delta h^{0}_{Y} q_{\rm M}^2 \omega 
    - 9 \delta h^{0}_{Y} q_{\rm E}^2 (q_{\rm E}^2 + q_{\rm M}^2)^2 \omega \\
    &\quad + e^{2\gamma} \Big(4i q_{\rm M} (q_{\rm E}^2 + q_{\rm M}^2) (-3 \delta h^{0}_{X} q_{\rm E} 
    + (4 \delta a^{0}_{X} + \delta h^{0}_{Y} q_{\rm M})(q_{\rm E}^2 + q_{\rm M}^2)) \\
    &\quad + \Big(4 (q_{\rm E}^2 + q_{\rm M}^2) (3 \delta a^{0}_{Y} q_{\rm E} 
    + q_{\rm E} (3 \delta h^{0}_{X} + \delta a^{0}_{X} q_{\rm E}) q_{\rm M} 
    + \delta a^{0}_{X} q_{\rm M}^3) \\
    &\quad + \delta h^{0}_{Y} (q_{\rm E}^4 q_{\rm M}^2 + q_{\rm M}^6 
    + q_{\rm E}^2 (-9 + 2 q_{\rm M}^4))\Big) \omega\Big) \\
    &\quad - 6 e^{\gamma} q_{\rm E} (q_{\rm E}^2 + q_{\rm M}^2) \Big(3 \delta h^{0}_{Y} q_{\rm E} \omega 
    - 2 \delta a^{0}_{Y} (q_{\rm E}^2 + q_{\rm M}^2) \omega 
    + \delta h^{0}_{X} q_{\rm M} (q_{\rm E}^2 + q_{\rm M}^2) (2i + \omega)\Big) \\
    &\quad - 6 e^{3\gamma} q_{\rm M} \Big(-3 \delta h^{0}_{X} q_{\rm E} \omega 
    + 2 \delta a^{0}_{X} (q_{\rm E}^2 + q_{\rm M}^2) \omega 
    + \delta h^{0}_{Y} q_{\rm M} (q_{\rm E}^2 + q_{\rm M}^2) (2i + \omega)\Big)
\Big)\,, \\
\delta_Y &= \frac{i e^{-2\gamma}}{16 q_{\rm M} (q_{\rm E}^2 + q_{\rm M}^2)^2} \Big(
    9 e^{4\gamma} \delta h^{0}_{X} q_{\rm M}^2 \omega 
    - 9 \delta h^{0}_{X} q_{\rm E}^2 (q_{\rm E}^2 + q_{\rm M}^2)^2 \omega \\
    &\quad + e^{2\gamma} \Big(4i q_{\rm M} (q_{\rm E}^2 + q_{\rm M}^2) (3 \delta h^{0}_{Y} q_{\rm E} 
    - (4 \delta a^{0}_{Y} - \delta h^{0}_{X} q_{\rm M})(q_{\rm E}^2 + q_{\rm M}^2)) \\
    &\quad + \Big(-4 (q_{\rm E}^2 + q_{\rm M}^2) (-3 \delta a^{0}_{X} q_{\rm E} 
    + q_{\rm E} (3 \delta h^{0}_{Y} + \delta a^{0}_{Y} q_{\rm E}) q_{\rm M} 
    + \delta a^{0}_{Y} q_{\rm M}^3) \\
    &\quad + \delta h^{0}_{X} (q_{\rm E}^4 q_{\rm M}^2 + q_{\rm M}^6 
    + q_{\rm E}^2 (-9 + 2 q_{\rm M}^4))\Big) \omega\Big) \\
    &\quad - 6 e^{3\gamma} q_{\rm M} \Big(3 \delta h^{0}_{Y} q_{\rm E} \omega 
    - 2 \delta a^{0}_{Y} (q_{\rm E}^2 + q_{\rm M}^2) \omega 
    + \delta h^{0}_{X} q_{\rm M} (q_{\rm E}^2 + q_{\rm M}^2) (2i + \omega)\Big) \\
    &\quad + 6 e^{\gamma} q_{\rm E} (q_{\rm E}^2 + q_{\rm M}^2) \Big(-3 \delta h^{0}_{X} q_{\rm E} \omega 
    + 2 \delta a^{0}_{X} (q_{\rm E}^2 + q_{\rm M}^2) \omega 
    + \delta h^{0}_{Y} q_{\rm M} (q_{\rm E}^2 + q_{\rm M}^2) (2i + \omega)\Big)
\Big)\,, \\
\gamma_X &= \frac{i e^{-\gamma}}{4 q_{\rm M} (q_{\rm E}^2 + q_{\rm M}^2)} \Big(
    3 e^{2\gamma} \delta h^{0}_{X} q_{\rm M} \omega 
    - 3 \delta h^{0}_{Y} q_{\rm E} (q_{\rm E}^2 + q_{\rm M}^2) \omega \\
    &\quad - e^{\gamma} \Big(3 \delta h^{0}_{Y} q_{\rm E} \omega 
    - 4 \delta a^{0}_{Y} (q_{\rm E}^2 + q_{\rm M}^2) \omega 
    + \delta h^{0}_{X} q_{\rm M} (q_{\rm E}^2 + q_{\rm M}^2) (4i + \omega)\Big)
\Big)\,, \\
\gamma_Y &= \frac{i e^{-\gamma}}{4 q_{\rm M} (q_{\rm E}^2 + q_{\rm M}^2)} \Big(
    3 e^{2\gamma} \delta h^{0}_{Y} q_{\rm M} \omega 
    + 3 \delta h^{0}_{X} q_{\rm E} (q_{\rm E}^2 + q_{\rm M}^2) \omega \\
    &\quad - e^{\gamma} \Big(-3 \delta h^{0}_{X} q_{\rm E} \omega 
    + 4 \delta a^{0}_{X} (q_{\rm E}^2 + q_{\rm M}^2) \omega 
    + \delta h^{0}_{Y} q_{\rm M} (q_{\rm E}^2 + q_{\rm M}^2) (4i + \omega)\Big) 
\Big)\,. 
\end{aligned}
\end{equation}


\bibliography{bib}

\begin{thebibliography}{161}
\expandafter\ifx\csname natexlab\endcsname\relax\def\natexlab#1{#1}\fi
\expandafter\ifx\csname bibnamefont\endcsname\relax
  \def\bibnamefont#1{#1}\fi
\expandafter\ifx\csname bibfnamefont\endcsname\relax
  \def\bibfnamefont#1{#1}\fi
\expandafter\ifx\csname citenamefont\endcsname\relax
  \def\citenamefont#1{#1}\fi
\expandafter\ifx\csname url\endcsname\relax
  \def\url#1{\texttt{#1}}\fi
\expandafter\ifx\csname urlprefix\endcsname\relax\def\urlprefix{URL }\fi
\providecommand{\bibinfo}[2]{#2}
\providecommand{\eprint}[2][]{\url{#2}}

\bibitem[{\citenamefont{'t~Hooft}(1993)}]{tHooft:1993dmi}
\bibinfo{author}{\bibfnamefont{G.}~\bibnamefont{'t~Hooft}}, \bibinfo{journal}{Conf. Proc. C} \textbf{\bibinfo{volume}{930308}}, \bibinfo{pages}{284} (\bibinfo{year}{1993}), \eprint{gr-qc/9310026}.

\bibitem[{\citenamefont{Susskind}(1995)}]{Susskind:1994vu}
\bibinfo{author}{\bibfnamefont{L.}~\bibnamefont{Susskind}}, \bibinfo{journal}{J. Math. Phys.} \textbf{\bibinfo{volume}{36}}, \bibinfo{pages}{6377} (\bibinfo{year}{1995}), \eprint{hep-th/9409089}.

\bibitem[{\citenamefont{Maldacena}(1998)}]{Maldacena:1997re}
\bibinfo{author}{\bibfnamefont{J.~M.} \bibnamefont{Maldacena}}, \bibinfo{journal}{Adv. Theor. Math. Phys.} \textbf{\bibinfo{volume}{2}}, \bibinfo{pages}{231} (\bibinfo{year}{1998}), \eprint{hep-th/9711200}.

\bibitem[{\citenamefont{Witten}(1998{\natexlab{a}})}]{Witten:1998qj}
\bibinfo{author}{\bibfnamefont{E.}~\bibnamefont{Witten}}, \bibinfo{journal}{Adv. Theor. Math. Phys.} \textbf{\bibinfo{volume}{2}}, \bibinfo{pages}{253} (\bibinfo{year}{1998}{\natexlab{a}}), \eprint{hep-th/9802150}.

\bibitem[{\citenamefont{Gubser et~al.}(1998)\citenamefont{Gubser, Klebanov, and Polyakov}}]{Gubser:1998bc}
\bibinfo{author}{\bibfnamefont{S.~S.} \bibnamefont{Gubser}}, \bibinfo{author}{\bibfnamefont{I.~R.} \bibnamefont{Klebanov}}, \bibnamefont{and} \bibinfo{author}{\bibfnamefont{A.~M.} \bibnamefont{Polyakov}}, \bibinfo{journal}{Phys. Lett. B} \textbf{\bibinfo{volume}{428}}, \bibinfo{pages}{105} (\bibinfo{year}{1998}), \eprint{hep-th/9802109}.

\bibitem[{\citenamefont{Hartnoll}(2009)}]{Hartnoll:2009sz}
\bibinfo{author}{\bibfnamefont{S.~A.} \bibnamefont{Hartnoll}}, \bibinfo{journal}{Class. Quant. Grav.} \textbf{\bibinfo{volume}{26}}, \bibinfo{pages}{224002} (\bibinfo{year}{2009}), \eprint{0903.3246}.

\bibitem[{\citenamefont{Zaanen et~al.}(2015)\citenamefont{Zaanen, Sun, Liu, and Schalm}}]{Zaanen:2015oix}
\bibinfo{author}{\bibfnamefont{J.}~\bibnamefont{Zaanen}}, \bibinfo{author}{\bibfnamefont{Y.-W.} \bibnamefont{Sun}}, \bibinfo{author}{\bibfnamefont{Y.}~\bibnamefont{Liu}}, \bibnamefont{and} \bibinfo{author}{\bibfnamefont{K.}~\bibnamefont{Schalm}}, \emph{\bibinfo{title}{{Holographic Duality in Condensed Matter Physics}}} (\bibinfo{publisher}{Cambridge Univ. Press}, \bibinfo{year}{2015}).

\bibitem[{\citenamefont{Hartnoll et~al.}(2016)\citenamefont{Hartnoll, Lucas, and Sachdev}}]{Hartnoll:2016apf}
\bibinfo{author}{\bibfnamefont{S.~A.} \bibnamefont{Hartnoll}}, \bibinfo{author}{\bibfnamefont{A.}~\bibnamefont{Lucas}}, \bibnamefont{and} \bibinfo{author}{\bibfnamefont{S.}~\bibnamefont{Sachdev}} (\bibinfo{year}{2016}), \eprint{1612.07324}.

\bibitem[{\citenamefont{Hartnoll et~al.}(2008{\natexlab{a}})\citenamefont{Hartnoll, Herzog, and Horowitz}}]{Hartnoll:2008vx}
\bibinfo{author}{\bibfnamefont{S.~A.} \bibnamefont{Hartnoll}}, \bibinfo{author}{\bibfnamefont{C.~P.} \bibnamefont{Herzog}}, \bibnamefont{and} \bibinfo{author}{\bibfnamefont{G.~T.} \bibnamefont{Horowitz}}, \bibinfo{journal}{Phys. Rev. Lett.} \textbf{\bibinfo{volume}{101}}, \bibinfo{pages}{031601} (\bibinfo{year}{2008}{\natexlab{a}}), \eprint{0803.3295}.

\bibitem[{\citenamefont{Hartnoll et~al.}(2008{\natexlab{b}})\citenamefont{Hartnoll, Herzog, and Horowitz}}]{Hartnoll:2008kx}
\bibinfo{author}{\bibfnamefont{S.~A.} \bibnamefont{Hartnoll}}, \bibinfo{author}{\bibfnamefont{C.~P.} \bibnamefont{Herzog}}, \bibnamefont{and} \bibinfo{author}{\bibfnamefont{G.~T.} \bibnamefont{Horowitz}}, \bibinfo{journal}{JHEP} \textbf{\bibinfo{volume}{12}}, \bibinfo{pages}{015} (\bibinfo{year}{2008}{\natexlab{b}}), \eprint{0810.1563}.

\bibitem[{\citenamefont{Horowitz and Roberts}(2008)}]{Horowitz:2008bn}
\bibinfo{author}{\bibfnamefont{G.~T.} \bibnamefont{Horowitz}} \bibnamefont{and} \bibinfo{author}{\bibfnamefont{M.~M.} \bibnamefont{Roberts}}, \bibinfo{journal}{Phys. Rev. D} \textbf{\bibinfo{volume}{78}}, \bibinfo{pages}{126008} (\bibinfo{year}{2008}), \eprint{0810.1077}.

\bibitem[{\citenamefont{Horowitz}(2011)}]{Horowitz:2010gk}
\bibinfo{author}{\bibfnamefont{G.~T.} \bibnamefont{Horowitz}}, \bibinfo{journal}{Lect. Notes Phys.} \textbf{\bibinfo{volume}{828}}, \bibinfo{pages}{313} (\bibinfo{year}{2011}), \eprint{1002.1722}.

\bibitem[{\citenamefont{Mann}(1997)}]{Mann:1997iz}
\bibinfo{author}{\bibfnamefont{R.~B.} \bibnamefont{Mann}}, \bibinfo{journal}{Annals Israel Phys. Soc.} \textbf{\bibinfo{volume}{13}}, \bibinfo{pages}{311} (\bibinfo{year}{1997}), \eprint{gr-qc/9709039}.

\bibitem[{\citenamefont{Horowitz and Roberts}(2009)}]{Horowitz:2009ij}
\bibinfo{author}{\bibfnamefont{G.~T.} \bibnamefont{Horowitz}} \bibnamefont{and} \bibinfo{author}{\bibfnamefont{M.~M.} \bibnamefont{Roberts}}, \bibinfo{journal}{JHEP} \textbf{\bibinfo{volume}{11}}, \bibinfo{pages}{015} (\bibinfo{year}{2009}), \eprint{0908.3677}.

\bibitem[{\citenamefont{Witten}(1998{\natexlab{b}})}]{Witten:1998zw}
\bibinfo{author}{\bibfnamefont{E.}~\bibnamefont{Witten}}, \bibinfo{journal}{Adv. Theor. Math. Phys.} \textbf{\bibinfo{volume}{2}}, \bibinfo{pages}{505} (\bibinfo{year}{1998}{\natexlab{b}}), \eprint{hep-th/9803131}.

\bibitem[{\citenamefont{Gubser}(2005)}]{Gubser:2005ih}
\bibinfo{author}{\bibfnamefont{S.~S.} \bibnamefont{Gubser}}, \bibinfo{journal}{Class. Quant. Grav.} \textbf{\bibinfo{volume}{22}}, \bibinfo{pages}{5121} (\bibinfo{year}{2005}), \eprint{hep-th/0505189}.

\bibitem[{\citenamefont{Gubser}(2008)}]{Gubser:2008px}
\bibinfo{author}{\bibfnamefont{S.~S.} \bibnamefont{Gubser}}, \bibinfo{journal}{Phys. Rev. D} \textbf{\bibinfo{volume}{78}}, \bibinfo{pages}{065034} (\bibinfo{year}{2008}), \eprint{0801.2977}.

\bibitem[{\citenamefont{Gregory et~al.}(2009)\citenamefont{Gregory, Kanno, and Soda}}]{Gregory:2009fj}
\bibinfo{author}{\bibfnamefont{R.}~\bibnamefont{Gregory}}, \bibinfo{author}{\bibfnamefont{S.}~\bibnamefont{Kanno}}, \bibnamefont{and} \bibinfo{author}{\bibfnamefont{J.}~\bibnamefont{Soda}}, \bibinfo{journal}{JHEP} \textbf{\bibinfo{volume}{10}}, \bibinfo{pages}{010} (\bibinfo{year}{2009}), \eprint{0907.3203}.

\bibitem[{\citenamefont{Zeng and Wu}(2014)}]{Zeng:2014uoa}
\bibinfo{author}{\bibfnamefont{H.~B.} \bibnamefont{Zeng}} \bibnamefont{and} \bibinfo{author}{\bibfnamefont{J.-P.} \bibnamefont{Wu}}, \bibinfo{journal}{Phys. Rev. D} \textbf{\bibinfo{volume}{90}}, \bibinfo{pages}{046001} (\bibinfo{year}{2014}), \eprint{1404.5321}.

\bibitem[{\citenamefont{Zhou et~al.}(2015)\citenamefont{Zhou, Wu, and Ling}}]{Zhou:2015dha}
\bibinfo{author}{\bibfnamefont{Z.}~\bibnamefont{Zhou}}, \bibinfo{author}{\bibfnamefont{J.-P.} \bibnamefont{Wu}}, \bibnamefont{and} \bibinfo{author}{\bibfnamefont{Y.}~\bibnamefont{Ling}}, \bibinfo{journal}{JHEP} \textbf{\bibinfo{volume}{08}}, \bibinfo{pages}{067} (\bibinfo{year}{2015}), \eprint{1504.00535}.

\bibitem[{\citenamefont{Jiang et~al.}(2017)\citenamefont{Jiang, Liu, Lu, and Pope}}]{Jiang:2017imk}
\bibinfo{author}{\bibfnamefont{W.-J.} \bibnamefont{Jiang}}, \bibinfo{author}{\bibfnamefont{H.-S.} \bibnamefont{Liu}}, \bibinfo{author}{\bibfnamefont{H.}~\bibnamefont{Lu}}, \bibnamefont{and} \bibinfo{author}{\bibfnamefont{C.~N.} \bibnamefont{Pope}}, \bibinfo{journal}{JHEP} \textbf{\bibinfo{volume}{07}}, \bibinfo{pages}{084} (\bibinfo{year}{2017}), \eprint{1703.00922}.

\bibitem[{\citenamefont{Cisterna and Oliva}(2018)}]{Cisterna:2017qrb}
\bibinfo{author}{\bibfnamefont{A.}~\bibnamefont{Cisterna}} \bibnamefont{and} \bibinfo{author}{\bibfnamefont{J.}~\bibnamefont{Oliva}}, \bibinfo{journal}{Class. Quant. Grav.} \textbf{\bibinfo{volume}{35}}, \bibinfo{pages}{035012} (\bibinfo{year}{2018}), \eprint{1708.02916}.

\bibitem[{\citenamefont{Cisterna et~al.}(2018{\natexlab{a}})\citenamefont{Cisterna, Fuenzalida, Lagos, and Oliva}}]{Cisterna:2018mww}
\bibinfo{author}{\bibfnamefont{A.}~\bibnamefont{Cisterna}}, \bibinfo{author}{\bibfnamefont{S.}~\bibnamefont{Fuenzalida}}, \bibinfo{author}{\bibfnamefont{M.}~\bibnamefont{Lagos}}, \bibnamefont{and} \bibinfo{author}{\bibfnamefont{J.}~\bibnamefont{Oliva}}, \bibinfo{journal}{Eur. Phys. J. C} \textbf{\bibinfo{volume}{78}}, \bibinfo{pages}{982} (\bibinfo{year}{2018}{\natexlab{a}}), \eprint{1810.02798}.

\bibitem[{\citenamefont{Cisterna et~al.}(2018{\natexlab{b}})\citenamefont{Cisterna, Erices, Kuang, and Rinaldi}}]{Cisterna:2018hzf}
\bibinfo{author}{\bibfnamefont{A.}~\bibnamefont{Cisterna}}, \bibinfo{author}{\bibfnamefont{C.}~\bibnamefont{Erices}}, \bibinfo{author}{\bibfnamefont{X.-M.} \bibnamefont{Kuang}}, \bibnamefont{and} \bibinfo{author}{\bibfnamefont{M.}~\bibnamefont{Rinaldi}}, \bibinfo{journal}{Phys. Rev. D} \textbf{\bibinfo{volume}{97}}, \bibinfo{pages}{124052} (\bibinfo{year}{2018}{\natexlab{b}}), \eprint{1803.07600}.

\bibitem[{\citenamefont{Cisterna et~al.}(2019{\natexlab{a}})\citenamefont{Cisterna, Guajardo, and Hassaine}}]{Cisterna:2019uek}
\bibinfo{author}{\bibfnamefont{A.}~\bibnamefont{Cisterna}}, \bibinfo{author}{\bibfnamefont{L.}~\bibnamefont{Guajardo}}, \bibnamefont{and} \bibinfo{author}{\bibfnamefont{M.}~\bibnamefont{Hassaine}}, \bibinfo{journal}{Eur. Phys. J. C} \textbf{\bibinfo{volume}{79}}, \bibinfo{pages}{418} (\bibinfo{year}{2019}{\natexlab{a}}), \bibinfo{note}{[Erratum: Eur.Phys.J.C 79, 710 (2019)]}, \eprint{1901.00514}.

\bibitem[{\citenamefont{Caldarelli et~al.}(2013)\citenamefont{Caldarelli, Charmousis, and Hassa\"\i{}ne}}]{Caldarelli:2013gqa}
\bibinfo{author}{\bibfnamefont{M.~M.} \bibnamefont{Caldarelli}}, \bibinfo{author}{\bibfnamefont{C.}~\bibnamefont{Charmousis}}, \bibnamefont{and} \bibinfo{author}{\bibfnamefont{M.}~\bibnamefont{Hassa\"\i{}ne}}, \bibinfo{journal}{JHEP} \textbf{\bibinfo{volume}{10}}, \bibinfo{pages}{015} (\bibinfo{year}{2013}), \eprint{1307.5063}.

\bibitem[{\citenamefont{Baggioli et~al.}(2021{\natexlab{a}})\citenamefont{Baggioli, Cisterna, and Pallikaris}}]{Baggioli:2021ejg}
\bibinfo{author}{\bibfnamefont{M.}~\bibnamefont{Baggioli}}, \bibinfo{author}{\bibfnamefont{A.}~\bibnamefont{Cisterna}}, \bibnamefont{and} \bibinfo{author}{\bibfnamefont{K.}~\bibnamefont{Pallikaris}}, \bibinfo{journal}{Phys. Rev. D} \textbf{\bibinfo{volume}{104}}, \bibinfo{pages}{104067} (\bibinfo{year}{2021}{\natexlab{a}}), \eprint{2106.07458}.

\bibitem[{\citenamefont{Hernandez-Vera}(2025)}]{Hernandez-Vera:2024zui}
\bibinfo{author}{\bibfnamefont{U.}~\bibnamefont{Hernandez-Vera}}, \bibinfo{journal}{Phys. Rev. D} \textbf{\bibinfo{volume}{111}}, \bibinfo{pages}{084035} (\bibinfo{year}{2025}), \eprint{2412.19388}.

\bibitem[{\citenamefont{Chen et~al.}(2010)\citenamefont{Chen, Kao, and Wen}}]{Chen:2009pt}
\bibinfo{author}{\bibfnamefont{J.-W.} \bibnamefont{Chen}}, \bibinfo{author}{\bibfnamefont{Y.-J.} \bibnamefont{Kao}}, \bibnamefont{and} \bibinfo{author}{\bibfnamefont{W.-Y.} \bibnamefont{Wen}}, \bibinfo{journal}{Phys. Rev. D} \textbf{\bibinfo{volume}{82}}, \bibinfo{pages}{026007} (\bibinfo{year}{2010}), \eprint{0911.2821}.

\bibitem[{\citenamefont{Faulkner et~al.}(2010{\natexlab{a}})\citenamefont{Faulkner, Horowitz, McGreevy, Roberts, and Vegh}}]{Faulkner:2009am}
\bibinfo{author}{\bibfnamefont{T.}~\bibnamefont{Faulkner}}, \bibinfo{author}{\bibfnamefont{G.~T.} \bibnamefont{Horowitz}}, \bibinfo{author}{\bibfnamefont{J.}~\bibnamefont{McGreevy}}, \bibinfo{author}{\bibfnamefont{M.~M.} \bibnamefont{Roberts}}, \bibnamefont{and} \bibinfo{author}{\bibfnamefont{D.}~\bibnamefont{Vegh}}, \bibinfo{journal}{JHEP} \textbf{\bibinfo{volume}{03}}, \bibinfo{pages}{121} (\bibinfo{year}{2010}{\natexlab{a}}), \eprint{0911.3402}.

\bibitem[{\citenamefont{Faulkner et~al.}(2010{\natexlab{b}})\citenamefont{Faulkner, Iqbal, Liu, McGreevy, and Vegh}}]{Faulkner:2010da}
\bibinfo{author}{\bibfnamefont{T.}~\bibnamefont{Faulkner}}, \bibinfo{author}{\bibfnamefont{N.}~\bibnamefont{Iqbal}}, \bibinfo{author}{\bibfnamefont{H.}~\bibnamefont{Liu}}, \bibinfo{author}{\bibfnamefont{J.}~\bibnamefont{McGreevy}}, \bibnamefont{and} \bibinfo{author}{\bibfnamefont{D.}~\bibnamefont{Vegh}} (\bibinfo{year}{2010}{\natexlab{b}}), \eprint{1003.1728}.

\bibitem[{\citenamefont{Faulkner and Polchinski}(2011)}]{Faulkner:2010tq}
\bibinfo{author}{\bibfnamefont{T.}~\bibnamefont{Faulkner}} \bibnamefont{and} \bibinfo{author}{\bibfnamefont{J.}~\bibnamefont{Polchinski}}, \bibinfo{journal}{JHEP} \textbf{\bibinfo{volume}{06}}, \bibinfo{pages}{012} (\bibinfo{year}{2011}), \eprint{1001.5049}.

\bibitem[{\citenamefont{Faulkner et~al.}(2011)\citenamefont{Faulkner, Iqbal, Liu, McGreevy, and Vegh}}]{Faulkner:2011tm}
\bibinfo{author}{\bibfnamefont{T.}~\bibnamefont{Faulkner}}, \bibinfo{author}{\bibfnamefont{N.}~\bibnamefont{Iqbal}}, \bibinfo{author}{\bibfnamefont{H.}~\bibnamefont{Liu}}, \bibinfo{author}{\bibfnamefont{J.}~\bibnamefont{McGreevy}}, \bibnamefont{and} \bibinfo{author}{\bibfnamefont{D.}~\bibnamefont{Vegh}}, \bibinfo{journal}{Phil. Trans. Roy. Soc.} \textbf{\bibinfo{volume}{A 369}}, \bibinfo{pages}{1640} (\bibinfo{year}{2011}), \eprint{1101.0597}.

\bibitem[{\citenamefont{Faulkner et~al.}(2013)\citenamefont{Faulkner, Iqbal, Liu, McGreevy, and Vegh}}]{Faulkner:2013bna}
\bibinfo{author}{\bibfnamefont{T.}~\bibnamefont{Faulkner}}, \bibinfo{author}{\bibfnamefont{N.}~\bibnamefont{Iqbal}}, \bibinfo{author}{\bibfnamefont{H.}~\bibnamefont{Liu}}, \bibinfo{author}{\bibfnamefont{J.}~\bibnamefont{McGreevy}}, \bibnamefont{and} \bibinfo{author}{\bibfnamefont{D.}~\bibnamefont{Vegh}}, \bibinfo{journal}{Phys. Rev. D} \textbf{\bibinfo{volume}{88}}, \bibinfo{pages}{045016} (\bibinfo{year}{2013}), \eprint{1306.6396}.

\bibitem[{\citenamefont{Grandi et~al.}(2021)\citenamefont{Grandi, Juri\v{c}i\'c, Salazar~Landea, and Soto-Garrido}}]{Grandi:2021bsp}
\bibinfo{author}{\bibfnamefont{N.}~\bibnamefont{Grandi}}, \bibinfo{author}{\bibfnamefont{V.}~\bibnamefont{Juri\v{c}i\'c}}, \bibinfo{author}{\bibfnamefont{I.}~\bibnamefont{Salazar~Landea}}, \bibnamefont{and} \bibinfo{author}{\bibfnamefont{R.}~\bibnamefont{Soto-Garrido}}, \bibinfo{journal}{JHEP} \textbf{\bibinfo{volume}{05}}, \bibinfo{pages}{123} (\bibinfo{year}{2021}), \eprint{2103.01690}.

\bibitem[{\citenamefont{Grandi et~al.}(2022)\citenamefont{Grandi, Juri\v{c}i\'c, Landea, and Soto-Garrido}}]{Grandi:2021jkj}
\bibinfo{author}{\bibfnamefont{N.}~\bibnamefont{Grandi}}, \bibinfo{author}{\bibfnamefont{V.}~\bibnamefont{Juri\v{c}i\'c}}, \bibinfo{author}{\bibfnamefont{I.~S.} \bibnamefont{Landea}}, \bibnamefont{and} \bibinfo{author}{\bibfnamefont{R.}~\bibnamefont{Soto-Garrido}}, \bibinfo{journal}{Phys. Rev. D} \textbf{\bibinfo{volume}{105}}, \bibinfo{pages}{L081902} (\bibinfo{year}{2022}), \eprint{2112.12198}.

\bibitem[{\citenamefont{Grandi et~al.}(2024)\citenamefont{Grandi, Juri\v{c}i\'c, Salazar~Landea, and Soto-Garrido}}]{Grandi:2023jna}
\bibinfo{author}{\bibfnamefont{N.}~\bibnamefont{Grandi}}, \bibinfo{author}{\bibfnamefont{V.}~\bibnamefont{Juri\v{c}i\'c}}, \bibinfo{author}{\bibfnamefont{I.}~\bibnamefont{Salazar~Landea}}, \bibnamefont{and} \bibinfo{author}{\bibfnamefont{R.}~\bibnamefont{Soto-Garrido}}, \bibinfo{journal}{JHEP} \textbf{\bibinfo{volume}{01}}, \bibinfo{pages}{030} (\bibinfo{year}{2024}), \eprint{2304.08603}.

\bibitem[{\citenamefont{Bahamondes et~al.}(2024)\citenamefont{Bahamondes, Salazar~Landea, and Soto-Garrido}}]{Bahamondes:2024zsm}
\bibinfo{author}{\bibfnamefont{S.}~\bibnamefont{Bahamondes}}, \bibinfo{author}{\bibfnamefont{I.}~\bibnamefont{Salazar~Landea}}, \bibnamefont{and} \bibinfo{author}{\bibfnamefont{R.}~\bibnamefont{Soto-Garrido}}, \bibinfo{journal}{JHEP} \textbf{\bibinfo{volume}{09}}, \bibinfo{pages}{080} (\bibinfo{year}{2024}), \eprint{2406.00156}.

\bibitem[{\citenamefont{Lopez-Arcos et~al.}(2014)\citenamefont{Lopez-Arcos, Nastase, Rojas, and Murugan}}]{Lopez-Arcos:2013uga}
\bibinfo{author}{\bibfnamefont{C.}~\bibnamefont{Lopez-Arcos}}, \bibinfo{author}{\bibfnamefont{H.}~\bibnamefont{Nastase}}, \bibinfo{author}{\bibfnamefont{F.}~\bibnamefont{Rojas}}, \bibnamefont{and} \bibinfo{author}{\bibfnamefont{J.}~\bibnamefont{Murugan}}, \bibinfo{journal}{JHEP} \textbf{\bibinfo{volume}{01}}, \bibinfo{pages}{036} (\bibinfo{year}{2014}), \eprint{1306.1263}.

\bibitem[{\citenamefont{Andrade et~al.}(2018)\citenamefont{Andrade, Krikun, Schalm, and Zaanen}}]{Andrade:2017ghg}
\bibinfo{author}{\bibfnamefont{T.}~\bibnamefont{Andrade}}, \bibinfo{author}{\bibfnamefont{A.}~\bibnamefont{Krikun}}, \bibinfo{author}{\bibfnamefont{K.}~\bibnamefont{Schalm}}, \bibnamefont{and} \bibinfo{author}{\bibfnamefont{J.}~\bibnamefont{Zaanen}}, \bibinfo{journal}{Nature Phys.} \textbf{\bibinfo{volume}{14}}, \bibinfo{pages}{1049} (\bibinfo{year}{2018}), \eprint{1710.05791}.

\bibitem[{\citenamefont{Alejo et~al.}(2019)\citenamefont{Alejo, Goulart, and Nastase}}]{Alejo:2019utd}
\bibinfo{author}{\bibfnamefont{L.}~\bibnamefont{Alejo}}, \bibinfo{author}{\bibfnamefont{P.}~\bibnamefont{Goulart}}, \bibnamefont{and} \bibinfo{author}{\bibfnamefont{H.}~\bibnamefont{Nastase}}, \bibinfo{journal}{JHEP} \textbf{\bibinfo{volume}{09}}, \bibinfo{pages}{003} (\bibinfo{year}{2019}), \eprint{1905.04898}.

\bibitem[{\citenamefont{Donos and Gauntlett}(2014{\natexlab{a}})}]{Donos:2013eha}
\bibinfo{author}{\bibfnamefont{A.}~\bibnamefont{Donos}} \bibnamefont{and} \bibinfo{author}{\bibfnamefont{J.~P.} \bibnamefont{Gauntlett}}, \bibinfo{journal}{JHEP} \textbf{\bibinfo{volume}{04}}, \bibinfo{pages}{040} (\bibinfo{year}{2014}{\natexlab{a}}), \eprint{1311.3292}.

\bibitem[{\citenamefont{Andrade and Withers}(2014)}]{Andrade:2013gsa}
\bibinfo{author}{\bibfnamefont{T.}~\bibnamefont{Andrade}} \bibnamefont{and} \bibinfo{author}{\bibfnamefont{B.}~\bibnamefont{Withers}}, \bibinfo{journal}{JHEP} \textbf{\bibinfo{volume}{05}}, \bibinfo{pages}{101} (\bibinfo{year}{2014}), \eprint{1311.5157}.

\bibitem[{\citenamefont{Donos and Gauntlett}(2014{\natexlab{b}})}]{Donos:2014uba}
\bibinfo{author}{\bibfnamefont{A.}~\bibnamefont{Donos}} \bibnamefont{and} \bibinfo{author}{\bibfnamefont{J.~P.} \bibnamefont{Gauntlett}}, \bibinfo{journal}{JHEP} \textbf{\bibinfo{volume}{06}}, \bibinfo{pages}{007} (\bibinfo{year}{2014}{\natexlab{b}}), \eprint{1401.5077}.

\bibitem[{\citenamefont{Baggioli et~al.}(2021{\natexlab{b}})\citenamefont{Baggioli, Kim, Li, and Li}}]{Baggioli:2021xuv}
\bibinfo{author}{\bibfnamefont{M.}~\bibnamefont{Baggioli}}, \bibinfo{author}{\bibfnamefont{K.-Y.} \bibnamefont{Kim}}, \bibinfo{author}{\bibfnamefont{L.}~\bibnamefont{Li}}, \bibnamefont{and} \bibinfo{author}{\bibfnamefont{W.-J.} \bibnamefont{Li}}, \bibinfo{journal}{Sci. China Phys. Mech. Astron.} \textbf{\bibinfo{volume}{64}}, \bibinfo{pages}{270001} (\bibinfo{year}{2021}{\natexlab{b}}), \eprint{2101.01892}.

\bibitem[{\citenamefont{Sorokin}(2022)}]{Sorokin:2021tge}
\bibinfo{author}{\bibfnamefont{D.~P.} \bibnamefont{Sorokin}}, \bibinfo{journal}{Fortsch. Phys.} \textbf{\bibinfo{volume}{70}}, \bibinfo{pages}{2200092} (\bibinfo{year}{2022}), \eprint{2112.12118}.

\bibitem[{\citenamefont{Born and Infeld}(1934)}]{Born:1934gh}
\bibinfo{author}{\bibfnamefont{M.}~\bibnamefont{Born}} \bibnamefont{and} \bibinfo{author}{\bibfnamefont{L.}~\bibnamefont{Infeld}}, \bibinfo{journal}{Proc. Roy. Soc. Lond. A} \textbf{\bibinfo{volume}{144}}, \bibinfo{pages}{425} (\bibinfo{year}{1934}).

\bibitem[{\citenamefont{Heisenberg and Euler}(1936)}]{Heisenberg:1936nmg}
\bibinfo{author}{\bibfnamefont{W.}~\bibnamefont{Heisenberg}} \bibnamefont{and} \bibinfo{author}{\bibfnamefont{H.}~\bibnamefont{Euler}}, \bibinfo{journal}{Z. Phys.} \textbf{\bibinfo{volume}{98}}, \bibinfo{pages}{714} (\bibinfo{year}{1936}), \eprint{physics/0605038}.

\bibitem[{\citenamefont{Gibbons and Rasheed}(1995)}]{Gibbons:1995cv}
\bibinfo{author}{\bibfnamefont{G.~W.} \bibnamefont{Gibbons}} \bibnamefont{and} \bibinfo{author}{\bibfnamefont{D.~A.} \bibnamefont{Rasheed}}, \bibinfo{journal}{Nucl. Phys. B} \textbf{\bibinfo{volume}{454}}, \bibinfo{pages}{185} (\bibinfo{year}{1995}), \eprint{hep-th/9506035}.

\bibitem[{\citenamefont{Fradkin and Tseytlin}(1985)}]{Fradkin:1985qd}
\bibinfo{author}{\bibfnamefont{E.~S.} \bibnamefont{Fradkin}} \bibnamefont{and} \bibinfo{author}{\bibfnamefont{A.~A.} \bibnamefont{Tseytlin}}, \bibinfo{journal}{Phys. Lett. B} \textbf{\bibinfo{volume}{163}}, \bibinfo{pages}{123} (\bibinfo{year}{1985}).

\bibitem[{\citenamefont{Miskovic and Olea}(2011{\natexlab{a}})}]{Miskovic:2010ey}
\bibinfo{author}{\bibfnamefont{O.}~\bibnamefont{Miskovic}} \bibnamefont{and} \bibinfo{author}{\bibfnamefont{R.}~\bibnamefont{Olea}}, \bibinfo{journal}{Phys. Rev. D} \textbf{\bibinfo{volume}{83}}, \bibinfo{pages}{064017} (\bibinfo{year}{2011}{\natexlab{a}}), \eprint{1012.4867}.

\bibitem[{\citenamefont{Miskovic and Olea}(2011{\natexlab{b}})}]{Miskovic:2010ui}
\bibinfo{author}{\bibfnamefont{O.}~\bibnamefont{Miskovic}} \bibnamefont{and} \bibinfo{author}{\bibfnamefont{R.}~\bibnamefont{Olea}}, \bibinfo{journal}{Phys. Rev. D} \textbf{\bibinfo{volume}{83}}, \bibinfo{pages}{024011} (\bibinfo{year}{2011}{\natexlab{b}}), \eprint{1009.5763}.

\bibitem[{\citenamefont{Dey et~al.}(2018)\citenamefont{Dey, Roy, and Sarkar}}]{Dey:2016pei}
\bibinfo{author}{\bibfnamefont{A.}~\bibnamefont{Dey}}, \bibinfo{author}{\bibfnamefont{P.}~\bibnamefont{Roy}}, \bibnamefont{and} \bibinfo{author}{\bibfnamefont{T.}~\bibnamefont{Sarkar}}, \bibinfo{journal}{JHEP} \textbf{\bibinfo{volume}{04}}, \bibinfo{pages}{098} (\bibinfo{year}{2018}), \eprint{1609.02290}.

\bibitem[{\citenamefont{Cano et~al.}(2022)\citenamefont{Cano, Murcia, Rivadulla~S\'anchez, and Zhang}}]{Cano:2022ord}
\bibinfo{author}{\bibfnamefont{P.~A.} \bibnamefont{Cano}}, \bibinfo{author}{\bibfnamefont{A.~J.} \bibnamefont{Murcia}}, \bibinfo{author}{\bibfnamefont{A.}~\bibnamefont{Rivadulla~S\'anchez}}, \bibnamefont{and} \bibinfo{author}{\bibfnamefont{X.}~\bibnamefont{Zhang}}, \bibinfo{journal}{JHEP} \textbf{\bibinfo{volume}{07}}, \bibinfo{pages}{010} (\bibinfo{year}{2022}), \eprint{2202.10473}.

\bibitem[{\citenamefont{Arenas-Henriquez et~al.}(2023{\natexlab{a}})\citenamefont{Arenas-Henriquez, Diaz, and Novoa}}]{Arenas-Henriquez:2022ntz}
\bibinfo{author}{\bibfnamefont{G.}~\bibnamefont{Arenas-Henriquez}}, \bibinfo{author}{\bibfnamefont{F.}~\bibnamefont{Diaz}}, \bibnamefont{and} \bibinfo{author}{\bibfnamefont{Y.}~\bibnamefont{Novoa}}, \bibinfo{journal}{JHEP} \textbf{\bibinfo{volume}{05}}, \bibinfo{pages}{072} (\bibinfo{year}{2023}{\natexlab{a}}), \eprint{2211.06355}.

\bibitem[{\citenamefont{Bravo-Gaete et~al.}(2022)\citenamefont{Bravo-Gaete, Guajardo, and Oliva}}]{Bravo-Gaete:2022mnr}
\bibinfo{author}{\bibfnamefont{M.}~\bibnamefont{Bravo-Gaete}}, \bibinfo{author}{\bibfnamefont{L.}~\bibnamefont{Guajardo}}, \bibnamefont{and} \bibinfo{author}{\bibfnamefont{J.}~\bibnamefont{Oliva}}, \bibinfo{journal}{Phys. Rev. D} \textbf{\bibinfo{volume}{106}}, \bibinfo{pages}{024017} (\bibinfo{year}{2022}), \eprint{2205.09282}.

\bibitem[{\citenamefont{Bueno et~al.}(2023)\citenamefont{Bueno, Cano, Moreno, and van~der Velde}}]{Bueno:2022ewf}
\bibinfo{author}{\bibfnamefont{P.}~\bibnamefont{Bueno}}, \bibinfo{author}{\bibfnamefont{P.~A.} \bibnamefont{Cano}}, \bibinfo{author}{\bibfnamefont{J.}~\bibnamefont{Moreno}}, \bibnamefont{and} \bibinfo{author}{\bibfnamefont{G.}~\bibnamefont{van~der Velde}}, \bibinfo{journal}{Phys. Rev. D} \textbf{\bibinfo{volume}{107}}, \bibinfo{pages}{064050} (\bibinfo{year}{2023}), \eprint{2212.00637}.

\bibitem[{\citenamefont{\'Alvarez et~al.}(2022)\citenamefont{\'Alvarez, Bravo-Gaete, Ju\'arez-Aubry, and Rodr\'\i{}guez}}]{Alvarez:2022upr}
\bibinfo{author}{\bibfnamefont{A.}~\bibnamefont{\'Alvarez}}, \bibinfo{author}{\bibfnamefont{M.}~\bibnamefont{Bravo-Gaete}}, \bibinfo{author}{\bibfnamefont{M.~M.} \bibnamefont{Ju\'arez-Aubry}}, \bibnamefont{and} \bibinfo{author}{\bibfnamefont{G.~V.} \bibnamefont{Rodr\'\i{}guez}}, \bibinfo{journal}{Phys. Rev. D} \textbf{\bibinfo{volume}{105}}, \bibinfo{pages}{084032} (\bibinfo{year}{2022}), \eprint{2202.11252}.

\bibitem[{\citenamefont{Santos et~al.}(2024)\citenamefont{Santos, Bravo-Gaete, Ferreira, and Casana}}]{Santos:2023mee}
\bibinfo{author}{\bibfnamefont{F.~F.} \bibnamefont{Santos}}, \bibinfo{author}{\bibfnamefont{M.}~\bibnamefont{Bravo-Gaete}}, \bibinfo{author}{\bibfnamefont{M.~M.} \bibnamefont{Ferreira}}, \bibnamefont{and} \bibinfo{author}{\bibfnamefont{R.}~\bibnamefont{Casana}}, \bibinfo{journal}{Fortsch. Phys.} \textbf{\bibinfo{volume}{72}}, \bibinfo{pages}{2400088} (\bibinfo{year}{2024}), \eprint{2310.17092}.

\bibitem[{\citenamefont{Bravo-Gaete et~al.}(2023)\citenamefont{Bravo-Gaete, Guajardo, and Santos}}]{Bravo-Gaete:2023iry}
\bibinfo{author}{\bibfnamefont{M.}~\bibnamefont{Bravo-Gaete}}, \bibinfo{author}{\bibfnamefont{L.}~\bibnamefont{Guajardo}}, \bibnamefont{and} \bibinfo{author}{\bibfnamefont{F.~F.} \bibnamefont{Santos}}, \bibinfo{journal}{Phys. Rev. D} \textbf{\bibinfo{volume}{107}}, \bibinfo{pages}{104032} (\bibinfo{year}{2023}), \eprint{2303.07493}.

\bibitem[{\citenamefont{Santos et~al.}(2023)\citenamefont{Santos, Bravo-Gaete, Sokoliuk, and Baransky}}]{Santos:2023flb}
\bibinfo{author}{\bibfnamefont{F.~F.} \bibnamefont{Santos}}, \bibinfo{author}{\bibfnamefont{M.}~\bibnamefont{Bravo-Gaete}}, \bibinfo{author}{\bibfnamefont{O.}~\bibnamefont{Sokoliuk}}, \bibnamefont{and} \bibinfo{author}{\bibfnamefont{A.}~\bibnamefont{Baransky}}, \bibinfo{journal}{Fortsch. Phys.} \textbf{\bibinfo{volume}{71}}, \bibinfo{pages}{2300008} (\bibinfo{year}{2023}), \eprint{2301.03121}.

\bibitem[{\citenamefont{Colip\'\i{}-Marchant et~al.}(2023)\citenamefont{Colip\'\i{}-Marchant, Corral, Flores-Alfonso, and Sanhueza}}]{Colipi-Marchant:2023awk}
\bibinfo{author}{\bibfnamefont{F.}~\bibnamefont{Colip\'\i{}-Marchant}}, \bibinfo{author}{\bibfnamefont{C.}~\bibnamefont{Corral}}, \bibinfo{author}{\bibfnamefont{D.}~\bibnamefont{Flores-Alfonso}}, \bibnamefont{and} \bibinfo{author}{\bibfnamefont{L.}~\bibnamefont{Sanhueza}}, \bibinfo{journal}{Phys. Rev. D} \textbf{\bibinfo{volume}{107}}, \bibinfo{pages}{104042} (\bibinfo{year}{2023}), \eprint{2302.09162}.

\bibitem[{\citenamefont{Bravo-Gaete et~al.}(2025)\citenamefont{Bravo-Gaete, Santos, Herrera-Mendoza, and Higuita-Borja}}]{Bravo-Gaete:2025vyd}
\bibinfo{author}{\bibfnamefont{M.}~\bibnamefont{Bravo-Gaete}}, \bibinfo{author}{\bibfnamefont{F.~F.} \bibnamefont{Santos}}, \bibinfo{author}{\bibfnamefont{J.~A.} \bibnamefont{Herrera-Mendoza}}, \bibnamefont{and} \bibinfo{author}{\bibfnamefont{D.~F.} \bibnamefont{Higuita-Borja}} (\bibinfo{year}{2025}), \eprint{2504.17081}.

\bibitem[{\citenamefont{Barbosa et~al.}(2025)\citenamefont{Barbosa, Fichet, and de~Souza}}]{Barbosa:2025smt}
\bibinfo{author}{\bibfnamefont{S.}~\bibnamefont{Barbosa}}, \bibinfo{author}{\bibfnamefont{S.}~\bibnamefont{Fichet}}, \bibnamefont{and} \bibinfo{author}{\bibfnamefont{L.}~\bibnamefont{de~Souza}} (\bibinfo{year}{2025}), \eprint{2503.20910}.

\bibitem[{\citenamefont{Soleng}(1995)}]{Soleng:1995kn}
\bibinfo{author}{\bibfnamefont{H.~H.} \bibnamefont{Soleng}}, \bibinfo{journal}{Phys. Rev. D} \textbf{\bibinfo{volume}{52}}, \bibinfo{pages}{6178} (\bibinfo{year}{1995}), \eprint{hep-th/9509033}.

\bibitem[{\citenamefont{Ayon-Beato and Garcia}(1998)}]{Ayon-Beato:1998hmi}
\bibinfo{author}{\bibfnamefont{E.}~\bibnamefont{Ayon-Beato}} \bibnamefont{and} \bibinfo{author}{\bibfnamefont{A.}~\bibnamefont{Garcia}}, \bibinfo{journal}{Phys. Rev. Lett.} \textbf{\bibinfo{volume}{80}}, \bibinfo{pages}{5056} (\bibinfo{year}{1998}), \eprint{gr-qc/9911046}.

\bibitem[{\citenamefont{Gonzalez et~al.}(2009)\citenamefont{Gonzalez, Hassaine, and Martinez}}]{Gonzalez:2009nn}
\bibinfo{author}{\bibfnamefont{H.~A.} \bibnamefont{Gonzalez}}, \bibinfo{author}{\bibfnamefont{M.}~\bibnamefont{Hassaine}}, \bibnamefont{and} \bibinfo{author}{\bibfnamefont{C.}~\bibnamefont{Martinez}}, \bibinfo{journal}{Phys. Rev. D} \textbf{\bibinfo{volume}{80}}, \bibinfo{pages}{104008} (\bibinfo{year}{2009}), \eprint{0909.1365}.

\bibitem[{\citenamefont{Bronnikov}(2001)}]{Bronnikov:2000vy}
\bibinfo{author}{\bibfnamefont{K.~A.} \bibnamefont{Bronnikov}}, \bibinfo{journal}{Phys. Rev. D} \textbf{\bibinfo{volume}{63}}, \bibinfo{pages}{044005} (\bibinfo{year}{2001}), \eprint{gr-qc/0006014}.

\bibitem[{\citenamefont{Ayon-Beato and Garcia}(2000)}]{Ayon-Beato:2000mjt}
\bibinfo{author}{\bibfnamefont{E.}~\bibnamefont{Ayon-Beato}} \bibnamefont{and} \bibinfo{author}{\bibfnamefont{A.}~\bibnamefont{Garcia}}, \bibinfo{journal}{Phys. Lett. B} \textbf{\bibinfo{volume}{493}}, \bibinfo{pages}{149} (\bibinfo{year}{2000}), \eprint{gr-qc/0009077}.

\bibitem[{\citenamefont{Cataldo et~al.}(2000)\citenamefont{Cataldo, Cruz, del Campo, and Garcia}}]{Cataldo:2000we}
\bibinfo{author}{\bibfnamefont{M.}~\bibnamefont{Cataldo}}, \bibinfo{author}{\bibfnamefont{N.}~\bibnamefont{Cruz}}, \bibinfo{author}{\bibfnamefont{S.}~\bibnamefont{del Campo}}, \bibnamefont{and} \bibinfo{author}{\bibfnamefont{A.}~\bibnamefont{Garcia}}, \bibinfo{journal}{Phys. Lett. B} \textbf{\bibinfo{volume}{484}}, \bibinfo{pages}{154} (\bibinfo{year}{2000}), \eprint{hep-th/0008138}.

\bibitem[{\citenamefont{Hassaine and Martinez}(2007)}]{Hassaine:2007py}
\bibinfo{author}{\bibfnamefont{M.}~\bibnamefont{Hassaine}} \bibnamefont{and} \bibinfo{author}{\bibfnamefont{C.}~\bibnamefont{Martinez}}, \bibinfo{journal}{Phys. Rev. D} \textbf{\bibinfo{volume}{75}}, \bibinfo{pages}{027502} (\bibinfo{year}{2007}), \eprint{hep-th/0701058}.

\bibitem[{\citenamefont{Hassaine and Martinez}(2008)}]{Hassaine:2008pw}
\bibinfo{author}{\bibfnamefont{M.}~\bibnamefont{Hassaine}} \bibnamefont{and} \bibinfo{author}{\bibfnamefont{C.}~\bibnamefont{Martinez}}, \bibinfo{journal}{Class. Quant. Grav.} \textbf{\bibinfo{volume}{25}}, \bibinfo{pages}{195023} (\bibinfo{year}{2008}), \eprint{0803.2946}.

\bibitem[{\citenamefont{Maeda et~al.}(2009)\citenamefont{Maeda, Hassaine, and Martinez}}]{Maeda:2008ha}
\bibinfo{author}{\bibfnamefont{H.}~\bibnamefont{Maeda}}, \bibinfo{author}{\bibfnamefont{M.}~\bibnamefont{Hassaine}}, \bibnamefont{and} \bibinfo{author}{\bibfnamefont{C.}~\bibnamefont{Martinez}}, \bibinfo{journal}{Phys. Rev. D} \textbf{\bibinfo{volume}{79}}, \bibinfo{pages}{044012} (\bibinfo{year}{2009}), \eprint{0812.2038}.

\bibitem[{\citenamefont{Cembranos et~al.}(2015)\citenamefont{Cembranos, de~la Cruz-Dombriz, and Jarillo}}]{Cembranos:2014hwa}
\bibinfo{author}{\bibfnamefont{J.~A.~R.} \bibnamefont{Cembranos}}, \bibinfo{author}{\bibfnamefont{A.}~\bibnamefont{de~la Cruz-Dombriz}}, \bibnamefont{and} \bibinfo{author}{\bibfnamefont{J.}~\bibnamefont{Jarillo}}, \bibinfo{journal}{JCAP} \textbf{\bibinfo{volume}{02}}, \bibinfo{pages}{042} (\bibinfo{year}{2015}), \eprint{1407.4383}.

\bibitem[{\citenamefont{Barrientos et~al.}(2016)\citenamefont{Barrientos, Gonz\'alez, and V\'asquez}}]{Barrientos:2016ubi}
\bibinfo{author}{\bibfnamefont{J.}~\bibnamefont{Barrientos}}, \bibinfo{author}{\bibfnamefont{P.~A.} \bibnamefont{Gonz\'alez}}, \bibnamefont{and} \bibinfo{author}{\bibfnamefont{Y.}~\bibnamefont{V\'asquez}}, \bibinfo{journal}{Eur. Phys. J. C} \textbf{\bibinfo{volume}{76}}, \bibinfo{pages}{677} (\bibinfo{year}{2016}), \eprint{1603.05571}.

\bibitem[{\citenamefont{Cisterna et~al.}(2016)\citenamefont{Cisterna, Hassaine, Oliva, and Rinaldi}}]{Cisterna:2016nwq}
\bibinfo{author}{\bibfnamefont{A.}~\bibnamefont{Cisterna}}, \bibinfo{author}{\bibfnamefont{M.}~\bibnamefont{Hassaine}}, \bibinfo{author}{\bibfnamefont{J.}~\bibnamefont{Oliva}}, \bibnamefont{and} \bibinfo{author}{\bibfnamefont{M.}~\bibnamefont{Rinaldi}}, \bibinfo{journal}{Phys. Rev. D} \textbf{\bibinfo{volume}{94}}, \bibinfo{pages}{104039} (\bibinfo{year}{2016}), \eprint{1609.03430}.

\bibitem[{\citenamefont{Eslam~Panah}(2023)}]{EslamPanah:2022ihg}
\bibinfo{author}{\bibfnamefont{B.}~\bibnamefont{Eslam~Panah}}, \bibinfo{journal}{Fortsch. Phys.} \textbf{\bibinfo{volume}{71}}, \bibinfo{pages}{2300012} (\bibinfo{year}{2023}), \eprint{2203.12619}.

\bibitem[{\citenamefont{Jing and Chen}(2010)}]{Jing:2010zp}
\bibinfo{author}{\bibfnamefont{J.}~\bibnamefont{Jing}} \bibnamefont{and} \bibinfo{author}{\bibfnamefont{S.}~\bibnamefont{Chen}}, \bibinfo{journal}{Phys. Lett. B} \textbf{\bibinfo{volume}{686}}, \bibinfo{pages}{68} (\bibinfo{year}{2010}), \eprint{1001.4227}.

\bibitem[{\citenamefont{Jing et~al.}(2011)\citenamefont{Jing, Pan, and Chen}}]{Jing:2011vz}
\bibinfo{author}{\bibfnamefont{J.}~\bibnamefont{Jing}}, \bibinfo{author}{\bibfnamefont{Q.}~\bibnamefont{Pan}}, \bibnamefont{and} \bibinfo{author}{\bibfnamefont{S.}~\bibnamefont{Chen}}, \bibinfo{journal}{JHEP} \textbf{\bibinfo{volume}{11}}, \bibinfo{pages}{045} (\bibinfo{year}{2011}), \eprint{1106.5181}.

\bibitem[{\citenamefont{Gangopadhyay and Roychowdhury}(2012)}]{Gangopadhyay:2012gx}
\bibinfo{author}{\bibfnamefont{S.}~\bibnamefont{Gangopadhyay}} \bibnamefont{and} \bibinfo{author}{\bibfnamefont{D.}~\bibnamefont{Roychowdhury}}, \bibinfo{journal}{JHEP} \textbf{\bibinfo{volume}{08}}, \bibinfo{pages}{104} (\bibinfo{year}{2012}), \eprint{1207.5605}.

\bibitem[{\citenamefont{Liu et~al.}(2016)\citenamefont{Liu, Gong, and Wang}}]{Liu:2015lit}
\bibinfo{author}{\bibfnamefont{Y.}~\bibnamefont{Liu}}, \bibinfo{author}{\bibfnamefont{Y.}~\bibnamefont{Gong}}, \bibnamefont{and} \bibinfo{author}{\bibfnamefont{B.}~\bibnamefont{Wang}}, \bibinfo{journal}{JHEP} \textbf{\bibinfo{volume}{02}}, \bibinfo{pages}{116} (\bibinfo{year}{2016}), \eprint{1505.03603}.

\bibitem[{\citenamefont{Baggioli and Pujolas}(2016)}]{Baggioli:2016oju}
\bibinfo{author}{\bibfnamefont{M.}~\bibnamefont{Baggioli}} \bibnamefont{and} \bibinfo{author}{\bibfnamefont{O.}~\bibnamefont{Pujolas}}, \bibinfo{journal}{JHEP} \textbf{\bibinfo{volume}{12}}, \bibinfo{pages}{107} (\bibinfo{year}{2016}), \eprint{1604.08915}.

\bibitem[{\citenamefont{Wang et~al.}(2019)\citenamefont{Wang, Wu, and Yang}}]{Wang:2018hwg}
\bibinfo{author}{\bibfnamefont{P.}~\bibnamefont{Wang}}, \bibinfo{author}{\bibfnamefont{H.}~\bibnamefont{Wu}}, \bibnamefont{and} \bibinfo{author}{\bibfnamefont{H.}~\bibnamefont{Yang}}, \bibinfo{journal}{Eur. Phys. J. C} \textbf{\bibinfo{volume}{79}}, \bibinfo{pages}{6} (\bibinfo{year}{2019}), \eprint{1805.07913}.

\bibitem[{\citenamefont{Chakraborty}(2020)}]{Chakraborty:2019vld}
\bibinfo{author}{\bibfnamefont{A.}~\bibnamefont{Chakraborty}}, \bibinfo{journal}{Class. Quant. Grav.} \textbf{\bibinfo{volume}{37}}, \bibinfo{pages}{065021} (\bibinfo{year}{2020}), \eprint{1903.00613}.

\bibitem[{\citenamefont{An et~al.}(2020)\citenamefont{An, Ji, and Li}}]{An:2020tkn}
\bibinfo{author}{\bibfnamefont{Y.-S.} \bibnamefont{An}}, \bibinfo{author}{\bibfnamefont{T.}~\bibnamefont{Ji}}, \bibnamefont{and} \bibinfo{author}{\bibfnamefont{L.}~\bibnamefont{Li}}, \bibinfo{journal}{JHEP} \textbf{\bibinfo{volume}{10}}, \bibinfo{pages}{023} (\bibinfo{year}{2020}), \eprint{2007.13918}.

\bibitem[{\citenamefont{Lai and Pan}(2022)}]{Lai:2021mxa}
\bibinfo{author}{\bibfnamefont{C.}~\bibnamefont{Lai}} \bibnamefont{and} \bibinfo{author}{\bibfnamefont{Q.}~\bibnamefont{Pan}}, \bibinfo{journal}{Nucl. Phys. B} \textbf{\bibinfo{volume}{974}}, \bibinfo{pages}{115615} (\bibinfo{year}{2022}).

\bibitem[{\citenamefont{Bandos et~al.}(2020)\citenamefont{Bandos, Lechner, Sorokin, and Townsend}}]{Bandos:2020jsw}
\bibinfo{author}{\bibfnamefont{I.}~\bibnamefont{Bandos}}, \bibinfo{author}{\bibfnamefont{K.}~\bibnamefont{Lechner}}, \bibinfo{author}{\bibfnamefont{D.}~\bibnamefont{Sorokin}}, \bibnamefont{and} \bibinfo{author}{\bibfnamefont{P.~K.} \bibnamefont{Townsend}}, \bibinfo{journal}{Phys. Rev. D} \textbf{\bibinfo{volume}{102}}, \bibinfo{pages}{121703} (\bibinfo{year}{2020}), \eprint{2007.09092}.

\bibitem[{\citenamefont{Kosyakov}(2020)}]{Kosyakov:2020wxv}
\bibinfo{author}{\bibfnamefont{B.~P.} \bibnamefont{Kosyakov}}, \bibinfo{journal}{Phys. Lett. B} \textbf{\bibinfo{volume}{810}}, \bibinfo{pages}{135840} (\bibinfo{year}{2020}), \eprint{2007.13878}.

\bibitem[{\citenamefont{Bandos et~al.}(2021)\citenamefont{Bandos, Lechner, Sorokin, and Townsend}}]{Bandos:2020hgy}
\bibinfo{author}{\bibfnamefont{I.}~\bibnamefont{Bandos}}, \bibinfo{author}{\bibfnamefont{K.}~\bibnamefont{Lechner}}, \bibinfo{author}{\bibfnamefont{D.}~\bibnamefont{Sorokin}}, \bibnamefont{and} \bibinfo{author}{\bibfnamefont{P.~K.} \bibnamefont{Townsend}}, \bibinfo{journal}{JHEP} \textbf{\bibinfo{volume}{03}}, \bibinfo{pages}{022} (\bibinfo{year}{2021}), \eprint{2012.09286}.

\bibitem[{\citenamefont{Nastase}(2022)}]{Nastase:2021ffq}
\bibinfo{author}{\bibfnamefont{H.}~\bibnamefont{Nastase}}, \bibinfo{journal}{Phys. Rev. D} \textbf{\bibinfo{volume}{105}}, \bibinfo{pages}{105024} (\bibinfo{year}{2022}), \eprint{2112.01234}.

\bibitem[{\citenamefont{Hartnoll and Herzog}(2007)}]{Hartnoll:2007ip}
\bibinfo{author}{\bibfnamefont{S.~A.} \bibnamefont{Hartnoll}} \bibnamefont{and} \bibinfo{author}{\bibfnamefont{C.~P.} \bibnamefont{Herzog}}, \bibinfo{journal}{Phys. Rev. D} \textbf{\bibinfo{volume}{76}}, \bibinfo{pages}{106012} (\bibinfo{year}{2007}), \eprint{0706.3228}.

\bibitem[{\citenamefont{Hartnoll and Kovtun}(2007)}]{Hartnoll:2007ai}
\bibinfo{author}{\bibfnamefont{S.~A.} \bibnamefont{Hartnoll}} \bibnamefont{and} \bibinfo{author}{\bibfnamefont{P.}~\bibnamefont{Kovtun}}, \bibinfo{journal}{Phys. Rev. D} \textbf{\bibinfo{volume}{76}}, \bibinfo{pages}{066001} (\bibinfo{year}{2007}), \eprint{0704.1160}.

\bibitem[{\citenamefont{Blake and Donos}(2015)}]{Blake:2014yla}
\bibinfo{author}{\bibfnamefont{M.}~\bibnamefont{Blake}} \bibnamefont{and} \bibinfo{author}{\bibfnamefont{A.}~\bibnamefont{Donos}}, \bibinfo{journal}{Phys. Rev. Lett.} \textbf{\bibinfo{volume}{114}}, \bibinfo{pages}{021601} (\bibinfo{year}{2015}), \eprint{1406.1659}.

\bibitem[{\citenamefont{Amoretti and Musso}(2015)}]{Amoretti:2015gna}
\bibinfo{author}{\bibfnamefont{A.}~\bibnamefont{Amoretti}} \bibnamefont{and} \bibinfo{author}{\bibfnamefont{D.}~\bibnamefont{Musso}}, \bibinfo{journal}{JHEP} \textbf{\bibinfo{volume}{09}}, \bibinfo{pages}{094} (\bibinfo{year}{2015}), \eprint{1502.02631}.

\bibitem[{\citenamefont{Amoretti et~al.}(2020)\citenamefont{Amoretti, Brattan, Magnoli, and Scanavino}}]{Amoretti:2020mkp}
\bibinfo{author}{\bibfnamefont{A.}~\bibnamefont{Amoretti}}, \bibinfo{author}{\bibfnamefont{D.~K.} \bibnamefont{Brattan}}, \bibinfo{author}{\bibfnamefont{N.}~\bibnamefont{Magnoli}}, \bibnamefont{and} \bibinfo{author}{\bibfnamefont{M.}~\bibnamefont{Scanavino}}, \bibinfo{journal}{JHEP} \textbf{\bibinfo{volume}{08}}, \bibinfo{pages}{097} (\bibinfo{year}{2020}), \eprint{2005.09662}.

\bibitem[{\citenamefont{Amoretti et~al.}(2021{\natexlab{a}})\citenamefont{Amoretti, Arean, Brattan, and Magnoli}}]{Amoretti:2021fch}
\bibinfo{author}{\bibfnamefont{A.}~\bibnamefont{Amoretti}}, \bibinfo{author}{\bibfnamefont{D.}~\bibnamefont{Arean}}, \bibinfo{author}{\bibfnamefont{D.~K.} \bibnamefont{Brattan}}, \bibnamefont{and} \bibinfo{author}{\bibfnamefont{N.}~\bibnamefont{Magnoli}}, \bibinfo{journal}{JHEP} \textbf{\bibinfo{volume}{05}}, \bibinfo{pages}{027} (\bibinfo{year}{2021}{\natexlab{a}}), \eprint{2101.05343}.

\bibitem[{\citenamefont{Amoretti et~al.}(2021{\natexlab{b}})\citenamefont{Amoretti, Arean, Brattan, and Martinoia}}]{Amoretti:2021lll}
\bibinfo{author}{\bibfnamefont{A.}~\bibnamefont{Amoretti}}, \bibinfo{author}{\bibfnamefont{D.}~\bibnamefont{Arean}}, \bibinfo{author}{\bibfnamefont{D.~K.} \bibnamefont{Brattan}}, \bibnamefont{and} \bibinfo{author}{\bibfnamefont{L.}~\bibnamefont{Martinoia}}, \bibinfo{journal}{JHEP} \textbf{\bibinfo{volume}{11}}, \bibinfo{pages}{011} (\bibinfo{year}{2021}{\natexlab{b}}), \eprint{2107.00519}.

\bibitem[{\citenamefont{Ge and Xu}(2024)}]{Ge:2023yom}
\bibinfo{author}{\bibfnamefont{X.-H.} \bibnamefont{Ge}} \bibnamefont{and} \bibinfo{author}{\bibfnamefont{Z.}~\bibnamefont{Xu}}, \bibinfo{journal}{JHEP} \textbf{\bibinfo{volume}{03}}, \bibinfo{pages}{069} (\bibinfo{year}{2024}), \eprint{2310.12067}.

\bibitem[{\citenamefont{Domenech et~al.}(2010)\citenamefont{Domenech, Montull, Pomarol, Salvio, and Silva}}]{Domenech:2010nf}
\bibinfo{author}{\bibfnamefont{O.}~\bibnamefont{Domenech}}, \bibinfo{author}{\bibfnamefont{M.}~\bibnamefont{Montull}}, \bibinfo{author}{\bibfnamefont{A.}~\bibnamefont{Pomarol}}, \bibinfo{author}{\bibfnamefont{A.}~\bibnamefont{Salvio}}, \bibnamefont{and} \bibinfo{author}{\bibfnamefont{P.~J.} \bibnamefont{Silva}}, \bibinfo{journal}{JHEP} \textbf{\bibinfo{volume}{08}}, \bibinfo{pages}{033} (\bibinfo{year}{2010}), \eprint{1005.1776}.

\bibitem[{\citenamefont{Montull et~al.}(2011)\citenamefont{Montull, Pujolas, Salvio, and Silva}}]{Montull:2011im}
\bibinfo{author}{\bibfnamefont{M.}~\bibnamefont{Montull}}, \bibinfo{author}{\bibfnamefont{O.}~\bibnamefont{Pujolas}}, \bibinfo{author}{\bibfnamefont{A.}~\bibnamefont{Salvio}}, \bibnamefont{and} \bibinfo{author}{\bibfnamefont{P.~J.} \bibnamefont{Silva}}, \bibinfo{journal}{Phys. Rev. Lett.} \textbf{\bibinfo{volume}{107}}, \bibinfo{pages}{181601} (\bibinfo{year}{2011}), \eprint{1105.5392}.

\bibitem[{\citenamefont{Montull et~al.}(2012)\citenamefont{Montull, Pujolas, Salvio, and Silva}}]{Montull:2012fy}
\bibinfo{author}{\bibfnamefont{M.}~\bibnamefont{Montull}}, \bibinfo{author}{\bibfnamefont{O.}~\bibnamefont{Pujolas}}, \bibinfo{author}{\bibfnamefont{A.}~\bibnamefont{Salvio}}, \bibnamefont{and} \bibinfo{author}{\bibfnamefont{P.~J.} \bibnamefont{Silva}}, \bibinfo{journal}{JHEP} \textbf{\bibinfo{volume}{04}}, \bibinfo{pages}{135} (\bibinfo{year}{2012}), \eprint{1202.0006}.

\bibitem[{\citenamefont{Salvio}(2012)}]{Salvio:2012at}
\bibinfo{author}{\bibfnamefont{A.}~\bibnamefont{Salvio}}, \bibinfo{journal}{JHEP} \textbf{\bibinfo{volume}{09}}, \bibinfo{pages}{134} (\bibinfo{year}{2012}), \eprint{1207.3800}.

\bibitem[{\citenamefont{Salvio}(2013)}]{Salvio:2013jia}
\bibinfo{author}{\bibfnamefont{A.}~\bibnamefont{Salvio}}, \bibinfo{journal}{JHEP} \textbf{\bibinfo{volume}{03}}, \bibinfo{pages}{136} (\bibinfo{year}{2013}), \eprint{1302.4898}.

\bibitem[{\citenamefont{Natsuume and Okamura}(2022)}]{Natsuume:2022kic}
\bibinfo{author}{\bibfnamefont{M.}~\bibnamefont{Natsuume}} \bibnamefont{and} \bibinfo{author}{\bibfnamefont{T.}~\bibnamefont{Okamura}}, \bibinfo{journal}{Phys. Rev. D} \textbf{\bibinfo{volume}{106}}, \bibinfo{pages}{086005} (\bibinfo{year}{2022}), \eprint{2207.07182}.

\bibitem[{\citenamefont{Hartnoll et~al.}(2007)\citenamefont{Hartnoll, Kovtun, Muller, and Sachdev}}]{Hartnoll:2007ih}
\bibinfo{author}{\bibfnamefont{S.~A.} \bibnamefont{Hartnoll}}, \bibinfo{author}{\bibfnamefont{P.~K.} \bibnamefont{Kovtun}}, \bibinfo{author}{\bibfnamefont{M.}~\bibnamefont{Muller}}, \bibnamefont{and} \bibinfo{author}{\bibfnamefont{S.}~\bibnamefont{Sachdev}}, \bibinfo{journal}{Phys. Rev. B} \textbf{\bibinfo{volume}{76}}, \bibinfo{pages}{144502} (\bibinfo{year}{2007}), \eprint{0706.3215}.

\bibitem[{\citenamefont{Kim et~al.}(2015)\citenamefont{Kim, Kim, Seo, and Sin}}]{Kim:2015wba}
\bibinfo{author}{\bibfnamefont{K.-Y.} \bibnamefont{Kim}}, \bibinfo{author}{\bibfnamefont{K.~K.} \bibnamefont{Kim}}, \bibinfo{author}{\bibfnamefont{Y.}~\bibnamefont{Seo}}, \bibnamefont{and} \bibinfo{author}{\bibfnamefont{S.-J.} \bibnamefont{Sin}}, \bibinfo{journal}{JHEP} \textbf{\bibinfo{volume}{07}}, \bibinfo{pages}{027} (\bibinfo{year}{2015}), \eprint{1502.05386}.

\bibitem[{\citenamefont{Chien et~al.}(1991)\citenamefont{Chien, Wang, and Ong}}]{PhysRevLett.67.2088}
\bibinfo{author}{\bibfnamefont{T.~R.} \bibnamefont{Chien}}, \bibinfo{author}{\bibfnamefont{Z.~Z.} \bibnamefont{Wang}}, \bibnamefont{and} \bibinfo{author}{\bibfnamefont{N.~P.} \bibnamefont{Ong}}, \bibinfo{journal}{Phys. Rev. Lett.} \textbf{\bibinfo{volume}{67}}, \bibinfo{pages}{2088} (\bibinfo{year}{1991}).

\bibitem[{\citenamefont{Tyler and Mackenzie}(1997)}]{TYLER19971185}
\bibinfo{author}{\bibfnamefont{A.}~\bibnamefont{Tyler}} \bibnamefont{and} \bibinfo{author}{\bibfnamefont{A.}~\bibnamefont{Mackenzie}}, \bibinfo{journal}{Physica C: Superconductivity} \textbf{\bibinfo{volume}{282-287}}, \bibinfo{pages}{1185} (\bibinfo{year}{1997}), \bibinfo{note}{proceedings of the International Conference on Materials and Mechanisms of Superconductivity High Temperature Superconductors V}.

\bibitem[{\citenamefont{Blauvelt et~al.}(2018)\citenamefont{Blauvelt, Cremonini, Hoover, Li, and Waskie}}]{Blauvelt:2017koq}
\bibinfo{author}{\bibfnamefont{E.}~\bibnamefont{Blauvelt}}, \bibinfo{author}{\bibfnamefont{S.}~\bibnamefont{Cremonini}}, \bibinfo{author}{\bibfnamefont{A.}~\bibnamefont{Hoover}}, \bibinfo{author}{\bibfnamefont{L.}~\bibnamefont{Li}}, \bibnamefont{and} \bibinfo{author}{\bibfnamefont{S.}~\bibnamefont{Waskie}}, \bibinfo{journal}{Phys. Rev. D} \textbf{\bibinfo{volume}{97}}, \bibinfo{pages}{061901} (\bibinfo{year}{2018}), \eprint{1710.01326}.

\bibitem[{\citenamefont{Ahn et~al.}(2023)\citenamefont{Ahn, Baggioli, Jeong, and Kim}}]{Ahn:2023ciq}
\bibinfo{author}{\bibfnamefont{Y.}~\bibnamefont{Ahn}}, \bibinfo{author}{\bibfnamefont{M.}~\bibnamefont{Baggioli}}, \bibinfo{author}{\bibfnamefont{H.-S.} \bibnamefont{Jeong}}, \bibnamefont{and} \bibinfo{author}{\bibfnamefont{K.-Y.} \bibnamefont{Kim}}, \bibinfo{journal}{Phys. Rev. B} \textbf{\bibinfo{volume}{108}}, \bibinfo{pages}{235104} (\bibinfo{year}{2023}), \eprint{2307.04433}.

\bibitem[{\citenamefont{Xu et~al.}(2000)\citenamefont{Xu, Ong, Wang, Kakeshita, and Uchida}}]{xu2000vortex}
\bibinfo{author}{\bibfnamefont{Z.}~\bibnamefont{Xu}}, \bibinfo{author}{\bibfnamefont{N.}~\bibnamefont{Ong}}, \bibinfo{author}{\bibfnamefont{Y.}~\bibnamefont{Wang}}, \bibinfo{author}{\bibfnamefont{T.}~\bibnamefont{Kakeshita}}, \bibnamefont{and} \bibinfo{author}{\bibfnamefont{S.-i.} \bibnamefont{Uchida}}, \bibinfo{journal}{Nature} \textbf{\bibinfo{volume}{406}}, \bibinfo{pages}{486} (\bibinfo{year}{2000}).

\bibitem[{\citenamefont{Wang et~al.}(2005)\citenamefont{Wang, Li, Naughton, Gu, Uchida, and Ong}}]{wang2005field}
\bibinfo{author}{\bibfnamefont{Y.}~\bibnamefont{Wang}}, \bibinfo{author}{\bibfnamefont{L.}~\bibnamefont{Li}}, \bibinfo{author}{\bibfnamefont{M.}~\bibnamefont{Naughton}}, \bibinfo{author}{\bibfnamefont{G.}~\bibnamefont{Gu}}, \bibinfo{author}{\bibfnamefont{S.}~\bibnamefont{Uchida}}, \bibnamefont{and} \bibinfo{author}{\bibfnamefont{N.}~\bibnamefont{Ong}}, \bibinfo{journal}{Physical review letters} \textbf{\bibinfo{volume}{95}}, \bibinfo{pages}{247002} (\bibinfo{year}{2005}).

\bibitem[{\citenamefont{Li et~al.}(2007)\citenamefont{Li, Checkelsky, Komiya, Ando, and Ong}}]{li2007low}
\bibinfo{author}{\bibfnamefont{L.}~\bibnamefont{Li}}, \bibinfo{author}{\bibfnamefont{J.}~\bibnamefont{Checkelsky}}, \bibinfo{author}{\bibfnamefont{S.}~\bibnamefont{Komiya}}, \bibinfo{author}{\bibfnamefont{Y.}~\bibnamefont{Ando}}, \bibnamefont{and} \bibinfo{author}{\bibfnamefont{N.}~\bibnamefont{Ong}}, \bibinfo{journal}{Nature Physics} \textbf{\bibinfo{volume}{3}}, \bibinfo{pages}{311} (\bibinfo{year}{2007}).

\bibitem[{\citenamefont{Wang et~al.}(2006)\citenamefont{Wang, Li, and Ong}}]{wang2006nernst}
\bibinfo{author}{\bibfnamefont{Y.}~\bibnamefont{Wang}}, \bibinfo{author}{\bibfnamefont{L.}~\bibnamefont{Li}}, \bibnamefont{and} \bibinfo{author}{\bibfnamefont{N.}~\bibnamefont{Ong}}, \bibinfo{journal}{Physical Review B—Condensed Matter and Materials Physics} \textbf{\bibinfo{volume}{73}}, \bibinfo{pages}{024510} (\bibinfo{year}{2006}).

\bibitem[{\citenamefont{Oganesyan and Ussishkin}(2004)}]{PhysRevB.70.054503}
\bibinfo{author}{\bibfnamefont{V.}~\bibnamefont{Oganesyan}} \bibnamefont{and} \bibinfo{author}{\bibfnamefont{I.}~\bibnamefont{Ussishkin}}, \bibinfo{journal}{Phys. Rev. B} \textbf{\bibinfo{volume}{70}}, \bibinfo{pages}{054503} (\bibinfo{year}{2004}).

\bibitem[{\citenamefont{Tartaglione}(2022)}]{Tartaglione:2022zye}
\bibinfo{author}{\bibfnamefont{G.}~\bibnamefont{Tartaglione}}, Master's thesis, \bibinfo{school}{Padua U.} (\bibinfo{year}{2022}).

\bibitem[{\citenamefont{Rathi and Roychowdhury}(2023)}]{Rathi:2023vhw}
\bibinfo{author}{\bibfnamefont{H.}~\bibnamefont{Rathi}} \bibnamefont{and} \bibinfo{author}{\bibfnamefont{D.}~\bibnamefont{Roychowdhury}}, \bibinfo{journal}{JHEP} \textbf{\bibinfo{volume}{07}}, \bibinfo{pages}{026} (\bibinfo{year}{2023}), \eprint{2303.14379}.

\bibitem[{\citenamefont{Damle and Sachdev}(1997)}]{damle1997nonzero}
\bibinfo{author}{\bibfnamefont{K.}~\bibnamefont{Damle}} \bibnamefont{and} \bibinfo{author}{\bibfnamefont{S.}~\bibnamefont{Sachdev}}, \bibinfo{journal}{Physical Review B} \textbf{\bibinfo{volume}{56}}, \bibinfo{pages}{8714} (\bibinfo{year}{1997}).

\bibitem[{\citenamefont{Flores-Alfonso et~al.}(2021{\natexlab{a}})\citenamefont{Flores-Alfonso, Gonz\'alez-Morales, Linares, and Maceda}}]{Flores-Alfonso:2020euz}
\bibinfo{author}{\bibfnamefont{D.}~\bibnamefont{Flores-Alfonso}}, \bibinfo{author}{\bibfnamefont{B.~A.} \bibnamefont{Gonz\'alez-Morales}}, \bibinfo{author}{\bibfnamefont{R.}~\bibnamefont{Linares}}, \bibnamefont{and} \bibinfo{author}{\bibfnamefont{M.}~\bibnamefont{Maceda}}, \bibinfo{journal}{Phys. Lett. B} \textbf{\bibinfo{volume}{812}}, \bibinfo{pages}{136011} (\bibinfo{year}{2021}{\natexlab{a}}), \eprint{2011.10836}.

\bibitem[{\citenamefont{Amirabi and Habib~Mazharimousavi}(2021)}]{Amirabi:2020mzv}
\bibinfo{author}{\bibfnamefont{Z.}~\bibnamefont{Amirabi}} \bibnamefont{and} \bibinfo{author}{\bibfnamefont{S.}~\bibnamefont{Habib~Mazharimousavi}}, \bibinfo{journal}{Eur. Phys. J. C} \textbf{\bibinfo{volume}{81}}, \bibinfo{pages}{207} (\bibinfo{year}{2021}), \eprint{2012.07443}.

\bibitem[{\citenamefont{Barrientos et~al.}(2025)\citenamefont{Barrientos, Cisterna, Hassaine, and Pallikaris}}]{Barrientos:2024umq}
\bibinfo{author}{\bibfnamefont{J.}~\bibnamefont{Barrientos}}, \bibinfo{author}{\bibfnamefont{A.}~\bibnamefont{Cisterna}}, \bibinfo{author}{\bibfnamefont{M.}~\bibnamefont{Hassaine}}, \bibnamefont{and} \bibinfo{author}{\bibfnamefont{K.}~\bibnamefont{Pallikaris}}, \bibinfo{journal}{Phys. Lett. B} \textbf{\bibinfo{volume}{860}}, \bibinfo{pages}{139214} (\bibinfo{year}{2025}), \eprint{2409.12336}.

\bibitem[{\citenamefont{de~Haro et~al.}(2001)\citenamefont{de~Haro, Solodukhin, and Skenderis}}]{deHaro:2000vlm}
\bibinfo{author}{\bibfnamefont{S.}~\bibnamefont{de~Haro}}, \bibinfo{author}{\bibfnamefont{S.~N.} \bibnamefont{Solodukhin}}, \bibnamefont{and} \bibinfo{author}{\bibfnamefont{K.}~\bibnamefont{Skenderis}}, \bibinfo{journal}{Commun. Math. Phys.} \textbf{\bibinfo{volume}{217}}, \bibinfo{pages}{595} (\bibinfo{year}{2001}), \eprint{hep-th/0002230}.

\bibitem[{\citenamefont{Ferrero et~al.}(2021)\citenamefont{Ferrero, Gauntlett, Ipi\~na, Martelli, and Sparks}}]{Ferrero:2020twa}
\bibinfo{author}{\bibfnamefont{P.}~\bibnamefont{Ferrero}}, \bibinfo{author}{\bibfnamefont{J.~P.} \bibnamefont{Gauntlett}}, \bibinfo{author}{\bibfnamefont{J.~M.~P.} \bibnamefont{Ipi\~na}}, \bibinfo{author}{\bibfnamefont{D.}~\bibnamefont{Martelli}}, \bibnamefont{and} \bibinfo{author}{\bibfnamefont{J.}~\bibnamefont{Sparks}}, \bibinfo{journal}{Phys. Rev. D} \textbf{\bibinfo{volume}{104}}, \bibinfo{pages}{046007} (\bibinfo{year}{2021}), \eprint{2012.08530}.

\bibitem[{\citenamefont{Myers}(1999)}]{Myers:1999psa}
\bibinfo{author}{\bibfnamefont{R.~C.} \bibnamefont{Myers}}, \bibinfo{journal}{Phys. Rev. D} \textbf{\bibinfo{volume}{60}}, \bibinfo{pages}{046002} (\bibinfo{year}{1999}), \eprint{hep-th/9903203}.

\bibitem[{\citenamefont{Ashtekar and Das}(2000)}]{Ashtekar:1999jx}
\bibinfo{author}{\bibfnamefont{A.}~\bibnamefont{Ashtekar}} \bibnamefont{and} \bibinfo{author}{\bibfnamefont{S.}~\bibnamefont{Das}}, \bibinfo{journal}{Class. Quant. Grav.} \textbf{\bibinfo{volume}{17}}, \bibinfo{pages}{L17} (\bibinfo{year}{2000}), \eprint{hep-th/9911230}.

\bibitem[{\citenamefont{Wald}(1974)}]{Wald:1974np}
\bibinfo{author}{\bibfnamefont{R.~M.} \bibnamefont{Wald}}, \bibinfo{journal}{Phys. Rev. D} \textbf{\bibinfo{volume}{10}}, \bibinfo{pages}{1680} (\bibinfo{year}{1974}).

\bibitem[{\citenamefont{Kastor et~al.}(2009)\citenamefont{Kastor, Ray, and Traschen}}]{Kastor:2009wy}
\bibinfo{author}{\bibfnamefont{D.}~\bibnamefont{Kastor}}, \bibinfo{author}{\bibfnamefont{S.}~\bibnamefont{Ray}}, \bibnamefont{and} \bibinfo{author}{\bibfnamefont{J.}~\bibnamefont{Traschen}}, \bibinfo{journal}{Class. Quant. Grav.} \textbf{\bibinfo{volume}{26}}, \bibinfo{pages}{195011} (\bibinfo{year}{2009}), \eprint{0904.2765}.

\bibitem[{\citenamefont{Caldarelli et~al.}(2000)\citenamefont{Caldarelli, Cognola, and Klemm}}]{Caldarelli:1999xj}
\bibinfo{author}{\bibfnamefont{M.~M.} \bibnamefont{Caldarelli}}, \bibinfo{author}{\bibfnamefont{G.}~\bibnamefont{Cognola}}, \bibnamefont{and} \bibinfo{author}{\bibfnamefont{D.}~\bibnamefont{Klemm}}, \bibinfo{journal}{Class. Quant. Grav.} \textbf{\bibinfo{volume}{17}}, \bibinfo{pages}{399} (\bibinfo{year}{2000}), \eprint{hep-th/9908022}.

\bibitem[{\citenamefont{Barrientos et~al.}(2022)\citenamefont{Barrientos, Cisterna, Kubiznak, and Oliva}}]{Barrientos:2022bzm}
\bibinfo{author}{\bibfnamefont{J.}~\bibnamefont{Barrientos}}, \bibinfo{author}{\bibfnamefont{A.}~\bibnamefont{Cisterna}}, \bibinfo{author}{\bibfnamefont{D.}~\bibnamefont{Kubiznak}}, \bibnamefont{and} \bibinfo{author}{\bibfnamefont{J.}~\bibnamefont{Oliva}}, \bibinfo{journal}{Phys. Lett. B} \textbf{\bibinfo{volume}{834}}, \bibinfo{pages}{137447} (\bibinfo{year}{2022}), \eprint{2205.15777}.

\bibitem[{\citenamefont{Chamblin et~al.}(1999)\citenamefont{Chamblin, Emparan, Johnson, and Myers}}]{Chamblin:1999tk}
\bibinfo{author}{\bibfnamefont{A.}~\bibnamefont{Chamblin}}, \bibinfo{author}{\bibfnamefont{R.}~\bibnamefont{Emparan}}, \bibinfo{author}{\bibfnamefont{C.~V.} \bibnamefont{Johnson}}, \bibnamefont{and} \bibinfo{author}{\bibfnamefont{R.~C.} \bibnamefont{Myers}}, \bibinfo{journal}{Phys. Rev. D} \textbf{\bibinfo{volume}{60}}, \bibinfo{pages}{064018} (\bibinfo{year}{1999}), \eprint{hep-th/9902170}.

\bibitem[{\citenamefont{Johnston}(2014)}]{10.1088/978-1-627-05532-1}
\bibinfo{author}{\bibfnamefont{D.~C.} \bibnamefont{Johnston}}, \emph{\bibinfo{title}{Advances in Thermodynamics of the van der Waals Fluid}} (\bibinfo{publisher}{Morgan \& Claypool Publishers}, \bibinfo{year}{2014}).

\bibitem[{\citenamefont{\"Okc\"u and Ayd\i{}ner}(2017)}]{Okcu:2016tgt}
\bibinfo{author}{\bibfnamefont{O.}~\bibnamefont{\"Okc\"u}} \bibnamefont{and} \bibinfo{author}{\bibfnamefont{E.}~\bibnamefont{Ayd\i{}ner}}, \bibinfo{journal}{Eur. Phys. J. C} \textbf{\bibinfo{volume}{77}}, \bibinfo{pages}{24} (\bibinfo{year}{2017}), \eprint{1611.06327}.

\bibitem[{\citenamefont{\"Okc\"u and Ayd\i{}ner}(2018)}]{Okcu:2017qgo}
\bibinfo{author}{\bibfnamefont{O.}~\bibnamefont{\"Okc\"u}} \bibnamefont{and} \bibinfo{author}{\bibfnamefont{E.}~\bibnamefont{Ayd\i{}ner}}, \bibinfo{journal}{Eur. Phys. J. C} \textbf{\bibinfo{volume}{78}}, \bibinfo{pages}{123} (\bibinfo{year}{2018}), \eprint{1709.06426}.

\bibitem[{\citenamefont{Mo and Li}(2020)}]{Mo:2018qkt}
\bibinfo{author}{\bibfnamefont{J.-X.} \bibnamefont{Mo}} \bibnamefont{and} \bibinfo{author}{\bibfnamefont{G.-Q.} \bibnamefont{Li}}, \bibinfo{journal}{Class. Quant. Grav.} \textbf{\bibinfo{volume}{37}}, \bibinfo{pages}{045009} (\bibinfo{year}{2020}), \eprint{1805.04327}.

\bibitem[{\citenamefont{Cisterna et~al.}(2019{\natexlab{b}})\citenamefont{Cisterna, Hu, and Kuang}}]{Cisterna:2018jqg}
\bibinfo{author}{\bibfnamefont{A.}~\bibnamefont{Cisterna}}, \bibinfo{author}{\bibfnamefont{S.-Q.} \bibnamefont{Hu}}, \bibnamefont{and} \bibinfo{author}{\bibfnamefont{X.-M.} \bibnamefont{Kuang}}, \bibinfo{journal}{Phys. Lett. B} \textbf{\bibinfo{volume}{797}}, \bibinfo{pages}{134883} (\bibinfo{year}{2019}{\natexlab{b}}), \eprint{1808.07392}.

\bibitem[{\citenamefont{Barrientos and Mena}(2022)}]{Barrientos:2022uit}
\bibinfo{author}{\bibfnamefont{J.}~\bibnamefont{Barrientos}} \bibnamefont{and} \bibinfo{author}{\bibfnamefont{J.}~\bibnamefont{Mena}}, \bibinfo{journal}{Phys. Rev. D} \textbf{\bibinfo{volume}{106}}, \bibinfo{pages}{044064} (\bibinfo{year}{2022}), \eprint{2206.06018}.

\bibitem[{\citenamefont{Reif}(1965)}]{REI65}
\bibinfo{author}{\bibfnamefont{F.}~\bibnamefont{Reif}}, \emph{\bibinfo{title}{Fundamentals of Statistical and Thermal Physics}} (\bibinfo{publisher}{McGraw Hill}, \bibinfo{address}{Tokyo}, \bibinfo{year}{1965}).

\bibitem[{\citenamefont{Donos and Gauntlett}(2014{\natexlab{c}})}]{Donos:2014cya}
\bibinfo{author}{\bibfnamefont{A.}~\bibnamefont{Donos}} \bibnamefont{and} \bibinfo{author}{\bibfnamefont{J.~P.} \bibnamefont{Gauntlett}}, \bibinfo{journal}{JHEP} \textbf{\bibinfo{volume}{11}}, \bibinfo{pages}{081} (\bibinfo{year}{2014}{\natexlab{c}}), \eprint{1406.4742}.

\bibitem[{\citenamefont{Witten}(2001)}]{Witten:2001ua}
\bibinfo{author}{\bibfnamefont{E.}~\bibnamefont{Witten}} (\bibinfo{year}{2001}), \eprint{hep-th/0112258}.

\bibitem[{\citenamefont{Papadimitriou}(2007)}]{Papadimitriou:2007sj}
\bibinfo{author}{\bibfnamefont{I.}~\bibnamefont{Papadimitriou}}, \bibinfo{journal}{JHEP} \textbf{\bibinfo{volume}{05}}, \bibinfo{pages}{075} (\bibinfo{year}{2007}), \eprint{hep-th/0703152}.

\bibitem[{\citenamefont{Anabalon et~al.}(2016)\citenamefont{Anabalon, Astefanesei, Choque, and Martinez}}]{Anabalon:2015xvl}
\bibinfo{author}{\bibfnamefont{A.}~\bibnamefont{Anabalon}}, \bibinfo{author}{\bibfnamefont{D.}~\bibnamefont{Astefanesei}}, \bibinfo{author}{\bibfnamefont{D.}~\bibnamefont{Choque}}, \bibnamefont{and} \bibinfo{author}{\bibfnamefont{C.}~\bibnamefont{Martinez}}, \bibinfo{journal}{JHEP} \textbf{\bibinfo{volume}{03}}, \bibinfo{pages}{117} (\bibinfo{year}{2016}), \eprint{1511.08759}.

\bibitem[{\citenamefont{Caceres et~al.}(2024)\citenamefont{Caceres, Corral, Diaz, and Olea}}]{Caceres:2023gfa}
\bibinfo{author}{\bibfnamefont{N.}~\bibnamefont{Caceres}}, \bibinfo{author}{\bibfnamefont{C.}~\bibnamefont{Corral}}, \bibinfo{author}{\bibfnamefont{F.}~\bibnamefont{Diaz}}, \bibnamefont{and} \bibinfo{author}{\bibfnamefont{R.}~\bibnamefont{Olea}}, \bibinfo{journal}{JHEP} \textbf{\bibinfo{volume}{05}}, \bibinfo{pages}{125} (\bibinfo{year}{2024}), \eprint{2311.04054}.

\bibitem[{\citenamefont{Kontani}(2002)}]{kontani2002nernst}
\bibinfo{author}{\bibfnamefont{H.}~\bibnamefont{Kontani}}, \bibinfo{journal}{arXiv preprint cond-mat/0204193}  (\bibinfo{year}{2002}).

\bibitem[{\citenamefont{Hernandez and Kovtun}(2017)}]{Hernandez:2017mch}
\bibinfo{author}{\bibfnamefont{J.}~\bibnamefont{Hernandez}} \bibnamefont{and} \bibinfo{author}{\bibfnamefont{P.}~\bibnamefont{Kovtun}}, \bibinfo{journal}{JHEP} \textbf{\bibinfo{volume}{05}}, \bibinfo{pages}{001} (\bibinfo{year}{2017}), \eprint{1703.08757}.

\bibitem[{\citenamefont{Amoretti and Brattan}(2022)}]{Amoretti:2022acb}
\bibinfo{author}{\bibfnamefont{A.}~\bibnamefont{Amoretti}} \bibnamefont{and} \bibinfo{author}{\bibfnamefont{D.~K.} \bibnamefont{Brattan}}, \bibinfo{journal}{Mod. Phys. Lett. A} \textbf{\bibinfo{volume}{37}}, \bibinfo{pages}{2230010} (\bibinfo{year}{2022}), \eprint{2209.11589}.

\bibitem[{\citenamefont{Behnia}(2022)}]{behnia2022nernst}
\bibinfo{author}{\bibfnamefont{K.}~\bibnamefont{Behnia}}, \bibinfo{journal}{Journal of Physics: Condensed Matter} \textbf{\bibinfo{volume}{35}}, \bibinfo{pages}{074003} (\bibinfo{year}{2022}).

\bibitem[{\citenamefont{Flores-Alfonso et~al.}(2021{\natexlab{b}})\citenamefont{Flores-Alfonso, Linares, and Maceda}}]{Flores-Alfonso:2020nnd}
\bibinfo{author}{\bibfnamefont{D.}~\bibnamefont{Flores-Alfonso}}, \bibinfo{author}{\bibfnamefont{R.}~\bibnamefont{Linares}}, \bibnamefont{and} \bibinfo{author}{\bibfnamefont{M.}~\bibnamefont{Maceda}}, \bibinfo{journal}{JHEP} \textbf{\bibinfo{volume}{09}}, \bibinfo{pages}{104} (\bibinfo{year}{2021}{\natexlab{b}}), \eprint{2012.03416}.

\bibitem[{\citenamefont{Ballon~Bordo et~al.}(2021)\citenamefont{Ballon~Bordo, Kubiz\v{n}\'ak, and Perche}}]{BallonBordo:2020jtw}
\bibinfo{author}{\bibfnamefont{A.}~\bibnamefont{Ballon~Bordo}}, \bibinfo{author}{\bibfnamefont{D.}~\bibnamefont{Kubiz\v{n}\'ak}}, \bibnamefont{and} \bibinfo{author}{\bibfnamefont{T.~R.} \bibnamefont{Perche}}, \bibinfo{journal}{Phys. Lett. B} \textbf{\bibinfo{volume}{817}}, \bibinfo{pages}{136312} (\bibinfo{year}{2021}), \eprint{2011.13398}.

\bibitem[{\citenamefont{Ay\'on-Beato et~al.}(2024)\citenamefont{Ay\'on-Beato, Flores-Alfonso, and Hassaine}}]{Ayon-Beato:2024vph}
\bibinfo{author}{\bibfnamefont{E.}~\bibnamefont{Ay\'on-Beato}}, \bibinfo{author}{\bibfnamefont{D.}~\bibnamefont{Flores-Alfonso}}, \bibnamefont{and} \bibinfo{author}{\bibfnamefont{M.}~\bibnamefont{Hassaine}}, \bibinfo{journal}{Phys. Rev. D} \textbf{\bibinfo{volume}{110}}, \bibinfo{pages}{064027} (\bibinfo{year}{2024}), \eprint{2404.08753}.

\bibitem[{\citenamefont{Meert and Nastase}(2024)}]{Meert:2024dud}
\bibinfo{author}{\bibfnamefont{P.}~\bibnamefont{Meert}} \bibnamefont{and} \bibinfo{author}{\bibfnamefont{H.}~\bibnamefont{Nastase}}, \bibinfo{journal}{Phys. Rev. D} \textbf{\bibinfo{volume}{110}}, \bibinfo{pages}{046004} (\bibinfo{year}{2024}), \eprint{2402.04194}.

\bibitem[{\citenamefont{Hale et~al.}(2025)\citenamefont{Hale, Kubiz\v{n}\'ak, Men\v{s}\'\i{}kov\'a, Mann, and Yang}}]{Hale:2025veb}
\bibinfo{author}{\bibfnamefont{T.}~\bibnamefont{Hale}}, \bibinfo{author}{\bibfnamefont{D.}~\bibnamefont{Kubiz\v{n}\'ak}}, \bibinfo{author}{\bibfnamefont{J.}~\bibnamefont{Men\v{s}\'\i{}kov\'a}}, \bibinfo{author}{\bibfnamefont{R.~B.} \bibnamefont{Mann}}, \bibnamefont{and} \bibinfo{author}{\bibfnamefont{J.}~\bibnamefont{Yang}}, \bibinfo{journal}{Phys. Rev. D} \textbf{\bibinfo{volume}{111}}, \bibinfo{pages}{104004} (\bibinfo{year}{2025}), \eprint{2501.13679}.

\bibitem[{\citenamefont{Arenas-Henriquez et~al.}(2022)\citenamefont{Arenas-Henriquez, Gregory, and Scoins}}]{Arenas-Henriquez:2022www}
\bibinfo{author}{\bibfnamefont{G.}~\bibnamefont{Arenas-Henriquez}}, \bibinfo{author}{\bibfnamefont{R.}~\bibnamefont{Gregory}}, \bibnamefont{and} \bibinfo{author}{\bibfnamefont{A.}~\bibnamefont{Scoins}}, \bibinfo{journal}{JHEP} \textbf{\bibinfo{volume}{05}}, \bibinfo{pages}{063} (\bibinfo{year}{2022}), \eprint{2202.08823}.

\bibitem[{\citenamefont{Arenas-Henriquez et~al.}(2023{\natexlab{b}})\citenamefont{Arenas-Henriquez, Cisterna, Diaz, and Gregory}}]{Arenas-Henriquez:2023hur}
\bibinfo{author}{\bibfnamefont{G.}~\bibnamefont{Arenas-Henriquez}}, \bibinfo{author}{\bibfnamefont{A.}~\bibnamefont{Cisterna}}, \bibinfo{author}{\bibfnamefont{F.}~\bibnamefont{Diaz}}, \bibnamefont{and} \bibinfo{author}{\bibfnamefont{R.}~\bibnamefont{Gregory}}, \bibinfo{journal}{JHEP} \textbf{\bibinfo{volume}{09}}, \bibinfo{pages}{122} (\bibinfo{year}{2023}{\natexlab{b}}), \eprint{2308.00613}.

\bibitem[{\citenamefont{Cisterna et~al.}(2023)\citenamefont{Cisterna, Diaz, Mann, and Oliva}}]{Cisterna:2023qhh}
\bibinfo{author}{\bibfnamefont{A.}~\bibnamefont{Cisterna}}, \bibinfo{author}{\bibfnamefont{F.}~\bibnamefont{Diaz}}, \bibinfo{author}{\bibfnamefont{R.~B.} \bibnamefont{Mann}}, \bibnamefont{and} \bibinfo{author}{\bibfnamefont{J.}~\bibnamefont{Oliva}}, \bibinfo{journal}{JHEP} \textbf{\bibinfo{volume}{11}}, \bibinfo{pages}{073} (\bibinfo{year}{2023}), \eprint{2309.05559}.

\bibitem[{\citenamefont{Tian and Lai}(2023)}]{Tian:2023ine}
\bibinfo{author}{\bibfnamefont{J.}~\bibnamefont{Tian}} \bibnamefont{and} \bibinfo{author}{\bibfnamefont{T.}~\bibnamefont{Lai}} (\bibinfo{year}{2023}), \eprint{2312.13718}.

\bibitem[{\citenamefont{Bunney and Mann}(2025)}]{Bunney:2024xic}
\bibinfo{author}{\bibfnamefont{C.~R.~D.} \bibnamefont{Bunney}} \bibnamefont{and} \bibinfo{author}{\bibfnamefont{R.~B.} \bibnamefont{Mann}}, \bibinfo{journal}{Class. Quant. Grav.} \textbf{\bibinfo{volume}{42}}, \bibinfo{pages}{075001} (\bibinfo{year}{2025}), \eprint{2410.19677}.

\bibitem[{\citenamefont{Arenas-Henriquez et~al.}(2025)\citenamefont{Arenas-Henriquez, Diaz, and Rivera-Betancour}}]{Arenas-Henriquez:2024ypo}
\bibinfo{author}{\bibfnamefont{G.}~\bibnamefont{Arenas-Henriquez}}, \bibinfo{author}{\bibfnamefont{F.}~\bibnamefont{Diaz}}, \bibnamefont{and} \bibinfo{author}{\bibfnamefont{D.}~\bibnamefont{Rivera-Betancour}}, \bibinfo{journal}{JHEP} \textbf{\bibinfo{volume}{02}}, \bibinfo{pages}{007} (\bibinfo{year}{2025}), \eprint{2411.12513}.

\bibitem[{\citenamefont{Luo}(2024)}]{Luo:2024cwm}
\bibinfo{author}{\bibfnamefont{S.}~\bibnamefont{Luo}} (\bibinfo{year}{2024}), \eprint{2412.21013}.

\bibitem[{\citenamefont{Li et~al.}(2025)\citenamefont{Li, Li, and Tian}}]{Li:2025rzl}
\bibinfo{author}{\bibfnamefont{Z.}~\bibnamefont{Li}}, \bibinfo{author}{\bibfnamefont{Z.}~\bibnamefont{Li}}, \bibnamefont{and} \bibinfo{author}{\bibfnamefont{J.}~\bibnamefont{Tian}}, \bibinfo{journal}{JHEP} \textbf{\bibinfo{volume}{05}}, \bibinfo{pages}{161} (\bibinfo{year}{2025}), \eprint{2502.11135}.

\bibitem[{\citenamefont{Roychowdhury}(2024)}]{Roychowdhury:2024oih}
\bibinfo{author}{\bibfnamefont{D.}~\bibnamefont{Roychowdhury}} (\bibinfo{year}{2024}), \eprint{2411.15474}.

\bibitem[{\citenamefont{Bel et~al.}(2003)\citenamefont{Bel, Behnia, Nakajima, Izawa, Matsuda, Shishido, Settai, and Onuki}}]{bel2003giant}
\bibinfo{author}{\bibfnamefont{R.}~\bibnamefont{Bel}}, \bibinfo{author}{\bibfnamefont{K.}~\bibnamefont{Behnia}}, \bibinfo{author}{\bibfnamefont{Y.}~\bibnamefont{Nakajima}}, \bibinfo{author}{\bibfnamefont{K.}~\bibnamefont{Izawa}}, \bibinfo{author}{\bibfnamefont{Y.}~\bibnamefont{Matsuda}}, \bibinfo{author}{\bibfnamefont{H.}~\bibnamefont{Shishido}}, \bibinfo{author}{\bibfnamefont{R.}~\bibnamefont{Settai}}, \bibnamefont{and} \bibinfo{author}{\bibfnamefont{Y.}~\bibnamefont{Onuki}}, \bibinfo{journal}{arXiv preprint cond-mat/0311473}  (\bibinfo{year}{2003}).

\end{thebibliography}
\end{document}